\documentclass[10pt,twocolumn,english,aps,showpacs,preprintnumbers,nofootinbib,prd,superscriptaddress,groupedaddress]{revtex4-2}
\usepackage{beramono}
\usepackage[T1]{fontenc}
\usepackage[latin9]{inputenc}
\usepackage{color}
\usepackage{babel}
\usepackage{verbatim}
\usepackage{mathrsfs}
\usepackage{mathtools}
\usepackage{bm}
\usepackage{amsmath}
\usepackage{amsthm}
\usepackage{amssymb}
\usepackage{graphicx}
\usepackage{esint}
\usepackage[unicode=true,
 bookmarks=false,
 breaklinks=false,pdfborder={0 0 1},backref=false,colorlinks=true]
 {hyperref}
\hypersetup{
 citecolor=darkblue,linkcolor=darkblue}

\makeatletter
\usepackage{babel}
\usepackage{amsthm}
\usepackage{amsfonts}
\usepackage{epsfig}
\usepackage{times}

\usepackage{epstopdf}
\usepackage{aas_macros}
\usepackage{tensor}
\usepackage{bm}
\usepackage{url}\usepackage{wasysym}

\@ifundefined{textcolor}{}{%
 \definecolor{BLACK}{gray}{0}
 \definecolor{WHITE}{gray}{1}
 \definecolor{RED}{rgb}{1,0,0}
 \definecolor{GREEN}{rgb}{0,1,0}
 \definecolor{BLUE}{rgb}{0,0,1}
 \definecolor{CYAN}{cmyk}{1,0,0,0}
 \definecolor{MAGENTA}{cmyk}{0,1,0,0}
 \definecolor{YELLOW}{cmyk}{0,0,1,0}
}
\definecolor{darkblue}{rgb}{0,0,0.5}
\usepackage{soul}

\usepackage{aecompl}

\def\rc{r}
\def\re{\varrho}
\def\rigid{\chi}

\makeatother

\begin{document}
\title{Reference frames in general relativity and the galactic rotation curves}
\author{L.~Filipe~O.~Costa}
\email{lfilipecosta@tecnico.ulisboa.pt}

\affiliation{CAMGSD, Departamento de Matemática, Instituto Superior Técnico, 1049-001
Lisboa, Portugal}
\author{F. Frutos-Alfaro}
\email{francisco.frutos@ucr.ac.cr}

\affiliation{Space Research Center (CINESPA), School of Physics, University of
Costa Rica, 11501 San José, Costa Rica.}
\author{José~Natário}
\email{jnatar@math.ist.utl.pt}

\affiliation{CAMGSD, Departamento de Matemática, Instituto Superior Técnico, 1049-001
Lisboa, Portugal}
\author{Michael Soffel}
\email{michael.soffel@tu-dresden.de}

\affiliation{Lohrmann Observatory, Dresden Technical University, 01062 Dresden,
Germany}
\date{\today}
\begin{abstract}
The physical interpretation of the exact solutions of the Einstein
field equations is, in general, a challenging task, part of the difficulties
lying in the significance of the coordinate system. We discuss the
extension of the International Astronomical Union (IAU) reference
system to the exact theory. It is seen that such an extension, retaining
some of its crucial properties, can be achieved in a special class
of spacetimes, admitting nonshearing congruences of observers which,
at infinity, have zero vorticity and acceleration. As applications,
we consider the Friedmann--Lemaître--Robertson--Walker (FLRW),
Kerr and Newman--Unti--Tamburino (NUT) spacetimes, the van Stockum
rotating dust cylinder, spinning cosmic strings and, finally, we debunk
the so-called Balasin-Grumiller (BG) model, and the claims that the
galaxies' rotation curves can be explained through gravitomagnetic
effects without the need for dark matter. The BG spacetime is shown
to be completely inappropriate as a galactic model: its dust is actually
\emph{static} with respect to the asymptotic inertial frame, its gravitomagnetic
effects arise from unphysical singularities along the axis (a pair
of NUT rods, combined with a spinning cosmic string), and the rotation
curves obtained are merely down to an invalid choice of reference
frame --- the congruence of zero angular momentum observers, which
are being dragged by the singularities. 
\end{abstract}
\maketitle
\tableofcontents{}

\section{Introduction}

As it replaced Newtonian mechanics as the state of the art theory
of gravitation, general relativity brought along equations allowing
for a more precise description of gravitational phenomena, at the
cost, however, of a mathematical complexity effectively preventing
their full use in actual astrophysical scenarios. Since then, two
fields have evolved parallelly in a largely separate way: one, finding
and mathematically characterizing exact solutions of the Einstein
field equations \cite{StephaniExact,GriffithsPodolsky2009,BicakSolutionsEFE},
which in most cases do not correspond to realistic physical systems
(or whose physical significance is not totally clear); and the other,
approximate methods such as post-Newtonian theory, for actual astrophysics.
Examples of the former are, besides the best understood black hole
and cosmological exact solutions, most instances of the Pleba\'{n}ski-Demia\'{n}sky
solutions, such as the \emph{C}-metric \cite{GriffithsKrtousPodolskyCMetric,GriffithsPodolsky2009,WyllemanBeke2010}
(interpreted as pair of accelerating black holes, but possessing both
radiative and static regions with a challenging interpretation); the
van Stockum \cite{Stockum1938}, Levi-Civita and Lewis metrics \cite{Lewis:1932,SantosGRG1995,MacCallumSantos1998,GriffithsPodolsky2009,Bronnikov:2019clf,Cilindros}
and their relationships (in some limits known to describe the field
of infinite cylinders, not so clear in others \cite{GriffithsPodolsky2009},
in all cases exhibiting perplexing features), the Gödel solution \cite{GodelRotatingUniverses}
(with its counterintuitive homogeneous rotation and anti-Machian features),
the NUT spacetime (with its gravitomagnetic monopole, and the different
versions of its line singularity, whose physical interpretation is
an open question) and, more recently, the so-called Balasin-Grumiller
(BG) model \cite{BG}, and the extraordinary claims that it can partially,
or even totally explain the galaxies' flat rotation curves without
the need for dark matter \cite{BG,Crosta2018,RuggieroBG}. Part of
the difficulties lie in the coordinate system and its interpretation,
as seen to be the case, for instance, in the Lewis-Weyl \cite{Cilindros}
or in the \emph{C-}Metric \cite{GriffithsKrtousPodolskyCMetric}.
Even in the simplest solutions, such as de Sitter universe, which
is maximally symmetric and represents a homogeneous isotropic expanding
universe, subtleties arise: it is well known to admit (within a cosmological
horizon) a coordinate system where it is explicitly time-independent
(being thus static therein), as well as to take an anisotropic form
in other standard coordinate systems \cite{GriffithsPodolsky2009}.
Setting up meaningful reference frames is crucial to make sense of
the solutions, and understand whether they can be reasonable models
for some astrophysical settings. We shall see that its misunderstanding
is indeed also at the origin the crucial misconception leading to
the BG galactic model.

These issues do not arise, on the other hand, in post-Newtonian theory
\cite{Damour:1990pi,Will:1993ns,WillPoissonBook,Soffel2009IAU,Kaplan:2009,SoffelBook2009},
equipped with reference frames (``PN frames'') as close as possible
to inertial, and having axes fixed with respect to distant reference
objects (either stars or quasars). By being a weak field and slow
motion approximation, its regime of validity is however limited, and
the way it is usually formulated relies moreover on asymptotic flatness.
Indeed, gravitational effects outside the realm of validity of such
approximation are important in an observational context. The state
of the art ICRS reference system uses quasars as reference objects,
which are at a distance where cosmological expansion becomes important.
The advent of gravitational wave detectors will also expose to observational
scrutiny strong field effects such as those in the latter stage of
black hole or neutron star inspirals, global gravitational effects
(sometimes regarded as topological defects \cite{KibbleCosmicStrings,VilenkinShellard_Book})
such as those produced by hypothetical cosmic strings, whose detection
can be within the reach of LISA and pulsar timing arrays \cite{EllisLewickiString,Auclair_LISAS_strings,Boileau_LISA_string},
or gravitomagnetic monopoles%
, which are not described by asymptotically flat metrics.

Here we discuss the possible generalization of the IAU system to the
exact theory. We shall see that, even though not possible generically,
suitable reference frames retaining important properties of the PN
frames --- namely being nonshearing, and having axes fixed do distant
reference objects --- can be achieved in a special class of spacetimes:
those admitting shear-free and \emph{asymptotically} vorticity-free
congruences of timelike curves (observers). It encompasses all stationary
spacetimes admitting asymptotically inertial Killing congruences,
but also expanding, nonstationary solutions. Crucially, asymptotic
flatness is not required.

\emph{Notation and conventions.---} We use the signature $(-+++)$;\textcolor{black}{{}
Greek letters $\alpha$, $\beta$, $\gamma$, ... denote 4D spacetime
indices, running 0-3; Roman letters $i,j,k,...$ are spatial indices,
running 1-3}; $\epsilon_{\alpha\beta\gamma\delta}\equiv\sqrt{-g}[\alpha\beta\gamma\delta]$
is the 4-D Levi-Civita tensor, with the orientation $[1230]=1$ (in
flat spacetime, $\epsilon_{1230}=1$); $\epsilon_{ijk}\equiv\sqrt{h}[ijk]$\textcolor{black}{{}
is the} Levi-Civita tensor in a 3-D Riemannian manifold of metric
$h_{ij}$. Our convention for the Riemann tensor is $R_{\ \beta\mu\nu}^{\alpha}=\Gamma_{\beta\nu,\mu}^{\alpha}-\Gamma_{\beta\mu,\nu}^{\alpha}+...$
. $\star$ denotes the Hodge dual (e.g. $\star F_{\alpha\beta}\equiv\epsilon_{\alpha\beta}^{\ \ \ \mu\nu}F_{\mu\nu}/2$,
for a 2-form $F_{\alpha\beta}=F_{[\alpha\beta]}$). The basis vector
corresponding to a coordinate $\phi$ is denoted by $\partial_{\phi}\equiv\partial/\partial\phi$,
and its $\alpha$-component by $\partial_{\phi}^{\alpha}\equiv\delta_{\phi}^{\alpha}$.

\section{PN approximation: the IAU reference systems}

The metric describing the gravitational field of a system of $N$
gravitationally interacting rotating bodies of arbitrary shape and
composition can be written, at first post-Newtonian (PN) order and
in geometrized units, as \cite{Damour:1990pi,Soffel2009IAU,Kaplan:2009}
\begin{align}
g_{00} & =-1+2w-2w^{2}+O(6)\ ;\nonumber \\
g_{i0} & =\mathcal{A}_{i}+O(5)\ ;\qquad g_{ij}=\delta_{ij}\left(1+2w\right)+O(4)\ ,\label{eq:PNmetric}
\end{align}
where $O(n)\equiv O(\epsilon^{n})$, $\epsilon$ is a small \emph{dimensionless}
parameter such that $U\sim\epsilon^{2}$, $U$ is minus the Newtonian
potential, and $w=U+O(4)$ consists of the sum of the Newtonian potential
$U$ plus \emph{nonlinear} terms of order $\epsilon^{4}$. The bodies'
velocities are assumed such that $v\lesssim\epsilon$ (since, for
bounded orbits, $v\sim\sqrt{U}$), and time derivatives increase the
degree of smallness of a quantity by a factor $\epsilon$; for example,
$\partial U/\partial t\sim Uv\sim\epsilon U$.

The coordinate system associated to the metric \eqref{eq:PNmetric}
is the basis of the IAU reference system \cite{Soffel2009IAU,Kopeikin2006RefFrames,KopeikinIAU2009,SoffelBook2009,soffel_Book_ReferenceSystems}.
The metric is assumed asymptotically flat, $\lim_{r\rightarrow\infty}g_{\alpha\beta}=\eta_{\alpha\beta}$,
so that the coordinate system is inertial at infinity, and the spatial
coordinate basis vectors $\partial_{i}$ have directions fixed (i.e.,
are ``rotationally'' locked \cite{Misner:1974qy}) to distant reference
objects, namely the ``fundamental'' stars (defining the axes of
the Astrographic Catalog of Reference Stars---ACRS) or extragalactic
radio sources (mainly quasars, which define the axis of the International
Celestial Reference System---ICRS) \cite{GalaticAstronomy,Soffel_et_al_RelativisticAstrometry1985,Soffel1989book,SoffelBook2009,soffel_Book_ReferenceSystems,kovalevskyetal_ReferenceFrames,Soffel_Brumberg_QA,Kopeikin2006RefFrames,KopeikinIAU2009,LindegrenGaiaFrame}.
The latter is the state of the art system, since, by being so far
away, such extragalactic sources exhibit no detectable angular motion
in the sky, to present accuracy.

It should be noted however that, at cosmological scales, the expansion
of the universe comes into play, and the asymptotic flatness assumption
breaks down. Modifications of the IAU system for accommodating cosmological
effects have been proposed in \cite{KlionerSoffel2005,Kopeikin2006RefFrames,KopeikinIAU2009},
by taking the cosmological expansion as a (tidal) perturbation around
the PN metric \eqref{eq:PNmetric} \cite{KlionerSoffel2005}, or by
considering perturbations around the flat ($k=1$) subcase of the
FLRW solution \cite{Kopeikin_et_al_Cosmological,Kopeikin2006RefFrames,KopeikinIAU2009}.

\section{Reference frames in the exact theory}

In the exact theory one needs to be more refined in some notions ---
observer family, system of axes, coordinate system --- and deal with
subtleties one is allowed to partially overlook in PN theory.

\subsection{Observers, observer congruences, and \textquotedblleft tetrad\textquotedblright{}
frames\label{subsec:Observers,-observer-congruences,}}

\begin{figure}
\includegraphics[width=1\columnwidth]{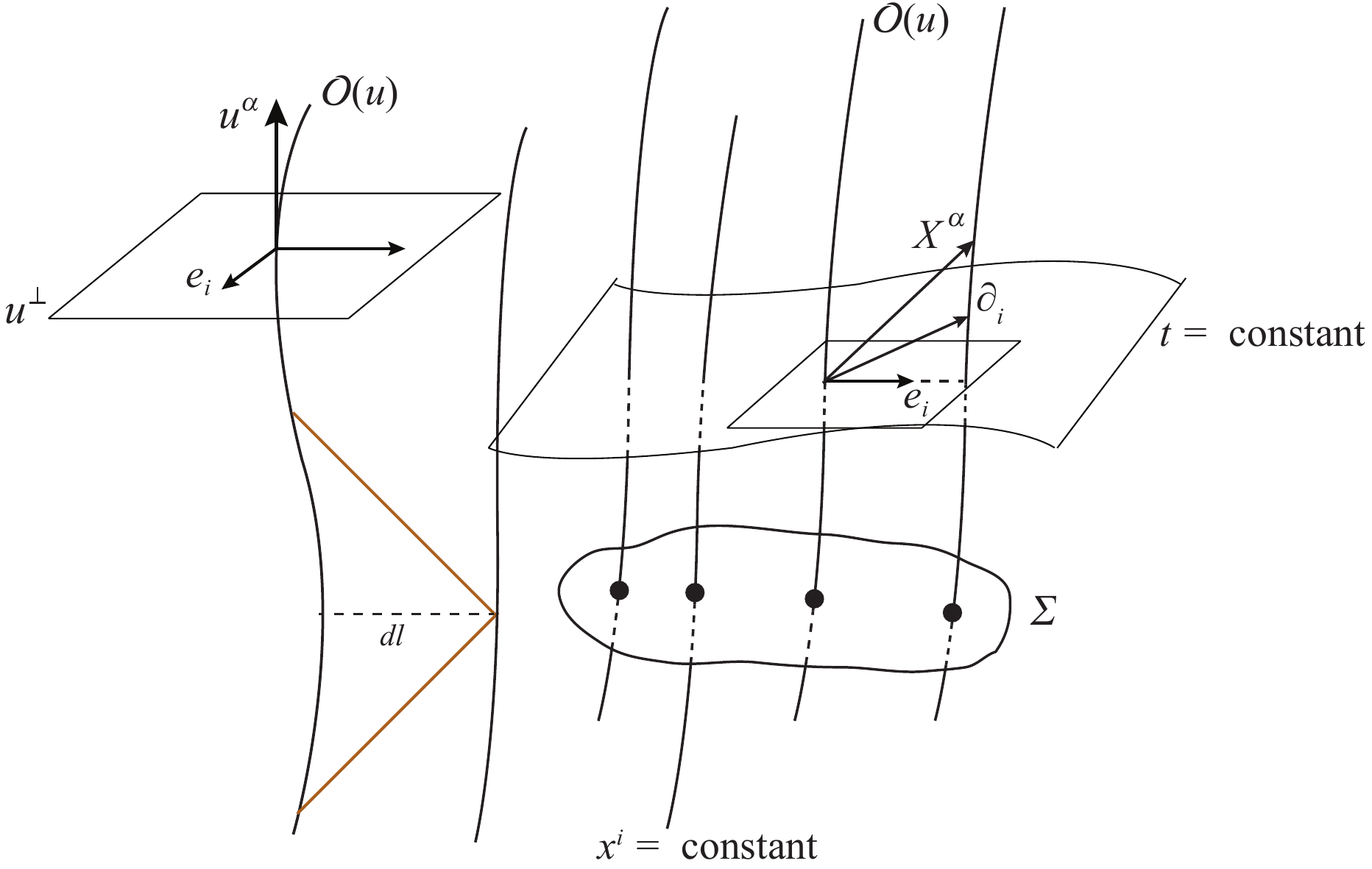}\caption{\label{fig:Frame}A reference frame. Observers are identified with
a congruence of timelike curves (observer worldlines) $\mathcal{O}(u)$
of tangent $u^{\alpha}$. The hyperplanes $u^{\perp}$ orthogonal
to $\mathcal{O}(u)$ are the observers' local rest spaces; therein
each observer sets up a triad of spatial axes $e_{i}$, which may
be orthonormal ($e_{i}=e_{\hat{\imath}}$), to perform measurements.
The associated coordinate system $\{t,x^{i}\}$, where the observers
are at rest, naturally embodies this construction, $\mathcal{O}(u)$
being the integral lines of $\partial_{t}$, and the projection $h_{\ \beta}^{\alpha}\partial_{i}^{\beta}$
of the coordinate basis vectors $\partial_{i}$ onto $u^{\perp}$
sets up therein a triad of axes pointing to the same\emph{ fixed neighboring
observer}. It moreover locates events in spacetime, the triplet $\{x^{i}\}$
labeling which observer, and $t$ where along its worldline. The quotient
$\Sigma=\mathcal{M}/\mathcal{O}(u)$ of the spacetime manifold $\mathcal{M}$
by the congruence $\mathcal{O}(u)$ is the \textquotedblleft space
manifold\textquotedblright{} --- a 3D space where each observer is
represented by a point.}
\end{figure}

An observer $\mathcal{O}(u)$ is identified with a worldline of tangent
vector (i.e., 4-velocity) $u^{\alpha}=dx^{\alpha}/d\tau_{u}$ \cite{deFeliceClarke,Soffel1989book,SachsWu1977,BolosIntrinsic,ManyFaces}.
A reference frame extended over some region requires a family observers
defined at every point therein, i.e., a \emph{congruence} of timelike
worldlines \cite{SachsWu1977} (sometimes called a ``platform''
\cite{Soffel1989book}), whose 4-velocity field we still denote by
$u^{\alpha}$, see Fig. \ref{fig:Frame}. Differentiation of $u^{\alpha}$
yields the congruence's ``kinematics,'' 
\begin{align}
 & \nabla_{\beta}u_{\alpha}\equiv u_{\alpha;\beta}=-u_{\beta}a_{\alpha}-\epsilon_{\alpha\beta\gamma\delta}\omega^{\gamma}u^{\delta}+\sigma_{\alpha\beta}+\frac{\theta}{3}h_{\alpha\beta}\ ,\label{eq:Kinematics-Decomp}\\
 & a^{\alpha}\equiv\nabla_{u}u^{\alpha}=u^{\beta}u_{\ ;\beta}^{\alpha}\ ;\qquad\omega^{\alpha}=\frac{1}{2}\epsilon^{\alpha\beta\gamma\delta}u_{\gamma;\beta}u_{\delta}\ ;\\
 & \sigma_{\alpha\beta}=h_{\alpha}^{\mu}h_{\beta}^{\nu}u_{(\mu;\nu)}-\theta h_{\alpha\beta}/3\ ;\qquad\theta\equiv u_{\ ;\alpha}^{\alpha}\ ,
\end{align}
where $a^{\alpha}$, $\omega^{\alpha}$, $\sigma_{\alpha\beta}$,
and $\theta$ are, respectively, the congruence's acceleration, vorticity,
shear, and expansion scalar. The hyperplanes $u^{\perp}$ orthogonal
to the observers' worldlines consist of the local rest spaces of each
observer (see Fig. \ref{fig:Frame}). Their distribution is nonintegrable
in the general case that the observer congruence has vorticity, $\omega^{\alpha}\ne0$;
i.e., there is in general no hypersurface orthogonal to such congruence.
The projection of a tensor in the observers' rest spaces $u^{\perp}$
is given by contraction with the tensor 

\begin{equation}
h_{\beta}^{\alpha}\equiv\delta_{\beta}^{\alpha}+u^{\alpha}u_{\beta}\ .\label{eq:SpaceProjector}
\end{equation}
This projector is also the metric induced in the distribution of rest
spaces $u^{\perp}$, with 
\begin{equation}
dl=\sqrt{h_{\alpha\beta}dx^{\alpha}dx^{\beta}}\label{eq:dl}
\end{equation}
yielding the distance between neighboring observers as measured by
Einstein's light signaling procedure (radar distance) \cite{LandauLifshitz}.
Each observer sets in its rest space a system of spatial axes $e_{i}$
(a set of rods), orthogonal to its worldline, to perform measurements.
The transport law for such axes along the observers' worldlines is
in principle arbitrary, and chosen by convenience. This construction
provides a vector (``tetrad'') basis $\{{\bf u},e_{i}\}$, sufficient
to measure tensor components; it is sometimes itself called a ``reference
frame''\footnote{Sometimes the reference frame is even defined as just the vector field
$u^{\alpha}$ (or, equivalently, the observer congruence), see e.g.
\cite{SachsWu1977,Gurses_2011}.} (e.g. \cite{Soffel1989book}, for orthonormal tetrads).

\subsection{Coordinate systems\label{subsec:Coordinate-systems}}

In order to locate events in spacetime (and, e.g., determine intervals
between them) one needs, however, a coordinate system $\{x^{\alpha}\}=\{t,x^{i}\}$.
Coordinate systems naturally embody the construction above: when the
basis vector $\partial_{t}$ is timelike, it is tangent to a congruence
of observers $\mathcal{O}(u)$ (the observers at rest in the given
coordinates: $u^{i}=0\Rightarrow u^{\alpha}\propto\partial_{t}^{\alpha}$);
and the projection of the spacelike coordinate basis vectors $\partial_{i}$
in the observer's rest space $u^{\perp}$, $h_{\ \beta}^{\alpha}\partial_{i}^{\beta}$
(see Fig. \ref{fig:Frame}), yields a system of spatial axis in $u^{\perp}$.
But it is additionally possible to label events; one can think of
the spatial triplet $\{x^{i}\}$ labeling which observer, and the
time coordinate $t$ where along the observer's worldline. In terms
of the coordinates adapted to an arbitrary congruence of observers
($u^{\alpha}\propto\partial_{t}^{\alpha}$, $g_{00}>0$), the metric
tensor can be generically written as 
\begin{equation}
ds^{2}=-e^{2\Phi(t,x^{k})}[dt-\mathcal{A}_{i}(t,x^{k})dx^{i}]^{2}+h_{ij}(t,x^{k})dx^{i}dx^{j}\label{eq:GenMetric}
\end{equation}
where $h_{ij}=g_{ij}+e^{2\Phi}\mathcal{A}_{i}\mathcal{A}_{j}$ equals
the space components of the projector \eqref{eq:SpaceProjector}.

The quotient of the spacetime manifold $\mathcal{M}$ by the congruence
of observer worldlines, $\Sigma=\mathcal{M}/\mathcal{O}(u)$, is the
``space manifold'', an abstract 3-dimensional space where each observer
worldline $\mathcal{O}(u)$ yields a point, and the triplet $\{x^{i}\}$
a coordinate system therein.

\subsection{Time and clock synchronization\label{subsec:Time-and-clock}}

\begin{figure}
\includegraphics[width=1\columnwidth]{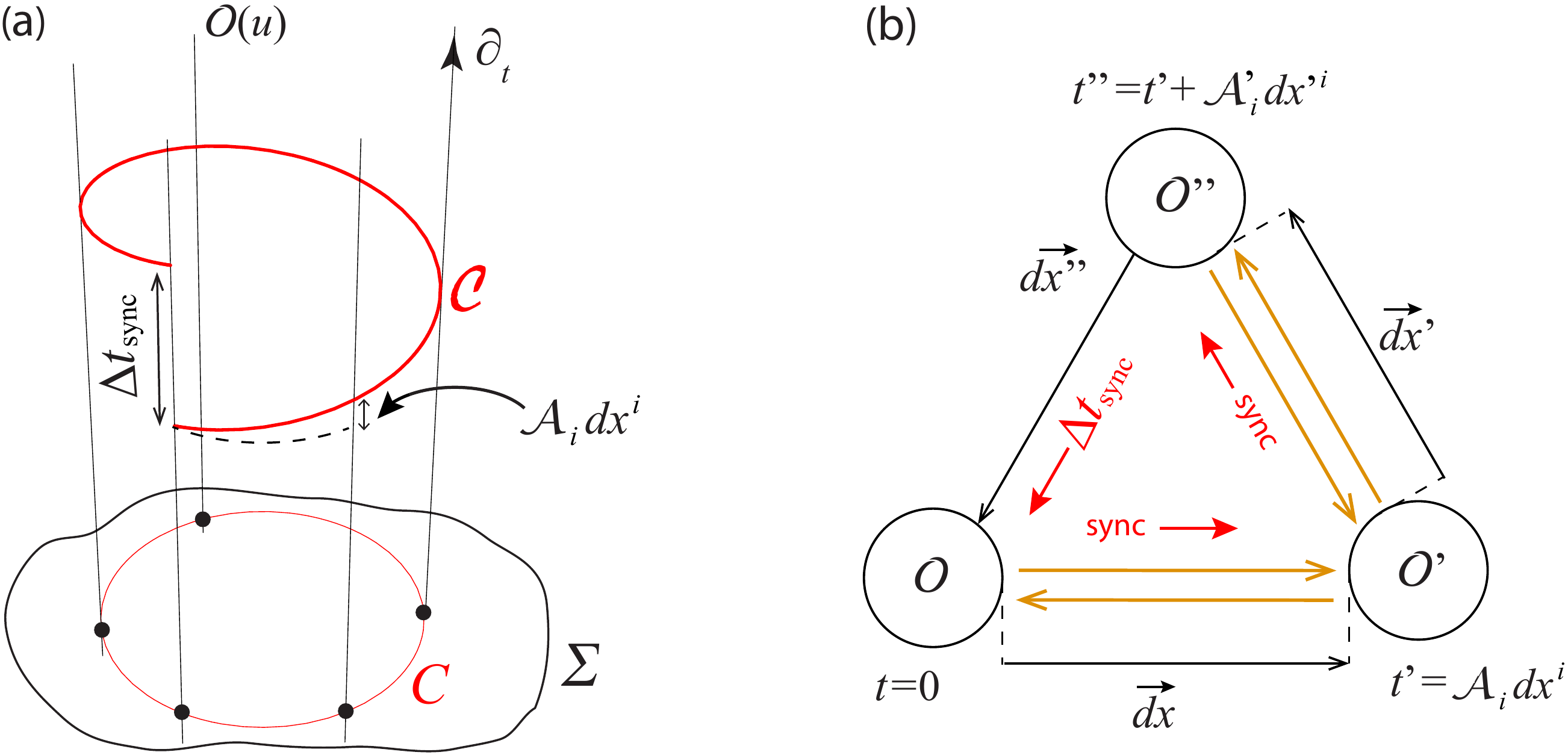}

\caption{\label{fig:SyncGap}(a) A synchronization curve $\mathcal{C}$ for
the congruence of rest observers $\mathcal{O}(u)$: ${\bf u}\propto\partial_{t}$,
which is \emph{spatially} closed (i.e., its projection $C$ on $\Sigma$
is closed), so that it starts and ends at the same observer. When
$\mathcal{A}_{i}=-g_{0i}/g_{00}\protect\ne0$, the \emph{spacetime}
curve $\mathcal{C}$ is not closed, leading to a synchronization gap
$\Delta t_{{\rm sync}}$: observers are unable to synchronize their
clocks along closed spatial loops $C$. (b) A pair of observers are
always able to synchronize their clocks, by exchanging light signals;
a triad of observers, in general, cannot: e.g., if $\mathcal{O}$
synchronizes its clock with $\mathcal{O}'$, and $\mathcal{O}'$ with
$\mathcal{O}''$, then the clocks of $\mathcal{O}$ and $\mathcal{O}''$
will not be synchronized. This is because while the events $t'\in\mathcal{O}'$
and $t\in\mathcal{O}$ are simultaneous, the same applying to $t''\in\mathcal{O}''$
and $t'\in\mathcal{O}'$, the event $t''\in\mathcal{O}''$ is not
simultaneous with $t\in\mathcal{O}$, but instead with the event $t'''=(t''+\mathcal{A}''_{i}dx''^{i})\in\mathcal{O}$,
ahead in time by a gap $\Delta t_{{\rm sync}}=t'''-t=\mathcal{A}''_{i}dx''^{i}+\mathcal{A}'_{i}dx'^{i}+\mathcal{A}_{i}dx^{i}$.}
\end{figure}

Consider a curve $\mathcal{C}:\ x^{\alpha}(\lambda)$ of tangent $dx^{\alpha}/d\lambda$.
The condition that the curve is of simultaneity for the observers
$\mathcal{O}(u)$ at rest in the coordinate system $\{t,x^{i}\}$,
or equivalently, that their clocks be synchronized along $\mathcal{C}$
through Einstein's light signaling procedure \cite{LandauLifshitz,Soffel1989book,Bazanski1997,MinguzziSimultaneity},
amounts to the curve being orthogonal (at every point) to $\partial_{t}$,
that is, $g_{0\beta}dx^{\beta}/d\lambda=0\Leftrightarrow dt=\mathcal{A}_{i}dx^{i}$,
where $\mathcal{A}_{i}=-g_{0i}/g_{00}$. Consider now the case that
$\mathcal{C}$ is spatially closed (i.e., its projection onto $\Sigma$,
$C=\pi_{\Sigma}(\mathcal{C})$, yields a closed curve $C$), so that
after each loop it re-intersects the worldline of the original observer,
at a coordinate time $t_{f}=t_{i}+\Delta t_{{\rm sync}}$, see Fig.
\ref{fig:SyncGap}. When $\Delta t_{{\rm sync}}\ne0$, $\mathcal{C}$
is not closed in spacetime, and the observer will find that his clock
is not synchronized with his preceding neighbor's. This is the case
generically when $\mathcal{A}_{i}\ne0$, as shown in Fig. \ref{fig:SyncGap}(b)
for infinitesimally close observers. In the \emph{special case that}
$\mathcal{A}_{i}$ is \emph{time-independent} (e.g., a stationary
spacetime), $\mathcal{A}_{i}=\mathcal{A}_{i}(x^{k})$, it becomes
a 1-form in $\Sigma$, and the synchronization gap \cite{LandauLifshitz,BiniJantzenMashhoon_Clock1,Cilindros,MinguzziSimultaneity,Gourgoulhon:SpecialRelativ}
follows by simply integrating it along $C$, 
\begin{equation}
\Delta t_{{\rm sync}}=\int_{C}\mathcal{A}_{i}dx^{i}\ .\label{eq:syncgap}
\end{equation}
Only when $\Delta t_{{\rm sync}}=0$ (case in which $\mathcal{C}$
is closed) are the observers able to fully synchronize their clocks
along a closed curve $C$ in space. In order for a congruence of observers
to be able to \emph{globally} synchronize their clocks, every synchronization
curve $\mathcal{C}$ whose projection $C=\pi_{\Sigma}(\mathcal{C})$
is closed, must be closed in spacetime. That requires the congruence
to be orthogonal to an hypersurface of global simultaneity; i.e.,
one that intersects each worldline of the congruence only once (called
a ``time-slice''). This amounts to $u_{\alpha}\propto\psi_{,\alpha}$
(or, equivalently, $(\partial_{t})_{\alpha}\propto\psi_{,\alpha}$)
for some globally defined (single-valued) function $\psi(t,x^{i})$.
An immediate consequence is that the congruence must not have vorticity,
$\omega^{\alpha}=0$, which, at 1PN order, amounts to the condition
$g_{0[i,j]}=0$. Notice however that this affects clocks around a
closed in space contour $C$; along an open contour in space, clocks
can always be synchronized. In particular, a pair of arbitrary observers
in an arbitrary spacetime can always synchronize their clocks (albeit,
if not infinitesimally close, the synchronization will be contour-dependent);
a triad of observers, however, in general cannot, see Fig. \ref{fig:SyncGap}(b).

The above notion of synchronization (sometimes called ``Einstein''
synchronization \cite{Soffel1989book}) only means that the observers
agree on the simultaneity events; in other words, the synchronization
is achieved at an instant. In general, their clocks will not remain
synchronized, since their proper times $d\tau_{u}=(-g_{00})^{\text{1/2}}dt$
elapse at different rates due to the nonconstant gravitational potential
in $g_{00}$, causing the interval of proper time between two events
$\mathscr{E}_{1}$ and $\mathscr{\mathscr{E}}_{2}$ along the worldline
of an observer $\mathcal{O}(u)$ to differ from that between the events
$\mathscr{E}'_{1}$ and $\mathscr{\mathscr{E}}'_{2}$ along the worldline
of another observer $\mathcal{O}'(u)$, that are simultaneous with
$\mathscr{E}_{1}$ and $\mathscr{\mathscr{E}}_{2}$ \cite{LandauLifshitz}.
Only a set of clocks subject to the same gravitational potential (e.g.,
those in a shell of constant radius in Schwarzschild spacetime) can
remain synchronized.

In an arbitrary spacetime, a coordinate system can however always
be found such that $g_{00}=-1$, $g_{0i}=0$: 
\begin{equation}
ds^{2}=-dt^{2}+h_{ij}(t,x^{k})dx^{i}dx^{j}\ ,\label{eq:Synchronous}
\end{equation}
called the ``synchronous'' reference system \cite{LandauLifshitz};
observers at rest in such coordinates are freely falling, and measure
the same proper time $\tau_{u}=t$, thus their clocks are all synchronized.

The IAU reference system is not synchronous; in general it has also
vorticity, hence a global ``instantaneous'' synchronization in the
sense of Einstein is not possible either. This is circumvented by
using instead a coordinate time synchronization, i.e., setting an
``official'' time (chosen as the $t$-coordinate). Common choices
for $t$ are the proper time measured by a rest observer at infinity
(in which case it does not correspond to the proper time measured
by any actual observer in the physical system), or the proper time
of observers at rest in the Earth's geoid \cite{SoffelBook2009,AshbyAllan_Sync1979,Soffel1989book}.
The pace and synchronization of every clock in the system is computer-corrected
for relativistic desynchronization effects according to its position
and motion \cite{AshbyAllan_Sync1979,KlionerSync,SoffelBook2009},
so that they read $t$. The clock pace is adjusted through the equation
\cite{KlionerSync,Soffel1989book,SoffelBook2009} 
\begin{equation}
\frac{d\tau}{dt}=\sqrt{-g_{00}-2g_{0i}\frac{dx^{i}}{dt}-g_{ij}\frac{dx^{i}}{dt}\frac{dx^{j}}{dt}}\ ;\label{eq:CoordSync}
\end{equation}
$g_{\alpha\beta}\equiv g_{\alpha\beta}(t,x^{k}(t)),$ which follows
from the definition of proper time along $\mathcal{O}(u)$, $d\tau^{2}=-g_{\alpha\beta}dx^{\alpha}dx^{\beta}$.
Solving this differential equation yields the correspondence between
$\tau$ and $t$ for an arbitrary observer moving with coordinate
velocity $dx^{i}/dt$, given the initial value $\tau(t=0)$. For the
congruence of rest observers ${\bf u}\propto\partial_{t}$, it reduces
to $d\tau/dt=(-g_{00})^{1/2}$, whose solution yields, up to a time-constant
function $\psi(x^{i})$, the correspondence between these observers'
proper time $\tau$ and $t$ everywhere. The function $\psi(x^{i})$
is fixed, up to a global constant, by synchronization of all clocks
in terms of $t$ , i.e., it is such that the ``computer-corrected''
time on each clock yields the same value along the hypersurfaces $t={\rm const.}$
(which represents simultaneous events \emph{according} to the chosen
time coordinate). This can be physically realized by transporting
clocks \cite{AshbyAllan_Sync1979,KlionerSync,Soffel1989book,SoffelBook2009}:
consider an observer $\mathcal{O}$ carrying two identical (synchronized)
clocks. At some instant one of the clocks departs from $\mathcal{O}$,
being transported along a known path, while it keeps solving and inverting
Eq. \eqref{eq:CoordSync}, so that its elapsed proper time is converted
into computer-corrected time $t(\tau)$. Then, every observer $\mathcal{O}'$
it meets along the way resets his computer-corrected time to $t'(\tau')=t(\tau)$;
thereby his clock will be synchronized (in the same time gauge) with
$\mathcal{O}$. Filling space with such paths realizes the ``coordinate
time grid'' \cite{AshbyAllan_Sync1979}; in practice, this is implemented
through light signals exchange \cite{AshbyAllan_Sync1979,KlionerSync,SoffelBook2009}.
This is an artificial synchronization, in the sense that the hypersurfaces
$t={\rm const}$ are not orthogonal to the worldlines of the observers
$\mathcal{O}(u)$ (thus their local hyperplanes of simultaneity, $u^{\perp}$,
in Fig. \ref{fig:Frame}, are not tangent to these hypersurfaces).
It requires also precise knowledge of the gravitational field and
of the clock's motion, being thus model dependent. It yields however
a practical global notion of time, with transitive simultaneity (e.g.,
for the triad of observers in Fig. \ref{fig:SyncGap}(b), if the clocks
of $\mathcal{O}$ and $\mathcal{O}'$, and $\mathcal{O}'$ and $\mathcal{O}''$,
are synchronized in terms of coordinate time, then so will be $\mathcal{O}''$
and $\mathcal{O}$), used in many applications, such as the GPS system.

The same prescription can be applied to the exact theory, with the
only difference that, if the spacetime is not asymptotically flat
(in particular, if $\lim_{r\rightarrow\infty}g_{00}\ne-1$), $t$
cannot be chosen to correspond to the proper time measured by an asymptotic
rest observer {[}an example is the exterior van Stockum metric in
Eq. \eqref{eq:StockumExt} below{]}\textcolor{blue}{. }This poses
no problems, since the choice of time coordinate is a gauge freedom
\cite{KlionerSync}.

\subsection{Generalization of the IAU system to the exact theory\label{subsec:Generalization-of-IAU}}

A crucial feature of the PN reference frame associated to the metric
in \eqref{eq:PNmetric} is that the basis vectors have fixed directions
with respect to inertial frames at infinity (where the distant reference
stars are assumed at rest). Let us dissect this property. Consider
a congruence of observers $\mathcal{O}(u)$ at rest in the given coordinate
system $\{t,x^{i}\}$, ${\bf u}=(-g_{00})^{-1/2}\partial_{t}$. The
coordinate basis vectors $\partial_{i}$ are connecting vectors between
the worldlines of neighboring observers (see Fig. \ref{fig:Frame}),
since they are Lie transported along the integral lines of $\partial_{t}$:
$\mathcal{L}_{\partial_{t}}\partial_{i}=[\partial_{t},\partial_{i}]=0$
(more precisely, they connect events with the same coordinate time
$t$). These vectors thus point to fixed neighboring observers. In
what follows it will be convenient considering vectors $X^{\alpha}$
connecting pairs of events with the same proper time ($\tau_{u}$)
interval, defined by the condition $\mathcal{L}_{{\bf u}}X^{\alpha}=0\Leftrightarrow X_{\ ,\beta}^{\alpha}u^{\beta}-u_{\ ,\beta}^{\alpha}X^{\beta}=0$;
these have constant spatial components in the coordinate basis $\{\partial_{i}\}$:
$X_{\ ,\beta}^{i}u^{\beta}\equiv dX^{i}/d\tau_{u}=0$.

Consider now an orthonormal frame $\{e_{\hat{\alpha}}\}$ such that
$e_{\hat{0}}={\bf u}$ (see Fig. \ref{fig:Frame}), and let $\Gamma_{\hat{\beta}\hat{\delta}}^{\hat{\alpha}}$
denote its connection coefficients. Using $\mathcal{L}_{{\bf u}}X^{\hat{\alpha}}=\nabla_{{\bf u}}X^{\hat{\alpha}}-u^{\hat{\alpha};\hat{\beta}}X_{\hat{\beta}}=0$,
substituting Eq. \eqref{eq:Kinematics-Decomp} in $u^{\hat{\alpha};\hat{\beta}}$,
noticing that $\nabla_{{\bf u}}X^{\hat{\imath}}=\dot{X}^{\hat{\imath}}+\Gamma_{\hat{0}\hat{0}}^{\hat{\imath}}X^{\hat{0}}+\Gamma_{\hat{0}\hat{\jmath}}^{\hat{\imath}}X^{\hat{\jmath}}$,
that $\Gamma_{\hat{0}\hat{0}}^{\hat{\imath}}=\nabla_{{\bf u}}u^{\hat{\imath}}=a^{\hat{\imath}}$
and that $\Gamma_{\hat{0}\hat{\jmath}}^{\hat{\imath}}=\epsilon_{\ \hat{k}\hat{\jmath}}^{\hat{\imath}}\Omega^{\hat{k}}$
(see e.g. Sec. 3 of \cite{Analogies}), where $\Omega^{\hat{k}}$
are the components of the angular velocity of rotation of the spatial
triad $\{e_{\hat{\imath}}\}$ relative to Fermi-Walker transport along
${\bf u}$, we have 
\begin{equation}
\dot{X}_{\hat{\imath}}=\frac{1}{3}\theta X_{\hat{\imath}}+\sigma_{\hat{\imath}\hat{\jmath}}X^{\hat{\jmath}}+\epsilon_{\hat{\imath}\hat{k}\hat{\jmath}}\left(\omega^{\hat{k}}-\Omega^{\hat{k}}\right)X^{\hat{\jmath}}.\label{eq:ConnectingVector}
\end{equation}
If the rotation of the spatial triad $\{e_{\hat{\imath}}\}$ is locked\footnote{Given a congruence of observers, the choice $\Omega^{\alpha}=\omega^{\alpha}$
can always be made (cf. Sec. \ref{subsec:Observers,-observer-congruences,});
it is actually the most natural transport law for the spatial triad,
corresponding to the case where it \emph{co-rotates} with the observers
(``congruence adapted''\emph{ }frame \cite{Analogies}), and arguably
the closest generalization of the Newtonian concept of reference frame
\cite{MassaZordan,MassaII}.} to the vorticity of the congruence, $\Omega^{\alpha}=\omega^{\alpha}$,
\emph{and the congruence is shear-free}, $\sigma_{\alpha\beta}=0$,
we have 
\begin{equation}
\dot{X}^{\hat{\imath}}=\frac{1}{3}\theta X^{\hat{\imath}}\ ,\label{eq:connectPN}
\end{equation}
telling us that the connecting vector's direction is fixed in the
triad $\{e_{\hat{\imath}}\}$. In other words, a set of orthogonal
spatial axes point to fixed neighboring rest observers $\mathcal{O}(u)$.
Therefore, the coordinate basis $\partial_{i}$ has likewise fixed
directions in the orthonormal triad $\{e_{\hat{\imath}}\}$ (more
precisely, the space projection of the coordinate basis vectors, $h_{\ \beta}^{\alpha}\partial_{i}^{\beta}$,
has fixed direction therein). This allows to define fixed directions
(hence rotations) with respect to distant reference objects. Observe
that $\mathcal{L}_{{\bf u}}h_{\alpha\beta}=2(\sigma_{\alpha\beta}+\theta h_{\alpha\beta}/3)$;
hence, the shear-free condition amounts to 
\begin{equation}
\mathcal{L}_{{\bf u}}h_{\alpha\beta}=\frac{2\theta}{3}h_{\alpha\beta}\ ,\label{eq:shear-freeLie}
\end{equation}
which is equivalent to the condition 
\begin{equation}
\mathcal{L}_{{\bf u}}\rigid_{\alpha\beta}=0\ ;\qquad\rigid_{\alpha\beta}\equiv h_{\alpha\beta}/f\ ,\label{eq:ConformalRigidity}
\end{equation}
for some function $f$ solving the first order differential equation
$\mathcal{L}_{{\bf u}}f-2\theta f/3=0$. The subcase $\mathcal{L}_{{\bf u}}f=0\Rightarrow\mathcal{L}_{{\bf u}}h_{\alpha\beta}=0$
corresponds to a \emph{rigid} congruence ($\theta=0=\sigma_{\alpha\beta}$,
see Sec. \ref{subsec:Stationary-spacetimes}); for this reason, condition
\eqref{eq:ConformalRigidity} is dubbed ``conformal rigidity'' \cite{Bel_Llosa_Meta_Rigid95,Llosa_shearfree1997}.

In a coordinate system $\{t,x^{i}\}$ where ${\bf u}=(-g_{00})^{-1/2}\partial_{t}$,
Eqs. \eqref{eq:ConformalRigidity} yield $\partial_{t}\rigid_{\alpha\beta}=0\Rightarrow\rigid_{\alpha\beta}\equiv\rigid_{\alpha\beta}(x^{i})$
and\footnote{This agrees with the earlier results obtained in \cite{ChrobokObukhovScherfnev,Gurses_2011},
for the special case of a coordinate system such that $g_{00}=-1$
(i.e., $t=\tau_{u}$).} $h_{\alpha\beta}=f(t,x^{i})\rigid_{\alpha\beta}(x^{i})$. From \eqref{eq:GenMetric},
the existence of shear-free observer congruences in a given spacetime
is thus equivalent to it admitting a coordinate system where the metric
takes the form 
\begin{equation}
ds^{2}=-e^{2\Phi}[dt-\mathcal{A}_{i}dx^{i}]^{2}+f(t,x^{k})\rigid_{ij}(x^{k})dx^{i}dx^{j}\ ,\label{eq:shear-freeMetric}
\end{equation}
where $\Phi\equiv\Phi(t,x^{k})$, $\mathcal{A}_{i}\equiv\mathcal{A}_{i}(t,x^{k})$.

This is the situation in a PN frame \eqref{eq:PNmetric}, identifying
$f=(1+2w)$, $\rigid_{ij}=\delta_{ij}$ {[}the underlying reason is
that, at 1PN accuracy, the observers at rest in \eqref{eq:PNmetric}
have no shear: $\sigma_{ij}=O(5)$, $\sigma_{0\alpha}=0${]}. However,
PN frames rely moreover on the assumption that the metric \eqref{eq:PNmetric}
is asymptotically flat. This implies, in particular, that, at infinity,
the reference frame is inertial, and therefore the basis vectors $\partial_{i}$
are anchored to inertial frames at infinity. Provided that they are
far enough, the distant stars have no detectable proper motion in
such frame; one can thus say that the $\partial_{i}$ point to fixed
stars; or in other words, are anchored to the ``stellar compass''
\cite{Soffel_et_al_RelativisticAstrometry1985,Soffel1989book,SchmidtStellarCompass,soffel_Book_ReferenceSystems}.
Physically, this materializes in that light rays from the distant
stars are received (to the accuracy at hand) at fixed directions in
the basis $\{\partial_{i}\}$; hence the frame can be set up by aiming
telescopes at the distant stars. This is a crucial property that should
be embodied in a suitable generalization of the IAU reference system.
For that, defining fixed directions over an extended spacetime region
{[}i.e. Eq. \eqref{eq:connectPN} being obeyed{]}, is not sufficient\footnote{Simple counterexamples are the coordinate system associated a rigidly
rotating frame in flat spacetime, or with the observers comoving with
the dust in the Gödel universe: indeed, they form a rigid grid in
space; however, such grid is not nonrotating at infinity (the observers
having constant nonvanishing vorticity in the second case, and not
even being defined past a certain radius in the first case, where
they would exceed the speed of light).}; they should be fixed to fundamental reference objects, representing
nonrotating axes at infinity (``compass of inertia'' at infinity,
see Sec. \ref{subsec:Stationary-spacetimes} below).

To generalize it to the exact theory, we start by noticing that (i)
the asymptotically flat condition is not necessary for the reference
frame to be anchored to inertial frames\footnote{\label{fn:Inertial-frame}An inertial frame is defined as a reference
frame where all the inertial forces vanish. While different definitions
of inertial force exist in the literature \cite{ManyFaces,Analogies}
(differing essentially in the connection/transport law for the spatial
frame), all agree in its vanishing in coordinate systems adapted to
a congruence of observers such that $\nabla_{\beta}u_{\alpha}=0$.} at infinity; it suffices the existence of a shear-free observer congruence
for which all the remaining kinematical quantities $a^{\alpha}$,
$\omega^{\alpha}$, and $\theta$ asymptotically vanish {[}so that
$\lim_{r\rightarrow\infty}\nabla_{\beta}u_{\alpha}=0${]}; an example
is the van Stockum exterior solution considered in Sec. \ref{subsec:The-van-Stockum}.
(ii) The reference frame does not even need to be asymptotically inertial.
As long as it does not shear, having expansion ($\theta\ne0$) does
not affect the property of defining directions fixed to distant objects,
cf. Eq. \eqref{eq:connectPN}; and in order for it to be nonrotating
at infinity, the necessary and sufficient condition is the vorticity
to asymptotically vanish, $\lim_{r\rightarrow\infty}\omega^{\alpha}=0$.
This ensures (see e.g. \cite{Cilindros} p.7, and footnote therein)
that the axes $\{e_{\hat{\imath}}\}$ are Fermi-Walker transported
at infinity, i.e. are fixed to the spin axes of gyroscopes (compass
of inertia) at infinity, cf. Eq. \eqref{eq:SpinPrec} below, the same
applying to the spatial directions of the coordinate basis vectors
$\partial_{i}$. Therefore, we have:

\emph{Lemma.}--- If a spacetime admits a nonshearing ($\sigma_{\alpha\beta}=0$)
congruence of observers which, at infinity, has zero vorticity ($\lim_{r\rightarrow\infty}\omega^{\alpha}=0$),
then a coordinate system where such observers are at rest has spatial
axes locked to nonrotating (i.e., Fermi-Walker transported) axes at
infinity.

This is important in a cosmological context, where asymptotically
inertial frames do not exist: the FLRW solution is of the form \eqref{eq:shear-freeMetric},
with $\Phi=\Phi(t)$ (usually set $\Phi=0$), $\mathcal{A}_{i}=0$,
and $f(t,x^{k})=f(t)\equiv a^{2}(t)$ the scale factor. Observers
at rest therein are comoving with the cosmological fluid (i.e., measure
an isotropic cosmic microwave background radiation). Their vorticity
and acceleration vanish, $\omega^{\alpha}=a^{\alpha}=0$, but they
have nonvanishing expansion $\theta=3\dot{a}/a$ everywhere; thus,
the congruence is not asymptotically inertial (cf. Footnote \ref{fn:Inertial-frame}).
It follows however from Eq. \eqref{eq:connectPN} that the connecting
vectors between neighboring particles of such fluid have directions
fixed with respect to orthonormal axes, hence preserving angles with
respect to each other. Therefore, the coordinate grid defined by them
has directions fixed to the distant quasars, which one may dub the
``quasar compass'' (generalizing the notion of stellar compass in
\cite{Soffel_et_al_RelativisticAstrometry1985,Soffel1989book,SchmidtStellarCompass,soffel_Book_ReferenceSystems},
whose definition relies on asymptotic flatness). This property extends
to the more general shear and vorticity-free (i.e., shearfree ``normal'')
cosmological fluids in e.g. \cite{KrazinskiShearfree,BarnesShearfree,Sussman_1993},
described by metrics of the form \eqref{eq:shear-freeMetric} with
$\mathcal{A}_{i}=0$.

In the IAU system, the distant reference objects are moreover assumed
geodesic (indeed, since their proper motions are negligible, so should
be their small nongeodesic part, discussed in e.g. \cite{Magnus}
Sec. VI); this property is retained by demanding the observers' acceleration
to asymptotically vanish, $\lim_{r\rightarrow\infty}a^{\alpha}=0$.
Therefore,

\emph{Proposition }\ref{subsec:Generalization-of-IAU}. --- If a
spacetime admits a non-shearing congruence of observers which, at
infinity, has zero vorticity and acceleration ($\lim_{r\rightarrow\infty}\omega^{\alpha}=\lim_{r\rightarrow\infty}a^{\alpha}=0$),
then a coordinate system where such observers are at rest has spatial
axes locked to the asymptotic rest frame of the distant quasars, being
the generalization of the IAU reference system to the exact theory.\\

The shear-free condition $\sigma_{\alpha\beta}=0$ is however restrictive,
as only special spacetimes admit timelike shear-free congruences of
curves. No classification for such spacetimes, or invariant conditions
for the existence of such congruences, are known in the literature\footnote{Only for congruences comoving with certain solutions' matter content.
Namely, it has been conjectured \cite{BerghRadu_shearfree,Bergh_ShearFree,EllisShearFree,Collins1986,Collins1985,SenovillaShearFree},
and proved in several special cases, that shear-free barytropic perfect
fluids can have vorticity \emph{or} expansion, but not both. This
has, in particular, been proven for dust \cite{EllisShearFree}, which
includes arbitrary \emph{geodesic} observer congruences in a vacuum.
But the reference frames we interested in are, in general, neither
geodesic or comoving with the celestial bodies/matter.}. One can see, however, from its degrees of freedom, that not every
metric can be written in the form \eqref{eq:shear-freeMetric}: the
most general metric tensor is composed of ten functions (corresponding
to the independent components in $g_{\alpha\beta}$) of four variables
(the coordinates); four of these functions are fixed by the coordinate
choice, leaving six ``free'' functions of four variables. These
are made explicit in the synchronous frame in \eqref{eq:Synchronous},
where all this gauge freedom has been used to eliminate the components
$g_{00}$ and $g_{0i}$, leaving out the six functions $h_{ij}(t,x^{k})$,
which can be arbitrary (as long they define a positive definite matrix).
The metric \eqref{eq:shear-freeMetric}, however, possesses only five
free functions of four variables ($\Phi$, $\mathcal{A}_{i}$, and
$f$), plus six functions ($\rigid_{ij}$) of the three variables
$x^{i}$ (notice that a function of four variables can be regarded
as infinitely many functions of three variables). Specifically, the
time dependence of \eqref{eq:shear-freeMetric} has only five degrees
of freedom, whereas a general metric has six \cite{FlanaganHughesWaves}.
For instance, the gravitational field of a set of $N$ celestial bodies
(even if no such field is known exactly) is expected to contain gravitational
radiation, which generically implies a nonvanishing shear, and so
cannot be described by \eqref{eq:shear-freeMetric}. In such general
case, reference frames with axes fixed to distant reference objects
can only be set up in the framework of some approximation; often,
radiation effects are negligible. In a cosmological setting, the gravitational
field of the celestial bodies can be considered as a perturbation
around the FLRW solution \cite{Stewart_1990,Kopeikin2006RefFrames,Kopeikin_et_al_Cosmological,Clifton_gauge},
the coordinates comoving with the FLRW fluid providing a frame with
axes fixed to the distant quasars. This extends to generic perturbations
without tensor modes in gauges complying with \eqref{eq:shear-freeMetric},
a well known example being the ``conformal Newtonian gauge\emph{}\footnote{It is also dubbed longitudinal \cite{BertschingerCourse,Stewart_1990,Clifton_gauge}
or ``zero-shear'' gauge \cite{Clifton_gauge}. It should not, however,
be confused with the (less restrictive) vanishing shear condition
described by the metric \eqref{eq:shear-freeMetric}: (i) the former
pertains to the shear $\sigma(n)_{\alpha\beta}$ of the vector field
$n^{\alpha}$ orthogonal to the $t=const.$ hypersurfaces {[}$n_{\alpha}=-\delta_{\alpha}^{0}/\sqrt{-g^{00}}${]};
not to the shear $\sigma_{\alpha\beta}\equiv\sigma(u)_{\alpha\beta}$
of the rest observers of 4-velocity $u^{\alpha}=\delta_{0}^{\alpha}/\sqrt{-g_{00}}$,
which vanishes in \eqref{eq:shear-freeMetric}. The tensor $\sigma(n)_{\alpha\beta}$
is, in general, nonzero for a perturbed cosmological metric of the
form \eqref{eq:shear-freeMetric}; for instance, for $e^{\Phi}=a^{2}(t)-\delta g_{00}$,
$f=a^{2}(t)+\delta f$, $\chi_{ij}=\delta_{ij}$, to first order in
the perturbations, $\sigma(n)_{ij}=a(t)[\delta_{ij}\mathcal{A}_{k}^{\ ,k}/3-\mathcal{A}_{(i,j)}]$.
(iii) The conformal Newtonian gauge requires moreover $\mathcal{A}_{i}=0$
{[}thereby excluding \cite{BertschingerCourse} vector perturbations,
that are allowed in \eqref{eq:shear-freeMetric}{]}, in which case
$u^{\alpha}=n^{\alpha}\Rightarrow\sigma(n)_{\alpha\beta}=\sigma(u)_{\alpha\beta}=0$
(i.e., both ``shears'' vanish).}'' for scalar perturbations \cite{Mukhanov_1992,BertschingerCourse,Clifton_gauge}. 

In the case of a stationary spacetime, light rays emitted by the distant
reference objects (quasars or stars) are received at fixed directions
in the basis $\{e_{\hat{\imath}}\}$ and $\{\partial_{i}\}$; the
reference frame can thus be physically set up through telescopes/radiotelescopes.
The same holds for conformally stationary spacetimes, whose metric
can be written in the form $ds^{2}=\psi(t,x^{i})\Psi(x^{i})_{\alpha\beta}dx^{\alpha}dx^{\beta}$,
since null geodesics are conformally invariant, see e.g. Appendix
D in \cite{Wald:1984}; these spacetimes include the FLRW models.

In general, however, this does not hold. Like in PN theory \cite{Soffel_et_al_RelativisticAstrometry1985,Soffel1989book},
in the exact theory null geodesics can be described in terms of inertial
spatial ``forces,'' Eqs. (10.5)-(10.6) of \cite{BiniIntrinsic}
(which are a generalization of Eq. \eqref{eq:QMGeo} below for null
geodesics and time-dependent settings). When the fields $\vec{G}$
and $\vec{H}$ therein are time-varying, so will be the deflection
of light by the gravitational field. This means that, even in the
case that the conditions in Proposition \ref{subsec:Generalization-of-IAU}
are met, in practice the axes of such frame only approximately coincide
with the directions of the light rays received from the reference
quasars.

\subsection{Stationary spacetimes\label{subsec:Stationary-spacetimes}}

Spacetimes admitting a timelike Killing vector field $\xi^{\alpha}$
are stationary. A coordinate system always exists where $\boldsymbol{\xi}=\partial_{t}$,
and the metric takes the explicitly time-independent form 
\begin{equation}
ds^{2}=-e^{2\Phi}(dt-\mathcal{A}_{i}dx^{i})^{2}+h_{ij}dx^{i}dx^{j}\ ,\label{eq:StatMetric}
\end{equation}
where $e^{2\Phi}=-g_{00}$, $\Phi\equiv\Phi(x^{j})$, $\mathcal{A}_{i}\equiv\mathcal{A}_{i}(x^{j})=-g_{0i}/g_{00}$,
and $h_{ij}\equiv h_{ij}(x^{k})$ is the ``space metric''. Considering
the congruence of observers $\mathcal{O}(u)$ whose worldlines are
tangent to $\xi^{\alpha}$, 
\begin{equation}
u^{\alpha}\equiv\frac{\xi^{\alpha}}{\sqrt{-\xi^{\alpha}\xi_{\alpha}}}=\frac{\partial_{t}^{\alpha}}{\sqrt{-g_{00}}}=e^{-\Phi}\partial_{t}^{\alpha}\equiv e^{-\Phi}\delta_{0}^{\alpha},\label{eq:uLab}
\end{equation}
which are, by definition, observers at rest in the coordinates of
\eqref{eq:StatMetric} (``static observers''), $h_{ij}$ equals
the components of the projector \eqref{eq:SpaceProjector} (which
it is identified with in spacetime). It follows from the Killing equation
$\xi_{(\alpha;\beta)}=0$ that the congruence of observers $\mathcal{O}(u)$
is \emph{rigid}, since it is both nonshearing and nonexpanding (vanishing
Born tensor \cite{MasonPooe_RigidMotion,Bel_Llosa_Meta_Rigid95,Soffel1989book}):
$h_{\alpha}^{\mu}h_{\beta}^{\nu}u_{(\mu;\nu)}=0\Leftrightarrow\sigma_{\alpha\beta}=\theta=0$.
Equivalently, the spatial metric $h_{\alpha\beta}$ has zero Lie derivative
along the congruence: $\mathcal{L}_{{\bf u}}h_{\alpha\beta}=0$ {[}implying
$\partial_{t}h_{ij}=0$ in the coordinates of \eqref{eq:StatMetric}{]}.
This just states that the spatial distance \eqref{eq:dl} between
neighboring observers as measured by Einstein's light signaling prescription
(see Fig. \ref{fig:Frame}) is constant. The spatial metric is therefore
independent of the time slice $t=t(x^{i})$, and can thus be regarded
as the Riemannian metric canonically associated with the quotient
$\Sigma$ of the spacetime by the congruence of worldlines $\mathcal{O}(u)$.
Together, they form the Riemannian manifold $(\Sigma,h)$, dubbed
the ``space manifold''.

Equation \eqref{eq:connectPN} is obeyed with $\dot{X}^{\hat{\imath}}=0$,
hence neighboring observers remain at fixed directions in an orthonormal
frame, the same occurring for the coordinate basis vectors $\partial_{i}$.
If the coordinate system is asymptotically inertial ($\lim_{r\rightarrow\infty}\nabla_{\beta}u_{\alpha}=0$),
then the distant stars are at rest in such coordinates, which are
the generalization of the IAU reference system for the given exact
stationary spacetime.

The observed spatial position vector of any (arbitrarily distant)
object $\mathcal{O}'$ with respect to an observer $\mathcal{O}$
is given by 
\begin{equation}
r_{P'}^{\alpha}=h_{\ \beta}^{\alpha}({\rm exp}_{P}^{-1}P')^{\beta}\label{eq:rP'}
\end{equation}
where $P\in\mathcal{O}$ is the event of observation, $P'$ is the
event where the past light cone of $P$ intersects the worldline of
$\mathcal{O}'$ (so that $P'$ and $P$ are connected by a null geodesic),
and ${\rm exp}_{P}^{-1}P'$ is the inverse by the exponential map\footnote{$({\rm exp}_{P}^{-1}P')^{\alpha}$ can be defined as the vector $k^{\alpha}=dx^{\alpha}/d\lambda$
at $P$, tangent to the null geodesic parametrized by the affine parameter
$\lambda$ such that $\lambda(P)=0$ and $\lambda(P')=1$. In the
optical coordinates in \cite{ELLISObsCosmology}, $({\rm exp}_{P}^{-1}P')^{\alpha}=(0,y,0,0)$
and $\|r_{P'}^{\alpha}\|=y$.} of $P'$ at $P$, see Fig. 3 in \cite{BolosIntrinsic} (cf. also
Fig. 1 in \cite{ELLISObsCosmology}). If $\mathcal{O}=\mathcal{O}(u)$
and $\mathcal{O}'=\mathcal{O}'(u)$ are at rest in the coordinates
of \eqref{eq:StatMetric}, then, since the metric is explicitly time-independent,
$r_{P'}^{\alpha}$, as well as the associated affine distance $\|r_{P'}^{\alpha}\|$,
are also constant.

\subsubsection{\textquotedblleft Gravitoelectromagnetic\textquotedblright{} (GEM)
fields}

Consider a (pointlike) test particle of worldline $x^{\alpha}(\tau)$,
4-velocity $dx^{\alpha}/d\tau\equiv U^{\alpha}$ and mass $m$. The
space components of the geodesic equation, $DU^{\alpha}/d\tau=0$,
yield,\footnote{\label{fn:Christoffel}Computing the Christoffel symbols $\Gamma_{00}^{i}=-e^{2\Phi}G^{i}$,
$\Gamma_{j0}^{i}=e^{2\Phi}\mathcal{A}_{j}G^{i}-e^{\Phi}H_{\ j}^{i}/2$,
and $\Gamma_{jk}^{i}=\Gamma(h)_{jk}^{i}-e^{\Phi}\mathcal{A}_{(k}H_{j)}^{\ i}-e^{2\Phi}G^{i}\mathcal{A}_{j}\mathcal{A}_{k}$,
where $H_{ij}\equiv e^{\Phi}[\mathcal{A}_{j,i}-\mathcal{A}_{i,j}]$.} for the line element \eqref{eq:StatMetric} \cite{LandauLifshitz,ZonozBell1998,NatarioQM2007,Analogies,Zonoz2019},
\begin{equation}
\frac{\tilde{D}\vec{U}}{d\tau}=\gamma\left[\gamma\vec{G}+\vec{U}\times\vec{H}\right]\equiv\frac{\vec{F}_{{\rm GEM}}}{m}\ ,\label{eq:QMGeo}
\end{equation}
where $\gamma=-U^{\alpha}u_{\alpha}=e^{\Phi}(U^{0}-U^{i}\mathcal{A}_{i})$
is the Lorentz factor between $U^{\alpha}$ and $u^{\alpha}$, 
\begin{align}
\left[\frac{\tilde{D}\vec{U}}{d\tau}\right]^{i} & =\frac{dU^{i}}{d\tau}+\Gamma(h)_{jk}^{i}U^{j}U^{k}\ ;\label{eq:3DAccel}\\
\Gamma(h)_{jk}^{i} & \equiv\frac{1}{2}h^{il}\left(h_{lj,k}+h_{lk,j}-h_{jk,l}\right)\ ,\label{eq:Christoffelh}
\end{align}
is a covariant derivative with respect to the spatial metric $h_{ij}$,
with $\Gamma(h)_{jk}^{i}$ the corresponding Christoffel symbols,
and 
\begin{equation}
G_{i}=-\Phi_{,i}\ ;\qquad\ H^{i}=e^{\Phi}\epsilon^{ijk}\mathcal{A}_{k,j}\quad(\epsilon_{ijk}\equiv\sqrt{h}[ijk])\label{eq:GEM1forms}
\end{equation}
are fields living on the space manifold $(\Sigma,h)$, dubbed, respectively,
``gravitoelectric'' and ``gravitomagnetic'' fields. These play
in Eq. (\ref{eq:QMGeo}) roles analogous to those of the electric
($\vec{E}$) and magnetic ($\vec{B}$) fields in the Lorentz force
equation, $DU^{i}/d\tau=(q/m)[\gamma\vec{E}+\vec{U}\times\vec{B}]^{i}$.
Equation (\ref{eq:3DAccel}) is the standard 3-D covariant acceleration,
thus Eq. (\ref{eq:QMGeo}) describes the acceleration of the curve
obtained by projecting the time-like geodesic onto the space manifold
$\Sigma$, being $\vec{U}$ its tangent vector. The latter is identified
in spacetime with the projection of $U^{\alpha}$ onto $\Sigma$:
$(\vec{U})^{\alpha}=h_{\ \beta}^{\alpha}U^{\beta}$ {[}so its space
components equal those of $U^{\alpha}$, $(\vec{U})^{i}=U^{i}${]}.

The physical interpretation of Eq. (\ref{eq:QMGeo}) is that, from
the point of view of the ``Killing'' or ``laboratory'' observers
of 4-velocity \eqref{eq:uLab}, the spatial trajectory of the test
particle will appear accelerated, as if acted upon by the fictitious
force $\vec{F}_{{\rm GEM}}$ (standing here for ``gravitoelectromagnetic''
force). In other words, these observers measure \emph{inertial} \emph{forces},
which arise from the fact that the laboratory frame is \emph{not inertial};
in fact, $\vec{G}$ and $\vec{H}$ are identified in spacetime, respectively,
with minus the acceleration and twice the vorticity of the laboratory
observers: 
\begin{equation}
G^{\alpha}=-\nabla_{\mathbf{u}}u^{\alpha}\equiv-u_{\ ;\beta}^{\alpha}u^{\beta}\;;\qquad H^{\alpha}=2\omega^{\alpha}=\epsilon^{\alpha\beta\gamma\delta}u_{\gamma;\beta}u_{\delta}\;.\label{eq:GEM Fields Cov}
\end{equation}

These fields are a generalization, to the exact theory, of the GEM
fields usually defined in post-Newtonian approximations, e.g.~\cite{Damour:1990pi,SoffelKlioner2008,Kaplan:2009,WillPoissonBook},
and (up to constant factors and sign conventions) in the linearized
theory approximations, e.g. \cite{CiufoliniWheeler,Harris_1992,ohanian_ruffini_2013,Ruggiero:2002GMeffects,Gralla:2010xg},
reducing to them in the corresponding limits \cite{PaperDragging,Invariants,ManyFaces,JantzenPN}.
They obey field equations resembling the Maxwell equations in a rotating
frame (cf. Table 2 of \cite{Analogies}), 
\begin{align}
 & \tilde{\nabla}\cdot\vec{G}=-4\pi(2\rho+T_{\ \alpha}^{\alpha})+{\vec{G}}^{2}+\frac{1}{2}{\vec{H}}^{2}\ ;\qquad\tilde{\nabla}\times\vec{G}=\ 0\ ;\label{eq:GFieldEq}\\
 & \tilde{\nabla}\cdot\vec{H}=-\vec{G}\cdot\vec{H}\;;\qquad\tilde{\nabla}\times\vec{H}=-16\pi\vec{J}+2\vec{G}\times\vec{H}\ ,\label{eq:HFieldEq}
\end{align}
where $\rho\equiv T^{\alpha\beta}u_{\alpha}u_{\beta}$ and $J^{\alpha}\equiv-T^{\alpha\beta}u_{\beta}$.
Here $\tilde{\nabla}$ denotes covariant differentiation with respect
to the spatial metric $h_{ij}$ {[}i.e., the Levi-Civita connection
of $(\Sigma,h)$, with Christoffel symbols \eqref{eq:Christoffelh}{]}.
The equations for $\tilde{\nabla}\cdot\vec{G}$ and $\tilde{\nabla}\times\vec{H}$
are, respectively, the time-time and time-space projections, with
respect to $u^{\alpha}$, of the Einstein field equations $R_{\alpha\beta}=8\pi(T_{\alpha\beta}^{\ }-\frac{1}{2}g_{\alpha\beta}^{\ }T_{\ \gamma}^{\gamma})$,
and the equations for $\tilde{\nabla}\cdot\vec{H}$ and $\tilde{\nabla}\times\vec{G}$
follow from \eqref{eq:GEM1forms}.

Another realization of the analogy is the ``precession'' of a gyroscope
(i.e., a spinning pole-dipole particle). According to the Mathisson-Papapetrou
equations \cite{Mathisson:1937zz,Papapetrou:1951pa}, under the Mathisson-Pirani
spin condition \cite{Costa:2012cy}, the spin vector $S^{\alpha}$
of a gyroscope of 4-velocity $U^{\alpha}$ is Fermi-Walker transported
along its center of mass worldline, $DS^{\alpha}/d\tau=S^{\mu}a_{\mu}U^{\alpha}$,
where $a^{\alpha}\equiv DU^{\alpha}/d\tau$. If the gyroscope's center
of mass is at rest in the coordinate system of (\ref{eq:StatMetric})
($U^{\alpha}=u^{\alpha}$), then the space part of this equation yields
\cite{Cilindros} 
\begin{equation}
\frac{d\vec{S}}{d\tau}=\frac{1}{2}\vec{S}\times\vec{H}\ ,\label{eq:SpinPrec}
\end{equation}
which is formally analogous to the precession of a magnetic dipole
in a magnetic field ($D\vec{S}/d\tau=\vec{\mu}\times\vec{B}$). The
physical meaning of this equation is that a system of spatial axes
fixed with respect to the basis vectors $\{\partial_{i}\}$ of the
coordinate system in \eqref{eq:StatMetric}, rotate, with respect
to axes fixed to the spin vectors of gyroscopes (which define the
local ``compass of inertia'' \cite{CiufoliniWheeler,MassaZordan,Analogies}),
with an angular velocity $\vec{\Omega}=\vec{H}/2$.

Notice, from Eq. \eqref{eq:QMGeo}, that $\vec{G}$ and $\vec{H}$
are the only inertial fields arising in the reference frame associated
to a ``Killing'' congruence of observers in a stationary spacetime;
should they vanish, that automatically implies such reference frame
to be inertial. If they asymptotically vanish, $\lim_{r\rightarrow\infty}\vec{G}=\lim_{r\rightarrow\infty}\vec{H}=\vec{0}$,
then the reference frame is \emph{asymptotically inertial}.

The 3-vector $\vec{\mathcal{A}}$, dubbed ``gravitomagnetic vector
potential\emph{,'' }manifests physically in effects like the synchronization
gap (Sec. \ref{subsec:Time-and-clock}) and the Sagnac effect. The
latter consists of the difference in arrival times of light-beams
propagating in opposite directions around a spatially closed path.
Along a photon worldline, $ds^{2}=0$; by (\ref{eq:StatMetric}),
this yields two solutions, the future-oriented one being $dt=\mathcal{A}_{i}dx^{i}+e^{-\Phi}dl$,
where $dl$ is the spatial distance element \eqref{eq:dl}. Consider
photons constrained to move within a closed loop $C$ in the space
manifold $\Sigma$ (for instance, along an optical fiber loop). Using
the + (-) sign to denote the anticlockwise (clockwise) directions,
the coordinate time it takes for a full loop is, respectively, $t_{\pm}=\oint_{\pm C}dt=\oint_{C}e^{-\Phi}dl\pm\oint_{C}\mathcal{A}_{i}dx^{i}$;
therefore, the Sagnac \emph{coordinate} time delay $\Delta t_{{\rm S}}$
is \cite{BiniJantzenMashhoon_Clock1,CiufoliniWheeler,Kajari:2009qy,Cilindros,Ruggiero_SagnacTestsReview,MinguzziSimultaneity,Gourgoulhon:SpecialRelativ}
\begin{equation}
\Delta t_{{\rm S}}\equiv t_{+}-t_{-}=2\oint_{C}\mathcal{A}_{i}dx^{i}=2\oint_{C}\bm{\mathcal{A}}\ .\label{eq:SagnacDiffForm}
\end{equation}
Observe that this is twice the synchronization gap \eqref{eq:syncgap}
along $C$, $\Delta t_{{\rm S}}=2\Delta t_{{\rm sync}}$.

\subsubsection{Zero angular momentum observers (ZAMOs)\label{subsec:ZAMOs}}

Consider an axisymmetric stationary spacetime, for which $\bm{\mathcal{A}}=\mathcal{A}_{\phi}{\bf d}\phi$.
In spite of being at rest, the observers \eqref{eq:uLab} have, in
general, non-zero angular momentum per unit mass, whose component
along the symmetry axis is $u_{\phi}=u^{0}g_{0\phi}=e^{\Phi}\mathcal{A}_{\phi}$
\cite{Misner:1974qy,SemerakGRG1998,Cilindros,BardeenPressTeukolsky}.
This manifests itself in the fact that these observers measure a Sagnac
effect, via Eq. \eqref{eq:SagnacDiffForm}. Another important class
of observers in these spacetimes are those in circular motion, $u_{{\rm Z}}^{\alpha}=u_{{\rm Z}}^{0}(\delta_{0}^{\alpha}+\Omega_{{\rm ZAMO}}\delta_{\phi}^{\alpha})$,
for which the angular momentum per unit mass vanishes, $(u_{{\rm Z}})_{\phi}=0$.
Their angular velocity is thus given by 
\begin{equation}
\Omega_{{\rm ZAMO}}\equiv\frac{u_{{\rm Z}}^{\phi}}{u_{{\rm Z}}^{0}}=-\frac{g_{0\phi}}{g_{\phi\phi}}\ .\label{eq:OmegaZamo}
\end{equation}
These are the only ``stationary'' observers that measure no Sagnac
effect \cite{Misner:1974qy,Semerak_Stationary,PaperDragging,Bardeen1970ApJ}.
If the coordinate system in \eqref{eq:StatMetric} is asymptotically
inertial (i.e., star-fixed), the fact that $\Omega_{{\rm ZAMO}}\ne0$
reflects a form of frame-dragging \cite{Misner:1974qy}, which one
may dub ``dragging of the ZAMOs'' \cite{PaperDragging}. The worldlines
of these observers are orthogonal to the hypersurfaces $t=const$;
their vorticity \eqref{eq:GEM Fields Cov} thus vanishes, $\omega^{\alpha}=0$.
Motivated by these properties, a sometimes confusing terminology,
where similar names mean different things, is commonly used for, or
in connection to the ZAMOs:
\begin{itemize}
\item nonrotating with respect to ``the local spacetime geometry'' \cite{Misner:1974qy}
(or ``locally nonrotating observers'' in \cite{Bardeen1970ApJ,BardeenPressTeukolsky}):
meant in the sense of the ZAMOs measuring no Sagnac effect, thus regarding
the $\pm\phi$ directions as geometrically equivalent. This property
pertains to circular loops around the symmetry axis; the word ``local''
(which can be misleading) means here points with the same $r$ and
$z$ (or $\theta$) \cite{Bardeen1970ApJ}. 
\item ``Nonrotating congruence'' of observers: commonly employed in the
literature on exact solutions (e.g. \cite{StephaniExact,WyllemanBeke2010})
to designate observers with vanishing vorticity $\omega^{\alpha}$
(i.e., being hypersurface orthogonal). This is a local, tensor property,
pertaining to a \emph{congruence} of timelike curves. Physically,
it means (see e.g. \cite{Cilindros} p.7, and footnote therein) that
their connecting vectors do not rotate with respect to axes fixed
to \emph{local} guiding gyroscopes. In any spacetime, there are infinitely
many such congruences; the ZAMOs are a special case of these.
\item The ``locally nonrotating frames'' of Bardeen \emph{et al} \cite{Bardeen1970ApJ,BardeenPressTeukolsky}:
orthonormal tetrad frames $\{{\bf u}_{{\rm Z}},e_{\hat{\imath}}\}$
carried by the ZAMOs, whose spatial axes are parallel to the coordinate
basis vectors, $e_{\hat{\imath}}=(g_{ii})^{-1/2}\partial_{i}$ (for
$g_{ij}|_{i\ne j}=0$). This concerns a system of axes, and is the
most misleading designation since, except when the spacetime is static,
it is not nonrotating either in a local sense, since they are not
Fermi-Walker transported (thus rotate with respect to local gyroscopes),
nor is it fixed to the distant stars. 
\end{itemize}
It is crucial to not confuse any of these notions with the ``kinematically
nonrotating local reference system'' used in astrometry \cite{KlionerSoffel_Nonrotating1998,Klioner_Microarcsecond_2003},
which is a local system of axes nonrotating with respect to distant
reference objects (such confusion leads to grave misunderstandings,
as we shall see in Sec. \ref{sec:Examples}). Since $\Omega_{{\rm ZAMO}}$
in Eq. \eqref{eq:OmegaZamo} is not, in general, constant {[}$\Omega_{{\rm ZAMO}}\equiv\Omega_{{\rm ZAMO}}(r,z)$,
in cylindrical coordinates{]}, they are a \emph{shearing congruence}
of observers. In order to see this explicitly, first notice that the
ZAMOs congruence has no expansion: $\theta_{{\rm Z}}=(u_{{\rm Z}})_{\ ;\alpha}^{\alpha}=0$;
it follows that {[}$(h_{{\rm Z}})_{\ \beta}^{\alpha}\equiv u_{{\rm Z}}^{\alpha}(u_{{\rm Z}})_{\beta}+g_{\ \beta}^{\alpha}${]}
\begin{equation}
\sigma_{{\rm Z}}^{\alpha\beta}=(h_{{\rm Z}})_{\ \mu}^{\alpha}(h_{{\rm Z}})_{\ \nu}^{\beta}u_{{\rm Z}}^{(\mu;\nu)}=u_{{\rm Z}}^{0}\Omega_{{\rm ZAMO}}^{,(\alpha}\delta_{\phi}^{\beta)}\ ,\label{eq:ShearZAMO}
\end{equation}
having thus generically nonzero components $\sigma_{{\rm Z}}^{r\phi}=\sigma_{{\rm Z}}^{\phi r}$
and $\sigma_{{\rm Z}}^{z\phi}=\sigma_{{\rm Z}}^{\phi z}$. Hence,
unless $\Omega_{{\rm ZAMO}}={\rm constant}$ (in which case the spacetime
is static\footnote{When $\Omega_{{\rm ZAMO}}=constant$, the metric \eqref{eq:StatMetric}
can be globally diagonalized through the coordinate rotation $\phi'=\phi-\Omega_{{\rm ZAMO}}t$.}), reference frames associated to the ZAMOs congruence are not viable
generalizations of the IAU reference systems; namely, Eq. \eqref{eq:connectPN}
is not obeyed, so they cannot define fixed directions over an extended
region. In particular, their connecting vectors are not anchored to
inertial frames at infinity, even in the case where $\lim_{r\rightarrow\infty}\Omega_{{\rm ZAMO}}=0$.

\subsection{Relative velocities\label{subsec:Relative-velocities}}

The relative velocity $v^{\alpha}$ of an object $\mathcal{O}(u')$
of 4-velocity $u'^{\alpha}$ with respect to a reference observer
$\mathcal{O}(u)$ of 4-velocity $u^{\alpha}$ \emph{momentarily located
at the same point} (i.e., at an event where their worldlines cross)
is a well-defined notion, given by the relation \cite{ManyFaces}
\begin{equation}
u'^{\alpha}=\gamma(u^{\alpha}+v^{\alpha})\ ;\qquad\gamma\equiv-u^{\alpha}u'_{\alpha}=\frac{1}{\sqrt{1-v^{\alpha}v_{\alpha}}}\ .\label{eq:u_u'}
\end{equation}
The vector $v^{\alpha}$ lies in the rest space $u^{\perp}$ orthogonal
to $u^{\alpha}$ (so that $u^{\alpha}v_{\alpha}=0$), and is interpreted
as the spatial velocity of $u'^{\alpha}$ relative to $u^{\alpha}$.
In a locally \emph{inertial} frame momentarily comoving with $\mathcal{O}(u)$
(where $u^{i}=0$), its components yield the ordinary 3-velocity of
$\mathcal{O}(u')$: $v^{0}=0$, $v^{i}=dx^{i}/dt$. An useful expression
for its magnitude follows from \eqref{eq:u_u'}, 
\begin{equation}
v\equiv\sqrt{v^{\alpha}v_{\alpha}}=\sqrt{\frac{\gamma^{2}-1}{\gamma^{2}}}\ .\label{eq:vMaggamma}
\end{equation}
Notice that $v^{\alpha}$ is proportional to the projection of $u'^{\alpha}$
orthogonal to $u^{\alpha}$: $v^{\alpha}=h_{\ \beta}^{\alpha}u'^{\beta}/\gamma$,
being thus zero when $u^{\alpha}$ and $u'^{\alpha}$ are parallel.

There is, however, no unique, natural definition of relative velocity
for distant objects in general. The relative velocity between inertial
observers in flat spacetime is well defined: at some arbitrary event
$P$ of the reference worldline $\mathcal{O}(u)$, it is the velocity
of $\mathcal{O}(u')$ with respect to the inertial rest frame of $\mathcal{O}(u)$.
This implicitly amounts to comparing $u^{\alpha}$ to the vector ${u'_{P}}^{\alpha}$
at $P$ which is parallel to $u'^{\alpha}$ --- which is well defined,
since $u'^{\alpha}$ is constant, and there is a global notion of
parallelism (i.e, parallel transport is path-independent). In the
general case of curved spacetime (or, in flat spacetime, when the
observers are accelerated), one first needs to choose, on each worldline,
the events at which $u'^{\alpha}$ and $u^{\alpha}$ are to be compared.
A natural way is to consider an event $P'\in\mathcal{O}(u')$ which,
from the point of view of $\mathcal{O}(u)$, is simultaneous with
the event $P\in\mathcal{O}(u)$. That is, parallel transporting $u'^{\alpha}$
from $P'$ to $P$ along the spatial geodesic orthogonal to $\mathcal{O}(u)$
that connects to $P'$ , resulting in the vector $u'^{\alpha}|_{P}$
(see Fig. 2 in \cite{BolosIntrinsic}), and then computing $v^{\alpha}$
by the expression analogous to (\ref{eq:u_u'}) with $u'^{\alpha}|_{P}$
in the place of $u'^{\alpha}$. This has been been dubbed ``kinematical''
relative velocity (based on ``spacelike simultaneity''). However,
the measurement of the velocities of celestial bodies is made through
light rays; it reports thus to events connected by null geodesics
(``lightlike simultaneity''). The usual means of determining radial
velocities of stars and galaxies is by spectroscopic measurements
of the Doppler effect \cite{LindegrenDravins,GalaticAstronomy,SoffelBook2009}.
Let $k^{\alpha}=dx^{\alpha}/d\lambda$ be the null vector tangent
to the photon's worldline. As is well known, the affine parameter
$\lambda$ can be chosen such that $-u_{\alpha}k^{\alpha}=\nu$ yields
the photon frequency as measured by an observer $\mathcal{O}(u)$.
Then 
\begin{equation}
\frac{\nu'}{\nu}=\frac{u'_{\alpha}k_{P'}^{\alpha}}{u_{\alpha}k_{P}^{\alpha}}\ ,\label{eq:Redshift}
\end{equation}
which can be written as \cite{BolosIntrinsic} (cf. also \cite{SoffelBook2009})
\begin{equation}
\frac{\nu'}{\nu}=\gamma_{{\rm s}}(1\pm v_{{\rm rad}})\ ;\qquad\gamma_{{\rm s}}=\frac{1}{\sqrt{1-v_{{\rm s}}^{\alpha}(v_{{\rm s}})_{\alpha}}}\equiv-u'^{\alpha}|_{P}u_{\alpha}\ ,\label{eq:Doppler}
\end{equation}
where the spectroscopic velocity $v_{{\rm s}}^{\alpha}$ follows from
an expression analogous to (\ref{eq:u_u'}), replacing $u'^{\alpha}$
by the vector $u'^{\alpha}|_{P}$ obtained by parallel transporting
$u'^{\alpha}$ from $P'$ to $P$ along the photon's worldline (see
Fig. 4 in \cite{BolosIntrinsic}): 
\begin{equation}
v_{{\rm s}}^{\alpha}=\frac{u'^{\alpha}|_{P}}{\gamma_{{\rm s}}}-u^{\alpha}\ ;\label{eq:vspec}
\end{equation}
$v_{{\rm rad}}^{\alpha}\equiv v_{{\rm s}}^{\beta}(v_{{\rm ph}})_{\beta}v_{{\rm ph}}^{\alpha}$
is its radial component along the line of sight (tangent to $v_{{\rm ph}}^{\alpha}$),
$v_{{\rm ph}}^{\alpha}=k^{\alpha}/\nu-u^{\alpha}$ is the relative
velocity of the photon with respect to $\mathcal{O}(u)$, $v_{{\rm rad}}\equiv\|v_{{\rm rad}}^{\alpha}\|$,
and the $+$ ($-$) sign in \eqref{eq:Doppler} applies when $v_{{\rm ph}}^{\alpha}(v_{{\rm s}})_{\alpha}<0$
($>0$) {[}in flat spacetime, when the object is moving away from
(toward) $\mathcal{O}(u)${]}. To first order in $v_{{\rm s}}$, we
have $\nu'/\nu\simeq1\pm v_{{\rm rad}}$, allowing to compute $v_{{\rm rad}}$
from the measured redshift. The exact computation of $v_{{\rm rad}}$
from \eqref{eq:Doppler} requires however full knowledge of the vector
$v_{{\rm s}}^{\alpha}$, namely of its transverse components. They
can be approximately determined \cite{LindegrenDravins,ELLISObsCosmology}
by measuring the object's proper motion and distance.\footnote{The redshift $\nu'/\nu$, together with the object's proper motion,
allows to determine all the components $u'^{\alpha}$ in ``observational
coordinates'', see Sec. 4.1 of \cite{ELLISObsCosmology}; however,
constructing such coordinate system {[}including the ``optical''
distance $\|r_{P'}^{\alpha}\|$ from Eq. \eqref{eq:rP'}{]}, and the
geodesic connecting $P'$ to $P$, in order to determine $v_{{\rm s}}^{\alpha}$
via \eqref{eq:vspec}, would require an exact knowledge of the gravitational
field.}

Another possible definition of relative velocity is that given by
the variation, with respect to some spatial frame, of the observed
relative spatial position vector $r_{P'}^{\alpha}$, as defined in
Eq. \eqref{eq:rP'}. This is dubbed in \cite{BolosIntrinsic} ``astrometric''
relative velocity $v_{{\rm ast}}^{\alpha}$. Considering a Fermi-Walker
transported frame (as is more natural), it reads 
\begin{equation}
v_{{\rm ast}}^{\alpha}=h_{\ \beta}^{\alpha}\nabla_{u}r_{P'}^{\beta}\equiv h_{\ \beta}^{\alpha}u^{\gamma}\nabla_{\gamma}r_{P'}^{\beta}\ .\label{eq:vast}
\end{equation}
It generically does not coincide with the spectroscopic velocity in
\eqref{eq:vspec}; actually, it does not even reduce to (\ref{eq:u_u'})
when $P$ and $P'$ coincide. In the simplest case of flat spacetime,
it is only when $\mathcal{O}(u)$ is inertial ($u_{\ ;\beta}^{\alpha}u^{\beta}=0$)
that $v_{{\rm ast}}^{\alpha}=v_{{\rm s}}^{\alpha}$, cf. Eq. (23)
of \cite{BolosIntrinsic}.

There is no addition rule for any of these velocities: knowing the
relative velocity ($v_{{\rm s}}^{\alpha}$ or $v_{{\rm ast}}^{\alpha}$)
of $\mathcal{O}(u')$ and $\mathcal{O}(u'')$ with respect to $\mathcal{O}(u)$
at some event $P\in\mathcal{O}(u)$ does not allow us to determine
the velocity of $\mathcal{O}(u')$ relative to $\mathcal{O}(u'')$
(at any instant). In fact, the comoving notion is not transitive:
$\mathcal{O}(u)$ being comoving with $\mathcal{O}(u')$, and $\mathcal{O}(u')$
being comoving with $\mathcal{O}(u'')$, does \emph{not} imply that
$\mathcal{O}(u)$ is comoving with $\mathcal{O}(u'')$. The notions
are not even symmetric: $\mathcal{O}(u')$ at $P'$ being at rest
with respect to $\mathcal{O}(u)$ at $P$ does not mean that $\mathcal{O}(u)$
is at rest with respect to $\mathcal{O}(u')$ at the point $P'_{2}$
where the photon emitted by $\mathcal{O}(u)$ intersects $\mathcal{O}(u')$
\cite{BolosIntrinsic}.

The spectroscopic velocity \eqref{eq:Doppler}-\eqref{eq:vspec} is
moreover unsatisfactory in that, when $P'$ and $P$ do not coincide,
it embodies not only the relativistic Doppler effect (that one would
naturally associate with actual motion), but also the gravitational
redshift. This leads to unnatural results. For instance, the \emph{static}
observers in the Schwarzschild spacetime, tangent to the timelike
Killing vector field $\partial_{t}$, and usually understood as being
at rest with respect to the black hole's asymptotic inertial rest
frame, would not be at rest with respect to each other according to
Eqs. \eqref{eq:Doppler}-\eqref{eq:vspec}. Their astrometric relative
velocity \eqref{eq:vast}, in turn, is zero in Schwarzschild; but
is nonzero in the Kerr spacetime.

In the framework of a post-Newtonian approximation, some of these
subtleties are circumvented: at any given instant of coordinate time
$t$, the velocity of $\mathcal{O}(u')$ with respect to $\mathcal{O}(u)$
is simply the velocity of $\mathcal{O}(u')$ with respect to the PN
frame where $\mathcal{O}(u)$ is at rest. This notion is totally symmetric
and transitive. However, it makes sense only at an approximate level
and in this special class of reference frames; it cannot be cogently
extended for general frames. In order to see that, consider, e.g.,
the following examples: (i) a rotating wheel in flat spacetime is
at rest in its co-rotating frame; it is however clear, from the point
view of an inertial frame, that all of its points have different velocity,
only the center remaining a rest; (ii) in the comoving FLRW coordinates,
galaxies are approximately at rest, while the Doppler effect shows
that they are in fact getting farther apart; (iii) if the reference
frame is allowed to have shear, then it is always possible to find
one where all the bodies of a given system are at rest.

In the special case of a stationary asymptotically flat spacetime,
where there is a timelike Killing vector field $\partial_{t}$ tangent
to observers which, at infinity, are inertial, such observers form
a \emph{rigid} grid anchored to the asymptotic inertial frame. We
consider such observers to be at rest relative to each other, and
to the asymptotic inertial frame. This can be extended to spacetimes
not necessarily asymptotically flat (e.g. the van Stockum exterior
solution considered in Sec. \ref{subsec:The-van-Stockum}) or stationary,
but simply admitting a rigid asymptotically inertial observer congruence.

\subsection{Rotation curves\label{subsec:Rotation-curves}}

The galactic rotation curves are a plot of the orbital speed of visible
stars vs. distance with respect to the center of the galaxy. It is
essentially a Newtonian concept, where Galilean velocity addition
rules are used to compute velocities relative to the center of the
galaxy from measurements made in the solar system. It applies as well
to a first post-Newtonian approximation. In the exact theory, the
construction does not hold in terms of the relative velocities between
distant objects discussed in Sec. \ref{subsec:Relative-velocities},
given the fact that these report to the point of observation $P$,
and the lack of transitivity and addition rules. In order to obtain
a cogent plot one would need to determine $v_{{\rm s}}^{\alpha}$
or $v_{{\rm ast}}^{\alpha}$ with respect to an observer $\mathcal{O}(u)$
at the center, which is not possible for a galaxy given the central
super-massive black hole therein. {[}As explained in Sec. \ref{subsec:Relative-velocities},
above, $v_{{\rm s}}^{\alpha}$, as given by Eq. \eqref{eq:vspec},
is not appropriate when gravitational effects are strong{]}.

On the other hand, given a spherical or cylindrical-type coordinate
system, the concept of angular velocity, $\Omega=d\phi/dt$ is always
well defined. If the spacetime is stationary and axisymmetric (which,
in rigor, is not the case of actual galaxies), then a constant $\Omega$
signals rigid rotation (see Sec. \ref{subsec:Stationary-spacetimes}),
since $\partial_{\phi}$ and $\partial_{t}$ are both Killing vectors
fields, and thus the corresponding 4-velocity $U^{\alpha}=U^{0}(\partial_{t}^{\alpha}+\Omega\partial_{\phi}^{\alpha})$
is proportional to the Killing vector field $\partial_{t}^{\alpha}+\Omega\partial_{\phi}^{\alpha}$.
If the spacetime is moreover asymptotically flat, and the coordinate
system such that $\lim_{r\rightarrow\infty}g_{00}=-1$, $\Omega$
has the interpretation of angular velocity as measured by an observer
$\mathcal{O}(u)$ at rest at infinity (whose proper time is $\tau_{u}=t$).
This corresponds to a curve of angular velocities measured with respect
to an universal time $t$, in agreement with the IAU framework, cf.
Sec. \ref{subsec:Time-and-clock}. We will henceforth consider rotation
curves based on the angular velocity with respect to the given coordinate
system where the metric is explicitly time-independent, or, alternatively
(for purposes of comparison with some notions in the literature) on
the ``coordinate speed''\footnote{Defining equatorial coordinates $x=r\cos\phi$, $y=r\sin\phi$, we
have $v_{{\rm c}}=\sqrt{(dx/dt)^{2}+(dy/dt)^{2}}$.} $v_{{\rm c}}(r)\equiv\Omega r$, so that rigid motion is a straight
line of slope $\Omega$.

An alternative definition of rotation curve would be to consider the
relative velocity $v=\sqrt{v^{\alpha}v_{\alpha}}$ as given in Eq.
(\ref{eq:u_u'}), with respect to the observers $u^{\alpha}=(-g_{00})^{-1/2}\partial_{t}^{\alpha}$
at rest in the given coordinate system. Observe that, in the Lorentz
frame momentarily comoving with $\mathcal{O}(u)$, $v^{i}=dx^{i}/d\tau_{u}$.
Hence, that would amount to plot velocities as measured by different
observers measuring different proper times $\tau_{u}=(-g_{00})^{1/2}t$
since the redshift factor $(-g_{00})^{1/2}$ is not a constant in
general. Even in the PN regime, they do not correspond to the velocities
$dx^{i}/dt$ with respect to a given PN coordinate system, and have
a counter-intuitive behavior; for instance, the function $v(r)$ is
not a straight line for rigid motion (see Fig. \ref{fig:Stockum}
below).

\section{Examples\label{sec:Examples}}

As seen in Secs. \ref{subsec:Stationary-spacetimes} and \ref{subsec:Generalization-of-IAU},
the Killing observers are a privileged class in stationary spacetimes,
since they form rigid congruences; when they are moreover asymptotically
inertial, their associated coordinate systems are a suitable generalization
of the IAU reference system, which we shall next exemplify with the
Kerr, NUT, van Stockum dust cylinder, and spinning cosmic string spacetimes.
In some recent literature proposing certain exact dust solutions \cite{BG,Crosta2018,RuggieroBG}
--- namely the Balasin-Grumiller (BG) solution \cite{BG} --- as
a galactic models, however, the ZAMOs congruence has been used instead
as reference observers for computing rotation curves. In what follows,
we will evince their inadequacy for this purpose in the above-mentioned
well-known physical examples, and, finally, dissect the BG solution
and the fundamental misconceptions leading to its use as a galactic
model.

\subsection{Kerr spacetime\label{subsec:Kerr-spacetime}}

The metric reads, in Boyer-Lindquist coordinates, 
\begin{eqnarray}
ds^{2} & = & -\frac{\Delta}{\Sigma}\left(dt-a\sin^{2}\theta d\phi\right)^{2}+\frac{\Sigma}{\Delta}d\re^{2}+\Sigma d\theta^{2}\nonumber \\
 &  & +\frac{\sin^{2}\theta}{\Sigma}\left[adt-(a^{2}+\re^{2})d\phi\right]^{2}\,,\label{eq:kerr}
\end{eqnarray}
where $\Sigma\equiv\re^{2}+a^{2}\cos^{2}\theta$ and ${\displaystyle \Delta\equiv\re^{2}-2M\re+a^{2}}$.
Observers with 4-velocity $u^{\alpha}=(-g_{00})^{-1/2}\partial_{t}^{\alpha}$,
tangent to the Killing vector field $\bm{\xi}=\partial_{t}$, are
known as ``static'' observers (e.g. \cite{Bardeen1970ApJ,Bini_KerrMetric}).
They form a rigid congruence, and the associated reference frame is
asymptotically inertial since $\lim_{\re\rightarrow\infty}g_{\alpha\beta}=\eta_{\alpha\beta}$.
The coordinate system $\{t,\re,\theta,\phi\}$ is therefore, by Proposition
\ref{subsec:Generalization-of-IAU}, the generalization of the IAU
reference system for this spacetime. The post-Newtonian limit of the
metric is moreover well defined (obtained by neglecting all terms
terms quadratic in $a$ and, in the case of $g_{ij}$, all nonlinear
terms), 
\begin{align*}
ds^{2}= & -\left(1-\frac{2M}{\re}\right)dt^{2}-\frac{4aM}{\re}\sin^{2}\theta d\phi dt\\
 & +\left(1+\frac{2M}{\re}\right)d\re^{2}+\re^{2}d\theta^{2}+\re^{2}\sin^{2}\theta d\phi^{2}\ ,
\end{align*}
which, through a suitable transformation $\re=r_{{\rm PN}}+M$ to
the ``post-Newtonian'' radial coordinate $r_{{\rm PN}}=\sqrt{x^{2}+y^{2}+z^{2}}$
(equaling, to this accuracy, the ``radial'' coordinate of either
harmonic or isotropic coordinates, cf. \cite{WillPoissonBook} pp.
269-270) yields indeed a metric in the form \eqref{eq:PNmetric},
with $w=M/r_{{\rm PN}}$, $\vec{\mathcal{A}}=-2\vec{J}\times\vec{r}_{{\rm PN}}/r_{{\rm PN}}^{3}$,
where $\vec{J}=aM\vec{e}_{z}$ is the black hole's angular momentum. 

This reference frame, however, is valid only outside the ergosphere
\cite{Bardeen1970ApJ,BardeenPressTeukolsky,Misner:1974qy}, since,
for $\re<M+\sqrt{M^{2}-a^{2}\cos^{2}\theta}\equiv r_{{\rm erg}}$,
$\partial_{t}$ becomes spacelike ($g_{00}>0$), i.e., observers fixed
with respect to the distant stars are no longer possible. Still, rigid
observer congruences, tangent to Killing vector fields of the form
$\zeta^{\alpha}=\partial_{t}^{\alpha}+\varpi\partial_{\phi}^{\alpha}$
(for $\varpi$ some constant such that $\zeta^{\alpha}\zeta_{\alpha}<0$
at the given radius) are possible inside the ergosphere $r_{+}<\re<r_{{\rm erg}}$.
Inside the horizon $\re<r_{+}$, the radial coordinate $\rho$ becomes
timelike, and the metric is no longer stationary (since no timelike
Killing vector fields exist therein).

Other reference observers, not corresponding to an extension of the
IAU reference system, prove sometimes suitable for some applications
\cite{Bardeen1970ApJ,BardeenPressTeukolsky,Semerak1994A&A,Semerak_CQG1996_GyroscopePrec}.
That is the case of the ZAMOs, sometimes suggested to be the generalization
of the Newtonian ``nonrotating observers'' \cite{Semerak1994A&A}
(and even having, somewhat misleadingly, been dubbed ``rest frame''
of asymptotically flat axisymmetric spacetimes \cite{Greene_et_al_1975}).
Caution, however, is needed in interpreting their properties, as well
as with extended reference frames based on them. Indeed, in recent
literature they have been incorrectly claimed to be at rest relative
to the ``asymptotic observer who is at rest with respect to the rotation
axis'' \cite{BG,Crosta2018}, and then used to compute rotation curves
--- an application for which they are in general unsuitable, as we
shall see.

The angular velocity of the ZAMOs is, cf. Eq. \eqref{eq:OmegaZamo},
\begin{figure}
\includegraphics[width=1\columnwidth]{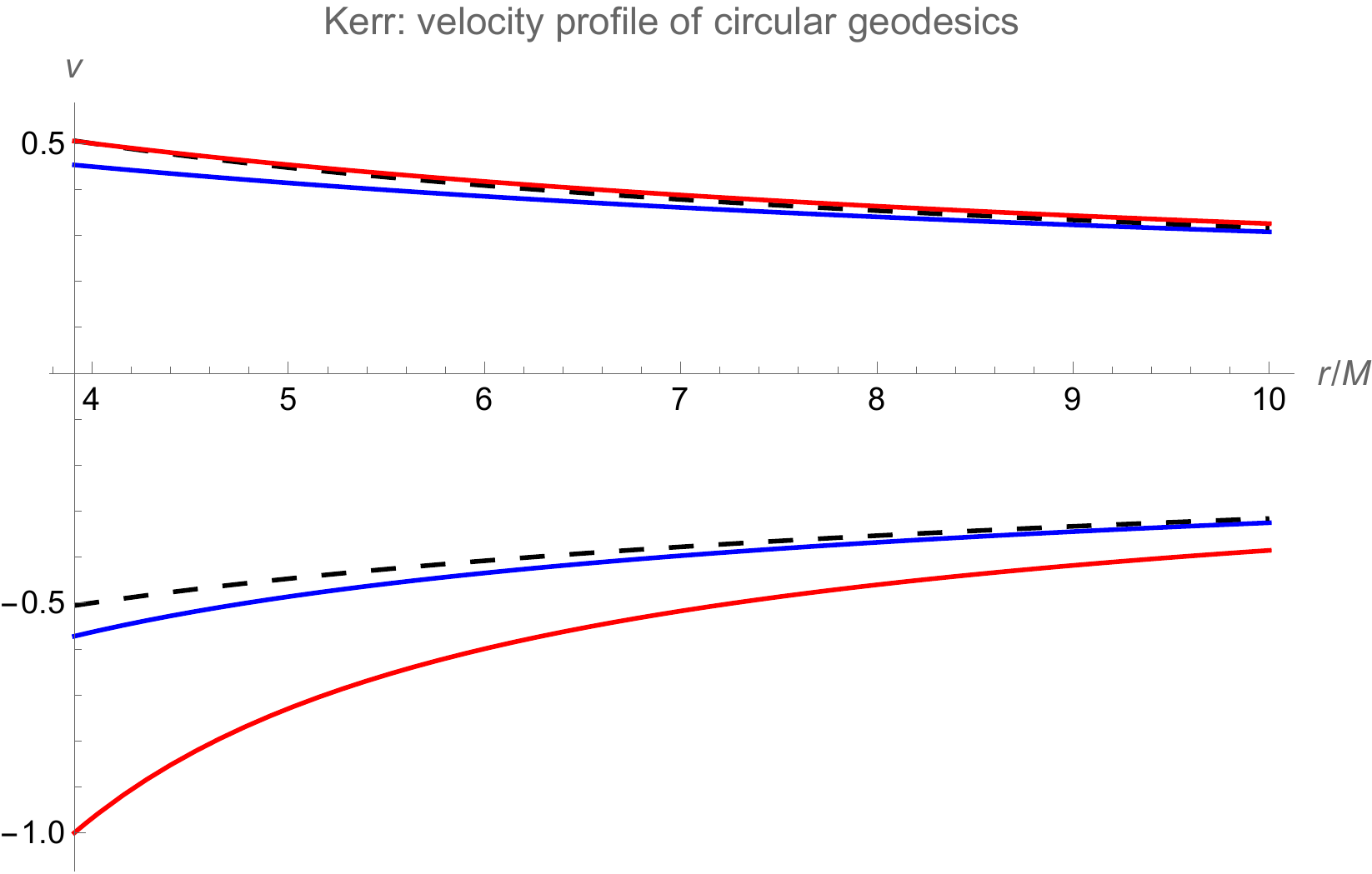} \caption{\label{fig:Kerr}Velocity profile for circular equatorial geodesics
in the Kerr spacetime. Dashed line is the Keplerian result $v_{{\rm K}}=\pm\sqrt{M/r}$,
blue solid lines are the general relativistic \textquotedblleft coordinate
velocity\textquotedblright{} $v_{{\rm c}}=\Omega_{{\rm geo}}r$, and
red solid lines represent the velocity with respect to the ZAMOs,
$v_{{\rm rZ}}=v_{{\rm rZ}}^{\hat{\phi}}$. As expected, $v_{{\rm c}}$
is lower (larger) than the Keplerian result for co-(counter-)rotating
geodesics, being, in both cases, small corrections, for not too small
$r$. Relative to the ZAMOs, however, there is a much larger difference
between co- and counter-rotating geodesics, and a larger deviation
from the Keplerian result, with $v_{{\rm rZ}}$ larger than the latter
\emph{even for co-rotating} geodesics.}
\end{figure}

\[
\Omega_{{\rm ZAMO}}(\re,\theta)=\frac{2aM\re}{(a^{2}+\re^{2})\Sigma+2a^{2}M\re\sin^{2}\theta}\ ,
\]
which is not constant. By Eq. \eqref{eq:ShearZAMO}, this implies
that the congruence shears, $\sigma_{\alpha\beta}\ne0$. These are
thus not observers attached to any rigid frame, and, in particular
to the black hole's asymptotic inertial rest frame. The closer to
the ergosphere, the faster they move with respect to such frame (thus
the more inadequate as reference observers for rotation curves). The
velocity $v_{{\rm rZ}}$ of any circular motion with respect to the
ZAMOs is readily computed from Eq. \eqref{eq:vMaggamma}, 
\begin{equation}
v_{{\rm rZ}}=\pm\sqrt{\frac{\gamma_{{\rm rZ}}^{2}-1}{\gamma_{{\rm rZ}}^{2}}}\:;\qquad\gamma_{{\rm rZ}}\equiv-U_{{\rm circ}}^{\alpha}(u_{{\rm Z}})_{\alpha}\ ,\label{eq:VrelZamoGamma}
\end{equation}
where $u_{{\rm Z}}^{\alpha}=u_{{\rm Z}}^{0}(\delta_{0}^{\alpha}+\Omega_{{\rm ZAMO}}\delta_{\phi}^{\alpha})$
is the ZAMOs 4-velocity, $U_{{\rm circ}}^{\alpha}=U^{0}(\delta_{0}^{\alpha}+\Omega_{{\rm circ}}\delta_{\phi}^{\alpha})$
is the 4-velocity of the circular motion, $\Omega_{{\rm circ}}$ its
angular velocity, and the $+$($-$) sign applies to prograde (retrograde)
motion. It can also be computed from Eq. \eqref{eq:u_u'}, identifying
$\{u'^{\alpha},u^{\alpha}\}\rightarrow\{U_{{\rm circ}}^{\alpha},u_{{\rm Z}}^{\alpha}\}$,
and recalling that $(u_{{\rm Z}})_{i}=0$, to obtain $v_{{\rm rZ}}^{0}=0\Rightarrow v_{{\rm rZ}}^{\alpha}=v_{{\rm rZ}}^{\phi}\delta_{\phi}^{\alpha}$
and 
\begin{equation}
v_{{\rm rZ}}^{\phi}=\sqrt{-g^{00}}(\Omega_{{\rm circ}}-\Omega_{{\rm ZAMO}});\qquad v_{{\rm rZ}}=v_{{\rm rZ}}^{\phi}\sqrt{g_{\phi\phi}}\equiv v_{{\rm rZ}}^{\hat{\phi}}\ .\label{eq:velwrtZamo}
\end{equation}
It is thus proportional to the difference between the angular velocities
of the circular motion and the ZAMOs, cf. \cite{BardeenPressTeukolsky,Crosta2018}.
Here $v_{{\rm rZ}}^{\hat{\phi}}$ is the azimuthal component in an
orthonormal tetrad having $e_{\hat{\phi}}=g_{\phi\phi}^{-1/2}\partial_{\phi}$
as one of its axis. One may check that Eqs. \eqref{eq:VrelZamoGamma}-\eqref{eq:velwrtZamo}
are equivalent to the definition used in \cite{BG,Crosta2018,RuggieroBG}
(where the tetrad $\{{\bf u}_{{\rm Z}},g_{\re\re}^{-1/2}\partial_{\re},g_{\theta\theta}^{-1/2}\partial_{\theta},g_{\phi\phi}^{-1/2}\partial_{\phi}\}$,
comoving with the ZAMOs, is chosen). 

The rotation curves obtained for circular equatorial geodesics in
the Kerr spacetime are shown in Fig. \ref{fig:Kerr}. Both $v_{{\rm rZ}}$
and the usual coordinate velocity $v_{{\rm c}}$ with respect to the
distant stars, as defined in Sec. \ref{subsec:Rotation-curves}, are
plotted. Whereas $v_{{\rm c}}$ exhibits the expected behavior ---
namely, the velocity of the co-(counter-)rotating geodesic is lower
(larger) than the Keplerian result, being, in both cases, a small
correction to the latter for not too small $r$ --- , $v_{{\rm rZ}}$,
in turn, exhibits a much larger deviation from the Keplerian result,
with the awkward feature that even the co-rotating geodesic is faster
than the Keplerian result, for arbitrarily large radius. This is down
to the fact that $v_{{\rm rZ}}^{\alpha}$ is a velocity in terms not
of an universal coordinate time, but of the proper times $\tau_{{\rm Z}}$
of the local ZAMOs.

More importantly, at the horizon $r_{+}=M+\sqrt{M^{2}-a^{2}}$, the
angular velocity of the ZAMOs coincides with that of the horizon:
\[
\Omega_{{\rm ZAMO}}(r_{+},\theta)=\Omega_{{\rm ZAMO}}(r_{+})=\frac{a}{r_{+}^{2}+a^{2}}=\Omega_{{\rm H}}
\]
i.e., such observers \emph{comove} with the horizon. Hence, by confusing
the ZAMOs with observers at rest with respect to the distant stars,
one would conclude by Eq. \eqref{eq:velwrtZamo} that Kerr black holes
do not rotate after all. This evinces the absurdity one may be led
to by using the ZAMOs as reference observers for rotation curves.

\subsection{NUT spacetime\label{subsec:NUT-spacetime}}

The NUT solution is a vacuum solution of the Einstein field equations
interpreted as describing a black hole endowed with a gravitomagnetic
monopole \cite{ZonozBell1998,ZonozNUTLine,BicakSolutionsEFE}. Different
versions of the line element can be found in the literature (see e.g.
\cite{GriffithsPodolsky2009}), one of them the following \cite{GriffithsPodolsky2009,BonnorNewNut,DowkerDyon1974}:
\begin{align}
ds^{2} & =-e^{2\Phi}(dt-\mathcal{A}_{\phi}d\phi)^{2}+e^{-2\Phi}d\re^{2}\nonumber \\
 & \ +(\re^{2}+l^{2})(d\theta^{2}+\sin^{2}\theta d\phi^{2})\ ;\label{eq:NUTmetric}
\end{align}
\begin{equation}
\mathcal{A}_{\phi}=-2l(\cos\theta-1)\ ;\qquad e^{2\Phi}=1-2\frac{M\re+l^{2}}{\re^{2}+l^{2}}\ ,\label{eq:PotentialsNUT}
\end{equation}
where $M$ is the total mass and $l$ is sometimes interpreted as
half the gravitomagnetic (or NUT) charge. The metric is locally (since
$\lim_{r\rightarrow\infty}R_{\alpha\beta\gamma\delta}=0$) but \emph{not}
globally asymptotically flat (as $g_{\alpha\beta}$ is not asymptotically
Minkowski when $l\ne0$). The component $g_{0\phi}$ is singular (nonvanishing)
along the semiaxis $\theta=\pi$; therefore, the metric \eqref{eq:NUTmetric}
is not defined therein. In the limit $l=0$, it reduces to the Schwarzschild
metric. The gravitoelectric and gravitomagnetic fields are, from Eqs.
\eqref{eq:GEM1forms}, 
\begin{eqnarray}
G^{i} & = & -h^{ij}\Phi_{,j}=\frac{l^{2}(M-2\re)-M\re^{2}}{(l^{2}+\re^{2})^{2}}\delta_{\re}^{i}\ ;\label{eq:GNUT}\\
H^{i} & = & e^{\Phi}\epsilon^{ijk}\mathcal{A}_{k,j}=-2l\frac{l^{2}+(2M-\re)\re}{(l^{2}+\re^{2})^{2}}\delta_{\re}^{i}\ ,\label{eq:HNut}
\end{eqnarray}
both being radial. Asymptotically, $\lim_{r\rightarrow\infty}\vec{G}=\lim_{r\rightarrow\infty}\vec{H}=\vec{0}$
{[}i.e., the acceleration $u_{\ ;\beta}^{\alpha}u^{\beta}=-G^{\alpha}$
and vorticity $\omega^{\alpha}=H^{\alpha}/2$ asymptotically vanish,
cf. Eq. \eqref{eq:GEM Fields Cov}{]}; hence, the reference frame
associated to the coordinate system in Eq. (\ref{eq:NUTmetric}) is
asymptotically inertial. In other words, fixed to the ``distant stars''.

The radial gravitomagnetic field \eqref{eq:HNut} corresponds to a
gravitomagnetic monopole, whose origin remains however an open question.
Inspired by a magnetic analogy, some authors have suggested that it
consists of a gravitomagnetic (or NUT) charge \cite{ZonozBell1998,ZonozNUTLine,BicakSolutionsEFE},
defined as follows. Let $\mathcal{S}$ be a closed 2-surface on the
space manifold $(\Sigma,h)$, and assume ${\bf d}\boldsymbol{\mathcal{A}}$
to be well defined along $\mathcal{S}$; the NUT charge enclosed in
$\mathcal{S}$ given by (cf. \cite{ZonozBell1998,ZonozNUTLine,GibbonsHawkingInstanton})
\begin{equation}
Q_{{\rm NUT}}=\frac{1}{4\pi}\int_{\mathcal{S}}{\bf d}\bm{\mathcal{A}}=\frac{1}{4\pi}\int_{\mathcal{S}}(\tilde{\nabla}\times\vec{\mathcal{A}})\cdot\vec{d}\mathcal{S}=\frac{1}{4\pi}\int_{\mathcal{S}}e^{-\Phi}\vec{H}\cdot\vec{d}\mathcal{S}\ ,\label{eq:QNUT}
\end{equation}
where $(\tilde{\nabla}\times\vec{\mathcal{A}})^{i}=\epsilon^{ijk}\mathcal{A}_{k,j}$
and $d\mathcal{S}_{k}\equiv\epsilon_{ijk}\mathbf{d}x^{i}\wedge\mathbf{d}x^{j}/2$
is an area element of $\mathcal{S}$ (``volume'' form of $\mathcal{S}$
\cite{Misner:1974qy}). Apart from the factor $e^{-\Phi}$ in the
integrand, Eq. \eqref{eq:QNUT} is formally analogous to the flat
spacetime definition of magnetic charge through the Gauss law, $Q_{{\rm M}}=\int_{\mathcal{S}}\vec{B}\cdot\vec{d}\mathcal{S}/4\pi$.
For a \emph{compact} 3-volume $\mathcal{V}$ with boundary $\mathcal{S}=\partial\mathcal{V}$
where\emph{ }${\bf d}\bm{\mathcal{A}}$ is well defined everywhere,
by virtue of the identity ${\bf d}({\bf d}\bm{\mathcal{A}})=0$ {[}$\Leftrightarrow$
$\tilde{\nabla}\cdot(\tilde{\nabla}\times\vec{\mathcal{A}})=0${]},
an application of the Stokes theorem yields $Q_{{\rm NUT}}=\int_{\mathcal{\mathcal{V}}}{\bf d}({\bf d}\bm{\mathcal{A}})/(4\pi)=0$.
The 2-form ${\bf d}\bm{\mathcal{A}}=2l\sin\theta{\bf d}\theta\wedge{\bf d}\phi$
is however singular at the origin $\re=0$, and has a well-defined
limit elsewhere along the axis {[}where the metric \eqref{eq:NUTmetric},
in rigor, is not defined{]}. Assuming it to be continuous therein,
and since it is well defined everywhere off the axis, by the Stokes
theorem (see e.g. \cite{Cilindros} Sec. 2.3) $Q_{{\rm NUT}}$ is
zero for any closed surface $\mathcal{S}$ not enclosing the singularity
at $\re=0$, and has the same (nonzero) value 
\begin{equation}
Q_{{\rm NUT}}=\frac{1}{4\pi}\int_{\mathcal{S}}2l\sin\theta{\bf d}\theta\wedge{\bf d}\phi=l\int_{0}^{\pi}\sin\theta d\theta=2l\label{eq:QNUT-NUT}
\end{equation}
when $\mathcal{S}$ encloses it. This is the justification for the
term NUT ``charge''.

Other authors suggested that the gravitomagnetic monopole arises from
a spinning cosmic string \cite{GriffithsPodolsky2009,BonnorNewNut,DowkerDyon1974}
--- since, from the electromagnetic analogue discussed in Appendix
\ref{subsec:EM2}, as well as the results from linearized theory in
\cite{BonnorNewNut}, one expects the tip of a semi-infinite spinning
string to also generate a monopole-like field. This leads however
to the possibility of a Dirac delta-type ${\bf d}\bm{\mathcal{A}}$
(thus $\vec{H}$) along the axis, canceling out the integral in \eqref{eq:QNUT},
analogously to the situation for a thin solenoid, Eq. \eqref{eq:FieldSemi}.
An interpretation in terms of spinning strings is indeed consistent,
as we shall now see. Let $\{t,\rc,\phi,z\}$ be a cylindrical coordinate
system such that $\re^{2}=\rc^{2}+z^{2}$ and $\cos\theta=z/\re$.
The Komar angular momentum \cite{Townsend,HansenWinicour1975,Cilindros,NatarioMathRel,BlackSaturn}
\begin{equation}
J=-\frac{1}{16\pi}\int_{\mathcal{\partial\mathcal{V}}}\star\mathbf{d}\bm{\zeta}\ ;\qquad\bm{\zeta}=\partial_{\phi}\ ,\label{eq:JKomarSurface}
\end{equation}
inside a cylinder of lateral surface $\mathcal{L}$, parametrized
by $\{\phi,z\}$, and top and bottom bases and $\mathcal{B}_{{\rm t}}$
and $\mathcal{B}_{{\rm b}}$, parametrized by $\{\rc,\phi\}$, is
\begin{equation}
J=-\frac{1}{16\pi}\left[\int_{\mathcal{B}_{{\rm t}}\sqcup\mathcal{B}_{{\rm b}}}(\star d\zeta)_{\rc\phi}{\bf d}\rc\wedge{\bf d}\phi+\int_{\mathcal{L}}(\star d\zeta)_{\phi z}{\bf d}\phi\wedge{\bf d}z\right].\label{eq:JCylinder}
\end{equation}
Here $(\star\mathbf{d}\bm{\zeta})_{\alpha\beta}\equiv\zeta_{\nu;\mu}\epsilon_{\ \ \alpha\beta}^{\mu\nu}$
is the 2-form dual to $\mathbf{d}\bm{\zeta}$. The explicit expressions
for the components $(\star d\zeta)_{\rc\phi}$ and $(\star d\zeta)_{\phi z}$
are given in the Supplemental Material \cite{EPAPSPRD}. Outside the
NUT black hole horizon $r_{{\rm H}}=M+\sqrt{M^{2}+l^{2}}$, the limit
$\lim_{\rc\rightarrow0}(\star d\zeta)_{\rc\phi}=0$ is well defined;
assuming $(\star d\zeta)_{\rc\phi}$ to be continuous at the axis
$\rc=0$ (namely, not of distribution-type therein, which is consistent
with both physical models), we can compute $J$ on cylinders whose
bases intersect the axis. Since moreover $\lim_{z\rightarrow\pm\infty}(\star d\zeta)_{\rc\phi}=0$,
for an infinitely long cylinder the integral \eqref{eq:JCylinder}
reduces to $J=-(1/8)\int_{-\infty}^{\infty}(\star d\zeta)_{\phi z}dz=-l\infty$.
For any cylinder above the horizon (i.e., a semi-infinite cylinder
$\infty>z>r_{{\rm H}}$) we have $J=0$; and for a semi-infinite cylinder
with $-\infty<z<r_{{\rm H}}$, $J=-l\infty$. This suggests the source
of $\bm{\mathcal{A}}$ to be a semi-infinite spinning string located
at $-\infty<z<0$, in agreement with the interpretation proposed in
\cite{GriffithsPodolsky2009,BonnorNewNut}.

Another form of the NUT metric given in the literature \cite{GriffithsPodolsky2009,NatarioQM2007}
is obtained by replacing, in Eqs. \eqref{eq:NUTmetric}-\eqref{eq:PotentialsNUT}
above, 
\begin{equation}
\mathcal{A}_{\phi}=-2l\cos\theta\ .\label{eq:ASemi}
\end{equation}
This corresponds to performing on \eqref{eq:NUTmetric}-\eqref{eq:PotentialsNUT}
the transformation $t'=t-2l\phi$. We first remark that such transformation,
while still assuming $\phi$ to be a periodic coordinate (i.e., having
closed integral lines), is a local but \emph{not global diffeomorphism};
i.e., not a globally valid coordinate transformation. As such, it
\emph{globally }changes the metric (for a detailed discussion of this
problem, we refer to Sec. 5.3.4 of \cite{Cilindros}). The metric
with \eqref{eq:ASemi} is now singular along the whole $z$-axis ($\theta=0\ \vee\ \theta=\pi$).
The NUT charge remains the same, $Q_{{\rm NUT}}=2l$, but the Komar
angular momentum is different: for an infinitely long cylinder, $J=-(1/8)\int_{-\infty}^{\infty}(\star d\zeta)_{\phi z}dz$
is now a nonconverging integral with zero principal value. Namely,
$\lim_{a\rightarrow\infty}\int_{-a}^{a}(\star d\zeta)_{\phi z}dz=0$,
but, e.g., $\lim_{a\rightarrow\infty}\int_{-a}^{2a}(\star d\zeta)_{\phi z}dz=l\infty$
and $\lim_{a\rightarrow\infty}\int_{-2a}^{a}(\star d\zeta)_{\phi z}dz=-l\infty$.
This suggests that the gravitomagnetic potential \eqref{eq:ASemi}
arises from \emph{a pair of} \emph{counter-rotating} \emph{semi-infinite
spinning strings}, one located at $-\infty<z<0$, and the other at
$0<z<\infty$. Computation of the Komar angular momentum contained
within finite cylinders supports this interpretation; and its analytical
value within a 2-sphere $\mathcal{S}$ of (arbitrary) radius $\rc=R$
is consistently {[}and contrary to the case for version \eqref{eq:PotentialsNUT}
of the metric{]} zero: 
\begin{align*}
J(R) & =-\frac{1}{16\pi}\int_{\mathcal{S}}(\star d\zeta)_{\theta\phi}d\theta d\phi\\
 & =l\frac{l^{2}(M-3R)+R^{2}(R-3M)}{4(R^{2}+l^{2})}\int_{0}^{\pi}\sin(2\theta)d\theta=0\ .
\end{align*}

Regardless of their physical interpretation (either as NUT charges
or semi-infinite spinning strings), it is important to note, in both
metrics \eqref{eq:PotentialsNUT} and \eqref{eq:ASemi}, that: (i)
the singularities drag the ZAMOs, causing them to have nonzero angular
velocity $\Omega_{{\rm ZAMO}}=-\mathcal{A}_{{\rm \phi}}e^{2\Phi}/g_{\phi\phi}$
with respect to the distant stars, cf. Eq. \eqref{eq:OmegaZamo};
for the metric version in \eqref{eq:ASemi}, it reads 
\[
\Omega_{{\rm ZAMO}}=-\left[2l\cos\theta+\frac{(l^{2}+\re^{2})\sin^{2}\theta}{2l\cos\theta(l^{2}+2M\re-\re^{2})}\right]^{-1}\ .
\]
(ii) They generate also a gravitomagnetic field $\vec{H}$ (i.e.,
the compass of inertia is also dragged), manifesting in the precession
of gyroscopes \eqref{eq:SpinPrec} and gravitomagnetic forces on test
particles \eqref{eq:QMGeo}. (iii) They change also the gravitoelectric
field comparing to that of the Schwarzschild solution; this is because
the gravitomagnetic field acts as a source for the gravitoelectric
field, via the first of Eqs. \eqref{eq:GFieldEq}, which in vacuum
reduces to $\tilde{\nabla}\cdot\vec{G}={\vec{G}}^{2}+{\vec{H}}^{2}/2$.

\subsection{The van Stockum rotating cylinder\label{subsec:The-van-Stockum}}

The van Stockum solution is an exact solution of the Einstein field
equations corresponding to an infinite, rigidly rotating cylinder
of dust. Let $R$ denote the cylinder's radius. In coordinates comoving
with the dust, the interior ($r<R$) metric takes the form \cite{Stockum1938,BonnorCQG1995,Cilindros}

\begin{equation}
ds^{2}=-(dt-w\rc^{2}d\phi)^{2}+e^{-w^{2}r^{2}}(d\rc^{2}+dz^{2})+\rc^{2}d\phi^{2}\ ,\label{eq:StockumInt}
\end{equation}
where $w$ is a positive constant. The exterior metric, in these coordinates,
is given by e.g. Eqs. (78)-(82) of \cite{Cilindros}, or Eqs. (2),
(11)-(14) of \cite{BonnorCQG1995}. The fact that the interior metric
\eqref{eq:StockumInt} is time-independent in comoving coordinates
reflects that the dust rotates rigidly (cf. Sec. \ref{subsec:Stationary-spacetimes}).
The vanishing gravitoelectric potential and field, $\Phi=0\Rightarrow\vec{G}=0$,
means that observers or particles at rest in these coordinates are
in fact following geodesics. This reflects that the dust is solely
driven by gravity which, in the co-rotating reference frame, can be
cast as the centrifugal inertial force exactly canceling out the gravitational
attraction. For $r>1/w$, the $\phi$ coordinate becomes timelike
($g_{\phi\phi}<0$), and so the integral lines of $\partial_{\phi}$
are closed timelike curves \cite{Tipler:1974gt} (in other words,
there is a ``time machine'' in that region).

Within the limit $wR<1/2$, the metric can be matched to an exterior
solution admitting a coordinate system fixed to the distant stars
(i.e., to the asymptotic inertial rest frame) \cite{Stockum1938,Cilindros}.
The angular velocity of the dust with respect to the star-fixed frame
is 
\begin{equation}
\Omega=\frac{w}{(1-2\lambda_{{\rm m}})^{2}}\ ,\label{eq:OmegadustStockum}
\end{equation}
were $\lambda_{{\rm m}}=[1-\sqrt{1-4w^{2}R^{2}}]/4$ is the cylinder's
Komar mass per unit $z$-length \cite{Cilindros}. In such star-fixed
coordinates, the interior metric takes the form in Eqs. (78), (88),
(101) and (102) of \cite{Cilindros}, whereas the exterior metric
reads 
\begin{figure*}
\includegraphics[width=0.9\columnwidth]{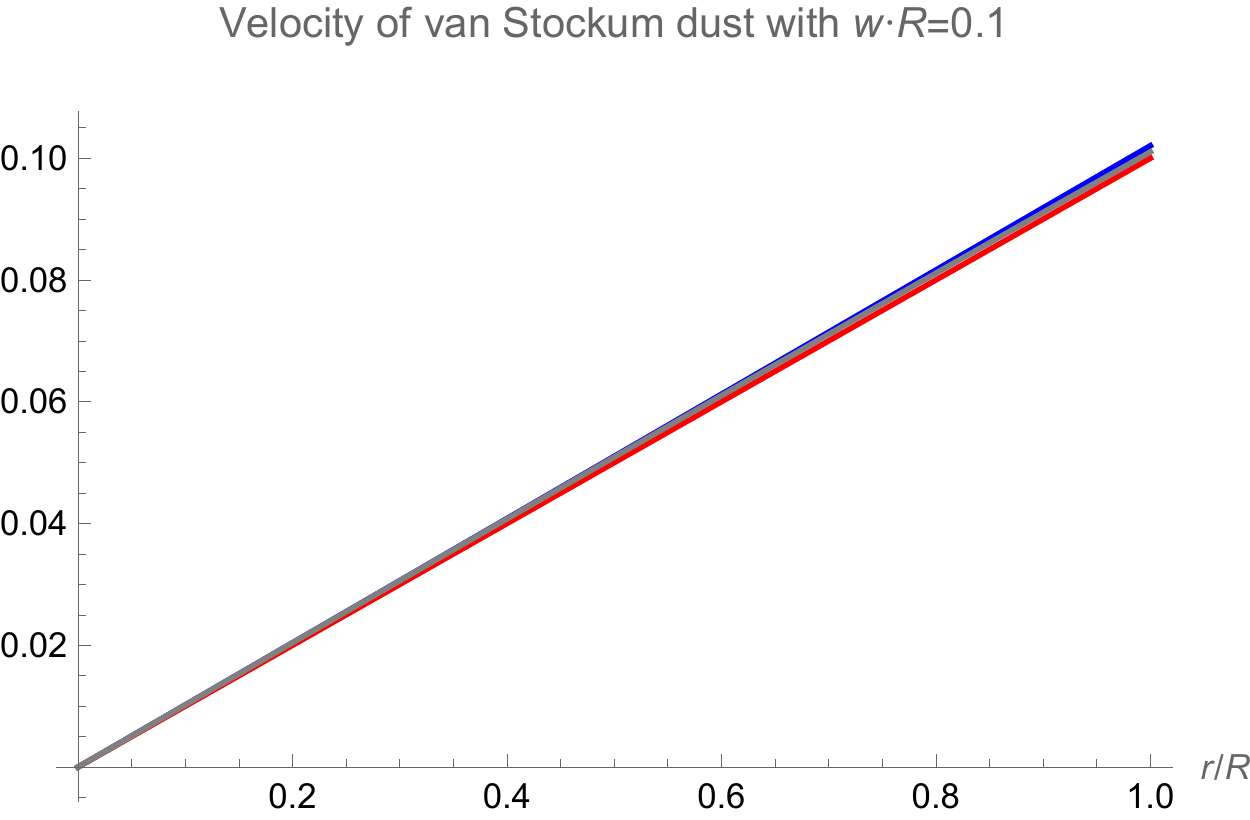}~~~\includegraphics[width=0.9\columnwidth]{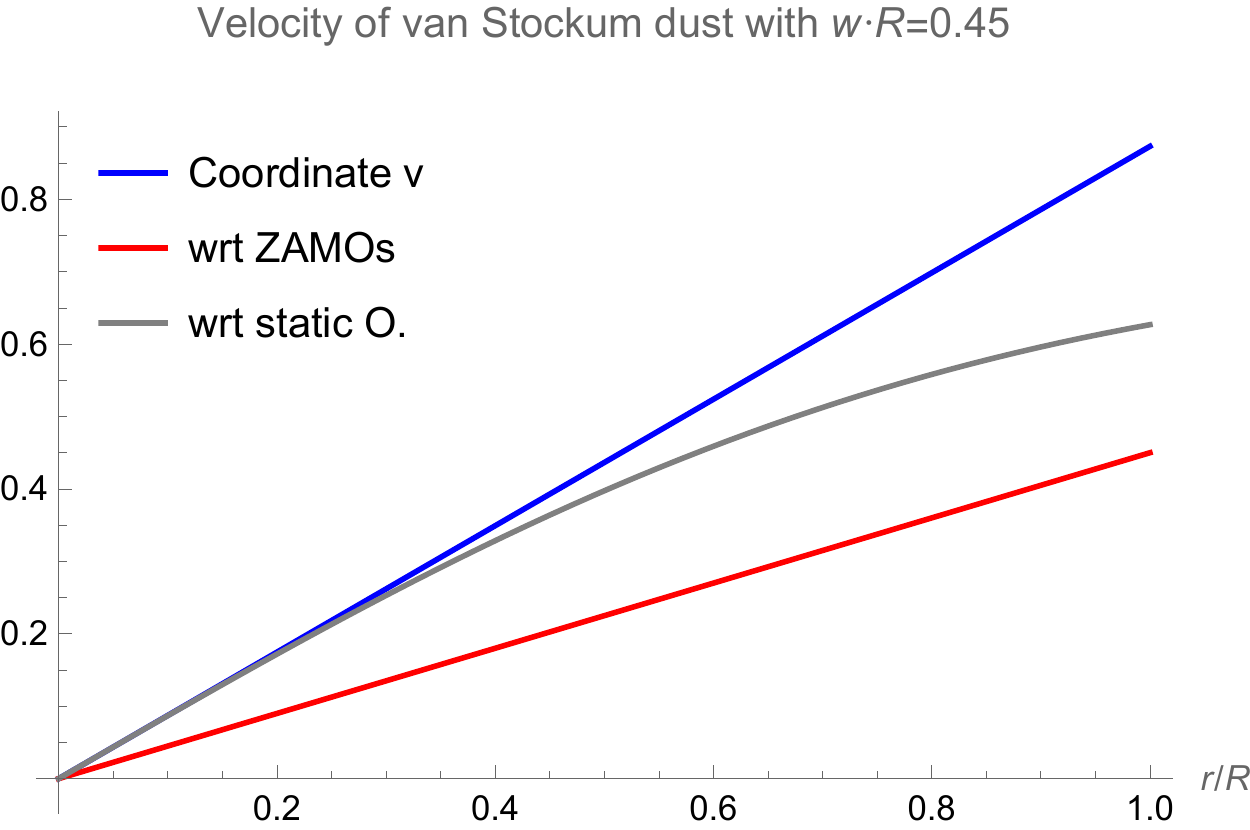}\caption{\label{fig:Stockum}Velocity profile for the van Stockum rigidly rotating
dust cylinder, according to three different definitions: the coordinate
velocity $v_{{\rm c}}=\Omega\rc$ with respect to star fixed coordinates
(blue line), the velocity $v_{{\rm rZ}}(\rc)$ relative to the ZAMOs
(red line), and the velocity $v(r)$ with respect to the static observers
{[}as given by Eqs. \eqref{eq:u_u'}-\eqref{eq:vMaggamma} with $u^{\alpha}=(-g_{00})^{-1/2}\partial_{t}^{\alpha}$,
gray line{]}. For slow cylinders (left panel), the three curves almost
coincide; for fast cylinders, where the dragging of the ZAMOs becomes
important, the velocity with respect to the ZAMOs greatly differs
from the other definitions --- even close to the axis $r=0$, where
the dust is slowly rotating.}
\end{figure*}

\begin{align}
ds^{2} & =-\frac{\rc^{4\lambda_{{\rm m}}}}{\alpha}\left(dt-\frac{j}{\lambda_{{\rm m}}-1/4}d\phi\right)^{2}+\alpha\rc^{2(1-2\lambda_{{\rm m}})}d\phi^{2}\nonumber \\
 & +\left[\frac{e^{1/2}r}{R}\right]^{-4\lambda_{{\rm m}}(1-2\lambda_{{\rm m}})}(d\rc^{2}+dz^{2})\ ,\label{eq:StockumExt}
\end{align}
where $\alpha\equiv R^{4\lambda_{{\rm m}}}(1-2\lambda_{{\rm m}})^{3}/(1-4\lambda_{{\rm m}})$
and $j=R^{4}w^{3}/4$ is the cylinder's Komar angular momentum per
unit $z$-length \cite{Cilindros}. The exterior gravitoelectric and
gravitomagnetic fields are, cf Eqs. \eqref{eq:GEM1forms}, 
\begin{equation}
G_{i}=-\frac{2\lambda_{{\rm m}}}{r}\delta_{i}^{\rc}\ ;\qquad\ \vec{H}=0\ .\label{eq:GEMStockumCan}
\end{equation}
Asymptotically, $\vec{G}\stackrel{r\rightarrow\infty}{=}\vec{0}$,
hence the reference frame associated to the coordinate system in (\ref{eq:StockumExt})
is \emph{asymptotically} inertial; it is also rigid (hence shear-free),
since the metric is explicitly time-independent in these coordinates
(cf. Sec. \ref{subsec:Stationary-spacetimes}). Therefore, it is indeed
a frame fixed to the ``distant stars'', and, by proposition \ref{subsec:Generalization-of-IAU},
the generalization of the IAU reference system to this solution.

The angular velocity of the ZAMOs with respect to the star-fixed frame
equals the sum of their angular velocity with respect to the dust
{[}i.e., to the coordinate system in \eqref{eq:StockumInt}{]} with
the angular velocity \eqref{eq:OmegadustStockum} of the dust relative
to the star-fixed frame, 
\begin{equation}
\Omega_{{\rm ZAMO}}(r)=\frac{w}{w^{2}\rc^{2}-1}+\Omega\qquad(\rc\le R)\ .\label{eq:OmegaZAMOCanonical}
\end{equation}
The ZAMOs are thus dragged by the cylinder's rotation, causing them
to describe circular motions. Notice that, since $\Omega_{{\rm ZAMO}}(\rc)$
depends on the radius, they are, by virtue of Eq. \eqref{eq:ShearZAMO},
a shearing congruence of observers. All this makes the ZAMOs unsuitable
as a reference frame for measuring the dust's rotation curve, as we
shall now see. Since the dust rotates rigidly with constant angular
velocity \eqref{eq:OmegadustStockum}, its rotation curve is, in star
fixed coordinates, the straight line $v_{{\rm c}}(r)=\Omega\rc$.
In Fig. \ref{fig:Stockum} this is plotted and compared with the curve
of the velocity relative to the ZAMOs, obtained from Eq. \eqref{eq:u_u'}
or \eqref{eq:vMaggamma} identifying $\{u'^{\alpha},u^{\alpha}\}\rightarrow\{u^{\alpha},u_{{\rm Z}}^{\alpha}\}$,
so that 
\begin{equation}
v_{{\rm rZ}}(r)=v_{{\rm rZ}}^{\hat{\phi}}(r)=w\rc\ .\label{eq:vZStockum}
\end{equation}
Interestingly, $v_{{\rm rZ}}(\rc)$ yields also a straight line.\footnote{In spite of being the relative velocity of a rigid dust with respect
to a shearing (thus nonrigid) congruence of observers. This can be
easily understood in the coordinates of \eqref{eq:StockumInt}, where
the dust is at rest and $\Omega_{{\rm ZAMO}}(\rc)=-w\rc^{2}/g_{\phi\phi}$.
Using $U_{{\rm dust}}^{i}=0$ and $u_{{\rm Z}}^{\phi}=\Omega_{{\rm ZAMO}}u_{{\rm Z}}^{0}$,
we have, from Eq. \eqref{eq:u_u'}, identifying $\{u'^{\alpha},u^{\alpha}\}\rightarrow\{U_{{\rm dust}}^{\alpha},u_{{\rm Z}}^{\alpha}\}$,
$v_{{\rm rZ}}^{\hat{\phi}}=v_{{\rm rZ}}^{\phi}\sqrt{g_{\phi\phi}}=-\Omega_{{\rm ZAMO}}u_{{\rm Z}}^{0}\sqrt{g_{\phi\phi}}$,
where $u_{{\rm Z}}^{0}=dt/d\tau_{{\rm Z}}=\sqrt{g_{\phi\phi}}/\rc$.
Thus, it is the radial dependence of the ZAMO's proper time (i.e.,
the fact that their clocks thick at different rates), combined with
that of $g_{\phi\phi}$, that exactly balances the radial variation
of $\Omega_{{\rm ZAMO}}(\rc)$.} Observe that, since the rotating dust is self-gravitating, the Komar
mass per unit length, $\lambda_{{\rm m}}$, is also a measure of the
cylinder's rotation speed; its allowed value range (for Weyl class
rotating cylinders \cite{Cilindros}) is $0<\lambda_{{\rm m}}<1/4$,
corresponding to $0<wR<1/2$. For slow cylinders (i.e., small $\lambda_{{\rm m}}$),
the dragging of the ZAMOs is small, and therefore $v_{{\rm c}}(r)=\Omega\rc$
is close to $v_{{\rm rZ}}(\rc)$, cf. Eqs \eqref{eq:OmegadustStockum}
and \eqref{eq:vZStockum}. For fast cylinders, however, they are very
different, as exemplified in the right panel of Fig. \ref{fig:Stockum}
(for $wR=0.4\Leftrightarrow\lambda_{{\rm m}}=0.1$).

\subsection{Cosmic string\label{subsec:Cosmic-string}}

The zero Komar mass ($\lambda_{{\rm m}}=0$) limit of \eqref{eq:StockumExt}
yields the exterior metric of a spinning cosmic string \cite{SantosGRG1995,Barros_Bezerra_Romero2003,JensenSoleng}
of angle deficit $2\pi(1-\alpha^{1/2})$, 
\begin{equation}
{\displaystyle {\displaystyle ds^{2}}=-\frac{1}{\alpha}\left[dt+4jd\phi\right]^{2}+d\rc^{2}+dz^{2}}+\alpha\rc^{2}d\phi^{2}\ .\label{eq:String}
\end{equation}
The spacetime is in this case locally flat ($R_{\alpha\beta\gamma\delta}=0$)
for $r\ne0$. The GEM inertial fields vanish, 
\[
\vec{G}=\vec{H}=0\ ,
\]
thus there are no gravitational forces of any kind. Only global gravitational
effects subsist, namely those governed by $\bm{\mathcal{A}}=-4j\mathbf{d}\phi$,
which include a Sagnac effect \eqref{eq:SagnacDiffForm} (thus dragging
of the ZAMOs) and a synchronization gap \eqref{eq:syncgap} along
closed loops $C$ enclosing the string (i.e., the axis $\rc=0$),
and those governed by $\alpha$, which include a holonomy \cite{Nouri-Zonoz:2013rfa,Ford_Vilenkin_1981}
along $C$ (vectors parallel transported along such loops turn out
rotated by an angle $-2\pi\alpha^{1/2}$ about the $z$-axis when
they return to the initial position), and double images of objects
located behind the string \cite{KibbleCosmicStrings,Ford_Vilenkin_1981}.

Since $\vec{G}=\vec{H}=0$, observers at rest in the coordinates of
\eqref{eq:String} are, as is well known, inertial, and fixed to the
distant stars. Such coordinates provide thus the generalization of
the IAU system for this spacetime. The ZAMOs, in turn, have angular
velocity 
\[
\Omega_{{\rm ZAMO}}=\frac{4j}{\alpha^{2}\rc^{2}-16j^{2}}
\]
with respect to the star-fixed frame. Relative to them, inertial bodies
which are at rest in the star fixed frame move along counter-rotating
circular trajectories with angular velocity $-\Omega_{{\rm ZAMO}}$,
and relative velocity
\begin{equation}
v_{{\rm rZ}}^{\phi}=-\sqrt{-g^{00}}\Omega_{{\rm ZAMO}}\ ;\qquad v_{{\rm rZ}}(r)=v_{{\rm rZ}}^{\hat{\phi}}=-\frac{4j}{\alpha\rc}\label{eq:vrZString}
\end{equation}
{[}the derivation is analogous to that of Eq. \eqref{eq:velwrtZamo},
with $u^{\alpha}=(-g_{00})^{-1/2}\delta_{0}^{\alpha}$ in the place
of $U_{{\rm circ}}^{\alpha}${]}. Thus, taking the perspective of
the ZAMOs, one would conclude that circular geodesics around a cosmic
string exist. Again, this hints on the absurdities one is led to by
using ZAMOs as reference observers for rotation curves (confusing
them with observers at rest with respect to asymptotic inertial frames):
circular orbits are impossible in this spacetime, since there is no
gravitational attraction to sustain them (cosmic strings exert no
gravitational attraction, as is well known).

The metric \eqref{eq:String} can be considered for all space excluding
the axis $\rc=0$; it describes, in this case, the exterior field
of an infinitely thin string. Along the axis $r=0$, where the gravitomagnetic
potential 1-form $\bm{\mathcal{A}}=-4j\mathbf{d}\phi$ {[}and thus
the metric \eqref{eq:String}{]} would be singular, the solution \eqref{eq:String}
is then assumed to be matched \cite{JensenSoleng} to the singular
limit of an interior solution with curvature $R_{\alpha\beta\gamma\delta}$
and energy momentum tensor $T^{\alpha\beta}$ in the form of Dirac
delta functions \cite{Nouri-Zonoz:2013rfa,Ford_Vilenkin_1981}. It
is in these singularities along the axis that \emph{all of the angular
momentum present in the spacetime is localized}, which can be seen
as follows. Using the identity $\mathbf{d}(\star\mathbf{d}\bm{\zeta})=-2R_{\alpha\beta}\zeta^{\beta}d\mathcal{V}^{\alpha}$,
where $\zeta^{\alpha}=\partial_{\phi}^{\alpha}$ and $d\mathcal{V}_{\alpha}=\epsilon_{\alpha\mu\nu\lambda}\mathbf{d}x^{\mu}\wedge\mathbf{d}x^{\nu}\wedge\mathbf{d}x^{\lambda}/6$
is the volume element 1-form of $\mathcal{V}$, to write the angular
momentum \eqref{eq:JKomarSurface} in terms of a volume integral,
we have \cite{Townsend,Cilindros} 
\begin{equation}
J=\frac{1}{8\pi}\int_{\mathcal{V}}R_{\ \beta}^{\alpha}\zeta^{\beta}d\mathcal{V}_{\alpha}\ .\label{eq:JKomarVolume}
\end{equation}
Since \eqref{eq:String} is a vacuum solution ($R_{\alpha\beta}=0$),
we see that the angular momentum inside any 2-surface $\partial\mathcal{V}\equiv\mathcal{S}$
not crossing the axis is zero. Consider now $\mathcal{V}$ to be a
simply connected tube, of arbitrary section, parallel to the $z$-axis,
and of length $\Delta z\equiv z_{{\rm t}}-z_{{\rm b}}$. Let $\partial\mathcal{V}=\mathcal{L}\cup\mathcal{B}_{1}\cup\mathcal{B}_{2}$
be the boundary of such tube, where $\mathcal{L}$ is the tube's lateral
surface, parametrized by $\{\phi,z\}$, and $\mathcal{B}_{{\rm t}}$
and $\mathcal{B}_{{\rm b}}$ are its top and bottom bases, parametrized
by $\{\rc,\phi\}$ and orthogonally crossing the $z$-axis at $z_{{\rm t}}$
and $z_{{\rm b}}$. From Eq. \eqref{eq:JCylinder}, noting that $(\star d\zeta)_{r\phi}=0$
(and assuming its continuity at $r=0$), and $(\star d\zeta)_{\phi z}=-8j$,
the Komar angular momentum inside $\partial\mathcal{V}$ is thus 
\begin{align}
J & =-\frac{1}{16\pi}\int_{z_{{\rm b}}}^{z_{{\rm t}}}dz\int_{0}^{2\pi}(\star d\zeta)_{\phi z}=j\Delta z\ \ .\label{eq:Jstring}
\end{align}
It follows from \eqref{eq:JKomarVolume} that, for any compact 3-volume
$\mathcal{V}'$ crossing the $z$-axis at the same points $z_{{\rm t}}$
and $z_{{\rm b}}$, the Komar angular momentum has the same value
\eqref{eq:Jstring}, regardless of the shape of $\mathcal{V}'$.

\subsection{The Balasin-Grumiller \textquotedblleft galactic\textquotedblright{}
toy-model\label{subsec:The-BG-galactic-model}}

In Ref. \cite{BG}, a metric in the form 
\begin{align}
 & ds^{2}=-(dt-Nd\phi)^{2}+r^{2}d\phi^{2}+e^{\nu}(dr^{2}+dz^{2})\ ,\label{eq:BGmetric}
\end{align}
where $\nu\equiv\nu(r,z)$ and 
\begin{align}
 & N(r,z)=V_{0}(R-r_{0})+\frac{V_{0}}{2}\left[d_{r_{0}}+d_{-r_{0}}-d_{R}-d_{-R}\right],\label{eq:N}\\
 & d_{R}\equiv\sqrt{r^{2}+(z-R)^{2}};\quad d_{-R}\equiv\sqrt{r^{2}+(z+R)^{2}};\label{eq:dR}\\
 & d_{r_{0}}\equiv\sqrt{r^{2}+(z-r_{0})^{2}};\quad d_{-r_{0}}\equiv\sqrt{r^{2}+(z+r_{0})^{2}},\label{eq:dr0}
\end{align}
has been claimed to describe, in comoving coordinates, a rotating
dust with a velocity profile similar to the observed galactic rotation
curves. Its parameters have the following interpretation: $r_{0}$
is the radius of the galactic bulge region, $R$ is the radial extension
of the galactic disk in the equatorial plane, and $V_{0}$ is claimed
to roughly represent the dust velocity, with respect to the ZAMOS,
in the ``flat regime.'' For the Milky Way, these values are taken
as $r_{0}=1{\rm kpc}$, $R=100{\rm kpc}$, $V_{0}=220{\rm km}/{\rm s}$.

\subsubsection{The dust is \emph{static\label{subsec:The-dust-is-static}}}

As correctly claimed in \cite{BG}, the ``dust'' is at rest in the
coordinates of \eqref{eq:BGmetric} (indeed, the mass-energy current
relative to such frame, $J^{i}=-T^{i\beta}u_{\beta}$, vanishes).
We note the formal similarities between \eqref{eq:BGmetric} and the
line element of the van Stockum interior solution \eqref{eq:StockumInt}.
Like in the latter, one is dealing with a \emph{rigid dust}, since
\emph{in comoving coordinates the metric is time-independent} (see
Sec. \ref{subsec:Stationary-spacetimes}). This alone would suffice
to immediately rule it out as a viable galactic model: the flat rotation
curves observed in galaxies are of course incompatible with rigid
motion: their angular velocity is nonconstant, which via \eqref{eq:ShearZAMO}
implies a shearing motion. The situation is, however, even worse,
as we shall see next.

Noticing that $\epsilon^{ijk}=[ijk]/\sqrt{h}$, where $h=r^{2}e^{2\nu}$
is the determinant of the space metric $h_{ij}$ as defined in Eq.
\eqref{eq:StatMetric}, the gravitomagnetic field \eqref{eq:GEM1forms}
is given by $H^{i}=\epsilon^{ijk}\mathcal{A}_{k,j}=\left[N_{,r}\delta_{z}^{i}-N_{,z}\delta_{r}^{i}\right]e^{-\nu}/r$,
reading, explicitly, 
\begin{figure*}
\includegraphics[width=1\columnwidth]{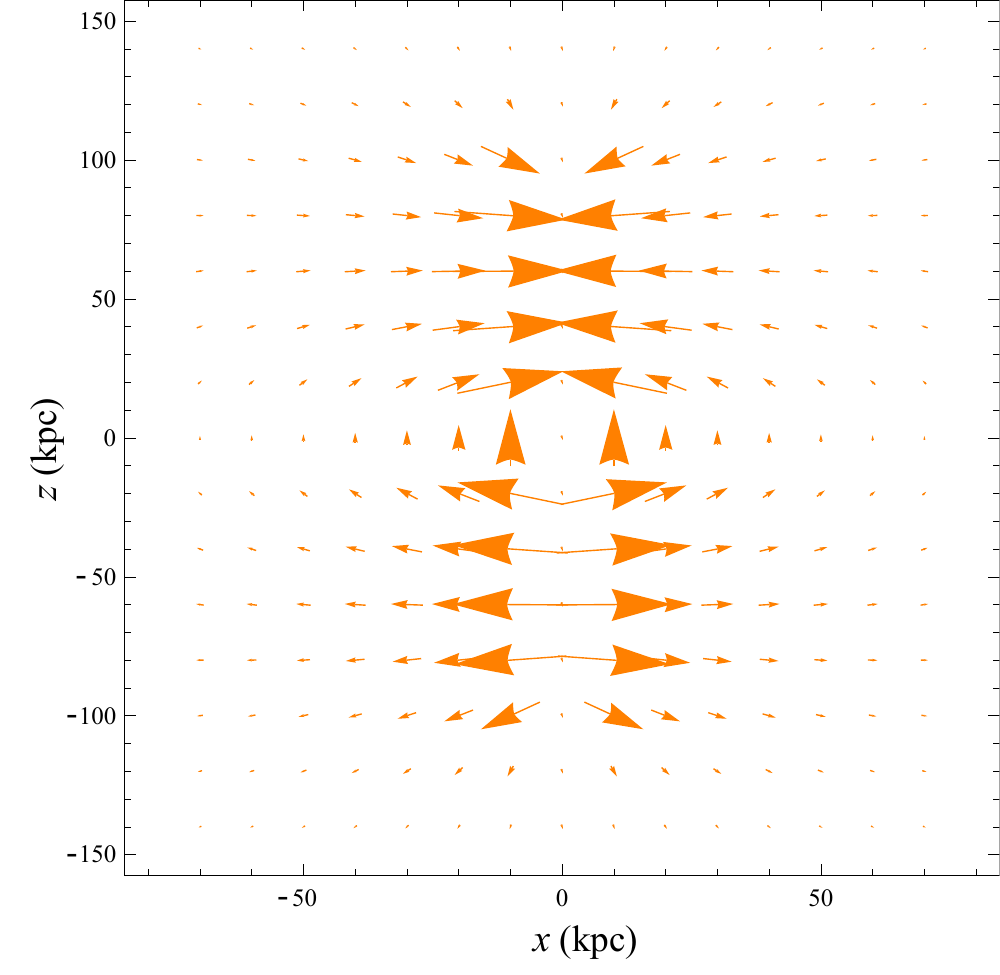}~~~\includegraphics[width=1\columnwidth]{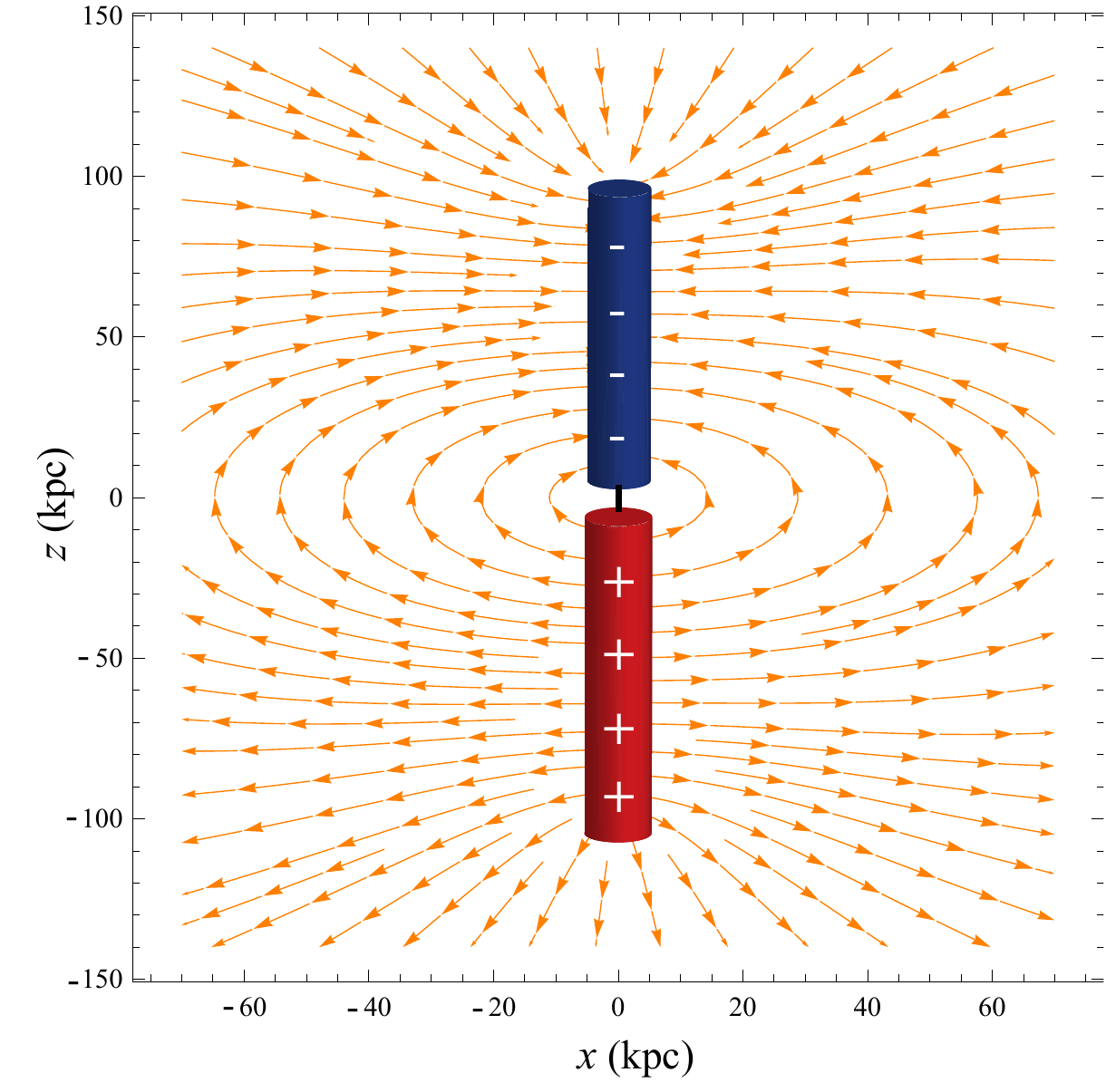}\caption{\label{fig:HBG}Plot of the gravitomagnetic $\vec{H}$ of the Balasin-Grumiller
solution for $r_{0}=1\,{\rm kpc}$ and $R=100\,{\rm kpc}$. Left panel:
arrow size reflects field strength; right panel: the corresponding
field lines. This corresponds to the gravitomagnetic field produced
by a pair of oppositely charged NUT rods along the $z$- axis, the
positive rod located at $-r_{0}>r>-R$, and the negative rod at $r_{0}<r<R$.
Apart from the factor $e^{-\nu}$, the field $\vec{H}$ has precisely
the same form as the electric field produced by a pair of oppositely
charged rods.}
\end{figure*}

\begin{align}
H^{i} & =\frac{e^{-\nu}V_{0}}{2}\left[\frac{1}{d_{r_{0}}}+\frac{1}{d_{-r_{0}}}-\frac{1}{d_{R}}-\text{\ensuremath{\frac{1}{d_{-R}}}}\right]\delta_{z}^{i}\nonumber \\
 & +\frac{e^{-\nu}V_{0}}{2r}\left\{ \frac{z-R}{d_{R}}+\frac{z+R}{d_{-R}}-\frac{z-r_{0}}{d_{r_{0}}}-\frac{z+r_{0}}{d_{-r_{0}}}\right\} \delta_{r}^{i}\label{eq:HBG}
\end{align}
(observe that the factor $e^{-\nu}$ drops out of its covariant components
$H_{i}=h_{ij}H^{j}=e^{\nu}H^{i}$.) This field is plotted in Fig.
\ref{fig:HBG}. Asymptotically, it vanishes, $\lim_{r\rightarrow\infty}H_{i}=\lim_{z\rightarrow\infty}H_{i}=0$;
since, moreover, the gravitoelectric field is zero everywhere, $G_{i}=-\Phi_{,i}=0$,
it follows that, in the limits $r\rightarrow\infty$ or $z\rightarrow\infty$,
no inertial forces are exerted on any test body, according to Eq.
\eqref{eq:QMGeo}. The reference frame is thus \emph{asymptotically
inertial}. It is the only rigid congruence of observers to be so in
this spacetime; in fact, it is the only globally defined Killing observer
congruence, since $\xi^{\alpha}=\partial_{t}^{\alpha}$ is the only
Killing vector field that is timelike at infinity.\footnote{Any Killing vector field of the metric \eqref{eq:BGmetric} can be
written in the form $\chi^{\alpha}=\partial_{t}^{\alpha}+\Omega\partial_{\phi}^{\alpha}$,
with $\Omega$ a constant. The norm of such vector is 
\[
\chi^{\alpha}g_{\alpha\beta}\chi^{\beta}=g_{00}+2g_{0\phi}\Omega+g_{\phi\phi}\Omega^{2}=-1+2N\Omega+(r^{2}-N^{2})\Omega^{2}.
\]
Since $\lim_{r\rightarrow\infty}N=\lim_{z\rightarrow\infty}N=V_{0}(R-r_{0})=const$,
then, for $r\rightarrow\infty$, the condition $\chi^{\alpha}g_{\alpha\beta}\chi^{\beta}<0$
is satisfied only if $\Omega=0$.} The coordinate system in \eqref{eq:BGmetric} corresponds thus to
the generalization of the IAU reference system for this solution.
The fact that the dust is at rest in such frame means that it is static
with respect the asymptotic inertial frame (thus nonrotating with
respect to the distant quasars).

The fact that $\vec{G}=0$ in this frame means that the ``galaxy''
would exert no gravitational attraction at all; this is consistent
with the fact that the Komar mass vanishes (Sec. \ref{subsec:Mass-and-angular}
below). Moreover, since $g_{00}=-1$ (i.e., $\Phi=0$), the dust 4-velocity
coincides with the time Killing vector field, $u^{\alpha}=\partial_{t}^{\alpha}\equiv\xi^{\alpha}$;
and since $k_{\alpha}\xi^{\alpha}$ is a conserved quantity along
a geodesic of tangent $k^{\alpha}$, this implies that an observer
sitting on one of the stars (at $P$) would measure no redshift \eqref{eq:Redshift}
from the other stars (at $P'$): $\nu'/\nu=\xi_{\alpha}k_{P'}^{\alpha}/\xi_{\alpha}k_{P}^{\alpha}=1$.
Then, according to Eq. \eqref{eq:Doppler}, which, in the nonrelativistic
limit, yields $\nu'/\nu\simeq1\pm v_{{\rm rad}}$, their relative
radial velocity $v_{{\rm rad}}$ would be zero. This would contradict
the well known measurements in the Milky Way: the measured redshift
between stars is of course not zero --- it is precisely from it that
the stars' radial velocities, and the galactic rotation curve, are
computed.

So, summarizing, according to this model: 
\begin{itemize}
\item Galaxies would be \emph{static} (i.e., would not rotate with respect
to the distant quasars); 
\item they would not generate gravitational attraction (since $\vec{G}=0$),
would have zero Komar mass, and the light emitted by stars would not
be redshifted (since $\Phi=0$). 
\end{itemize}
Of course, all of this is preposterous, and contrary to measurement;
in what follows we will merely dissect the model, the actual origin
of its gravitomagnetic field and potential, the mechanism by which
the dust can remain static, and the basic misunderstanding that led
to the claimed rotation curves.

\subsubsection{Mass and angular momentum\label{subsec:Mass-and-angular}}

If in a stationary spacetime the timelike Killing vector field $\partial_{t}^{\alpha}=\xi^{\alpha}$
is tangent to inertial observers at infinity (corresponding to the
source's asymptotic inertial ``rest'' frame), and is moreover normalized
so that $\xi^{\alpha}\xi_{\alpha}\stackrel{r\rightarrow\infty}{\rightarrow}-1$,
then the Komar mass contained in a compact spacelike hypersurface
(i.e., 3-volume) $\mathcal{V}$ with boundary $\partial\mathcal{V}$
is defined as \cite{Wald:1984,HansenWinicour1975,Whittaker1935,Cilindros,NatarioMathRel}
\begin{equation}
M=\frac{1}{8\pi}\int_{\partial\mathcal{V}}\star\mathbf{d}\bm{\xi}\ ,\label{eq:KomarMass}
\end{equation}
where $(\star\mathbf{d}\bm{\xi})_{\alpha\beta}\equiv\xi_{\nu;\mu}\epsilon_{\ \ \alpha\beta}^{\mu\nu}$
is the 2-form dual to $\mathbf{d}\bm{\xi}$. It is interpreted as
the ``active gravitational mass,'' or total mass/energy present
in the spacetime \cite{Wald:1984,Whittaker1935,Komar1959,HansenWinicour1975}.
Noting that $\mathbf{d}(\star\mathbf{d}\bm{\xi})=-2R_{\alpha\beta}\xi^{\beta}d\mathcal{V}^{\alpha}$,
where $d\mathcal{V}_{\alpha}=\epsilon_{\alpha\mu\nu\lambda}\mathbf{d}x^{\mu}\wedge\mathbf{d}x^{\nu}\wedge\mathbf{d}x^{\lambda}/6=-n_{\alpha}d\mathcal{V}$
is the volume element 1-form of $\mathcal{V}$ and $n^{\alpha}$ its
unit future-pointing normal vector, and since $\mathcal{V}$ is compact,
one can use the Stokes theorem to write \eqref{eq:KomarMass} as a
volume integral \cite{Wald:1984,Townsend,Cilindros}, 
\begin{equation}
M=-\frac{1}{4\pi}\int_{\mathcal{V}}R_{\ \beta}^{\alpha}\xi^{\beta}d\mathcal{V}_{\alpha}=\frac{1}{4\pi}\int_{\mathcal{V}}R_{\ \beta}^{\alpha}\xi^{\beta}n_{\alpha}d\mathcal{V}\ .\label{eq:KomarMassVolume}
\end{equation}
Considering spherical coordinates $\{\varrho,\theta,\phi\}$, so that
$r=\varrho\sin\theta$, $z=\varrho\cos\theta$, and taking $\partial\mathcal{V}$
as a 2-sphere of radius $\varrho=\varrho_{{\rm s}}$, parametrized
by $\{\theta,\phi\}$, Eq. \eqref{eq:KomarMass} yields
\begin{equation}
M=\frac{1}{8\pi}\int_{\partial\mathcal{V}}(\star d\xi)_{\theta\phi}{\bf d}\theta\wedge{\bf d}\phi=\frac{1}{4}\int(\star d\xi)_{\theta\phi}|_{\varrho=\varrho_{{\rm s}}}d\theta,\label{eq:KomarBGSpherical}
\end{equation}
where
\begin{align}
(\star d\xi)_{\theta\phi}= & \frac{V_{0}^{2}}{4}(2R-2r_{0}-d_{R}-d_{-R}+d_{r_{0}}+d_{-r_{0}})\nonumber \\
 & \times\left\{ \frac{R}{\tan\theta}\left[\frac{1}{d_{R}}-\frac{1}{d_{-R}}\right]+\frac{r_{0}}{\tan\theta}\left[\frac{1}{d_{-r_{0}}}-\frac{1}{d_{r_{0}}}\right]\right.\nonumber \\
 & \ \left.+\frac{\rho}{\sin\theta}\left[\frac{1}{d_{r_{0}}}+\frac{1}{d_{-r_{0}}}-\frac{1}{d_{R}}-\frac{1}{d_{-R}}\right]\right\} .\label{eq:KomarBGShperical}
\end{align}
Taking the sphere to be infinitely large (i.e., $\varrho_{{\rm s}}\rightarrow\infty$)
leads to the Komar mass of the whole spacetime. Since, asymptotically,
\begin{equation}
(\star d\xi)_{\theta\phi}\stackrel{\varrho\rightarrow\infty}{=}\frac{V_{0}^{2}\sin\theta}{2\varrho^{2}}(R-r_{0})^{2}(R+r_{0})\ ,\label{eq:KomarSphericalLimit}
\end{equation}
we have $\lim_{\varrho\rightarrow\infty}(\star d\xi)_{\theta\phi}=0$,
and therefore, by \eqref{eq:KomarBGSpherical}, the spacetime has
zero total Komar mass, $M_{{\rm total}}=0$.

The dust however has positive energy density $\rho=T_{\alpha\beta}u^{\alpha}u^{\beta}=T_{00}=R_{00}/(4\pi)$:
\begin{figure}
\includegraphics[width=1\columnwidth]{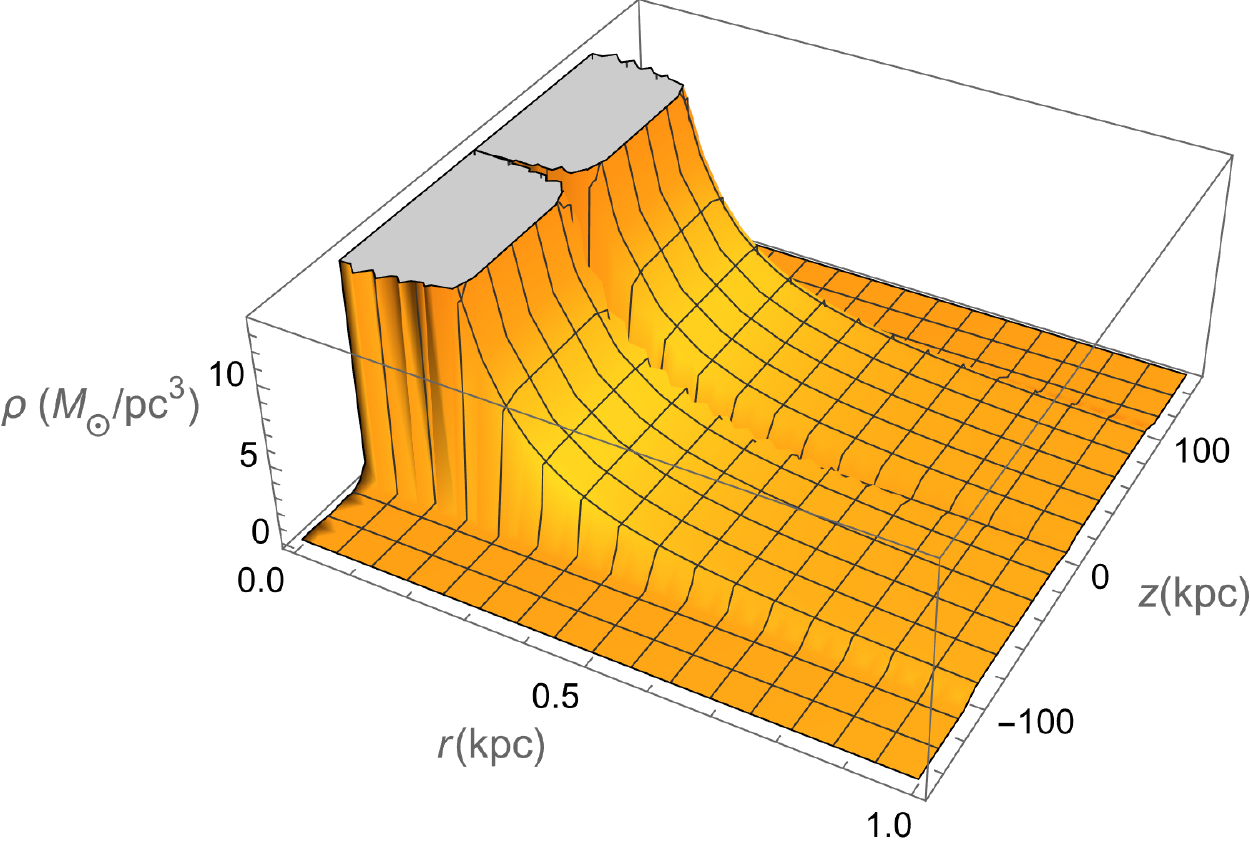}

\caption{\label{fig:Rho3D}Dust mass density $\rho$ in solar masses ($M_{\odot}$)
per cubic parsec (${\rm pc}$), for $r_{0}=1\,{\rm kpc}$, $R=100\,{\rm kpc}$,
$V_{0}=220\,{\rm km/s}$, and assuming, \emph{in Eq.} \eqref{eq:RhoBG},
$\nu\approx0$. $\rho$ is positive everywhere, being infinite along
the axial rods defined by $r=0$, $r_{0}<|z|<R$, and finite elsewhere.}
\end{figure}

\begin{align}
 & \rho(r,z)=\frac{V_{0}^{2}e^{-\nu}}{32\pi}\left[\left(\frac{1}{d_{R}}+\frac{1}{d_{-R}}-\frac{1}{d_{r_{0}}}-\frac{1}{d_{-r_{0}}}\right)^{2}\right.\nonumber \\
 & \left.+\frac{1}{r^{2}}\left(\frac{R-z}{d_{R}}-\frac{R+z}{d_{-R}}+\frac{r_{0}+z}{d_{-r_{0}}}-\frac{r_{0}-z}{d_{r_{0}}}\right)^{2}\right],\label{eq:RhoBG}
\end{align}
plotted in Fig. \eqref{fig:Rho3D}. The Komar mass within any hollow
cylinder $r_{{\rm in}}<r<r_{{\rm c}}$ can be computed by the volume
integral \eqref{eq:KomarMassVolume}, 
\begin{equation}
M_{{\rm dust}}=\frac{1}{4\pi}\int_{\mathcal{V}}R_{\ 0}^{0}n_{0}d\mathcal{V}=2\pi\int_{r_{{\rm in}}}^{r_{{\rm c}}}\int_{z_{{\rm b}}}^{z_{{\rm t}}}\rho e^{\nu}rdrdz\ ,\label{eq:MassBGVol}
\end{equation}
{[}which, after substituting \eqref{eq:RhoBG}, does not depend on
$\nu(r,z)${]} where in the second equality we used the energy-momentum
tensor of dust, $T^{\alpha\beta}=\rho u^{\alpha}u^{\beta}=\rho\delta_{0}^{\alpha}\delta_{0}^{\beta}$,
$\mathcal{V}$ is a hollow cylinder in a hypersurface $t=const$,
so that $n_{\alpha}=-r(r^{2}-N^{2})^{-1/2}\delta_{\alpha}^{0}$, $g_{ij}$
is the metric induced therein {[}which follows from \eqref{eq:BGmetric}
by taking $dt=0${]}, and $d\mathcal{V}=\sqrt{|g_{ij}|}drd\phi dz$,
with $|g_{ij}|=e^{\nu}(r^{2}-N^{2})^{1/2}$. Since $\rho>0$, $M_{{\rm dust}}$
is positive; since this is valid for any hollow cylinder, the vanishing
total mass $M_{{\rm total}}$ implies the axis $r=0$ to have nonvanishing
\emph{negative} Komar mass. 

\begin{figure*}
\includegraphics[width=1\textwidth]{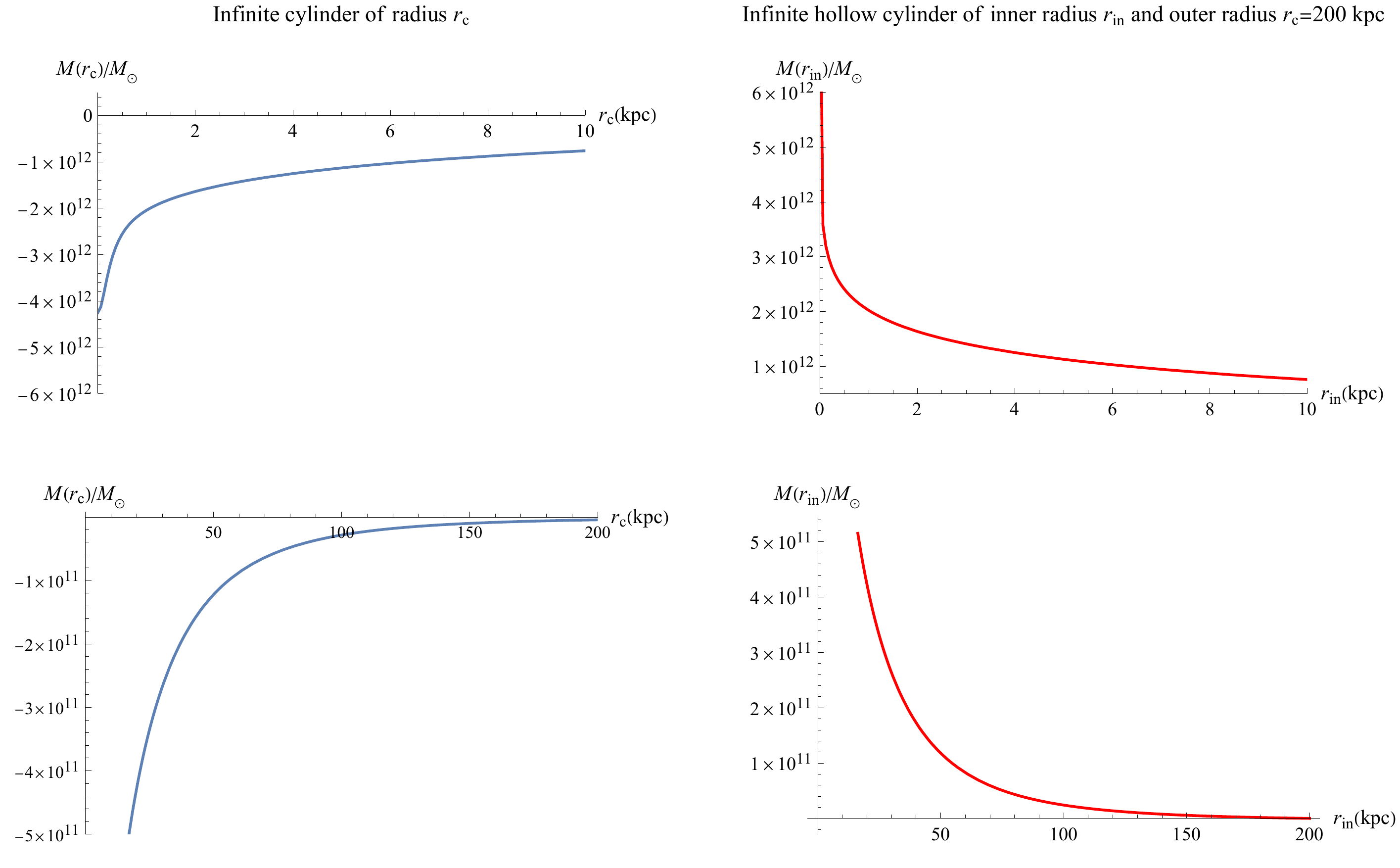}\caption{\label{fig:BGKomarvsRho}Left panel: Komar mass contained within an
infinitely long solid cylinder plotted as a function of its radius
$r_{{\rm c}}$, measured in solar masses ($M_{\odot}$). Right panel:
Komar mass within an infinitely long hollow cylinder of outer radius
$r_{{\rm c}}=200\,{\rm kpc}$, plotted as a function of its inner
radius $r_{{\rm in}}$. It is assumed, in both cases, $r_{0}=1\,{\rm kpc}$,
$R=100\,{\rm kpc}$, $V_{0}=220\,{\rm km/s}$. The mass of the solid
cylinder is always finite and negative, approaching $M(r_{{\rm c}})=0$
for $r_{{\rm c}}\rightarrow\infty$. The mass of the hollow cylinder
is always positive, as expected for dust; until the bulge $r_{{\rm in}}\approx2\ {\rm kpc}$,
its mass is in line with known estimates for the Milky Way's mass
($\sim10^{12}M_{\odot}$); it peaks to infinity however as $r_{{\rm in}}\rightarrow0$,
implying the axis $r=0$ itself to have an infinite negative mass.}
\end{figure*}

\begin{figure}
\includegraphics[width=1\columnwidth]{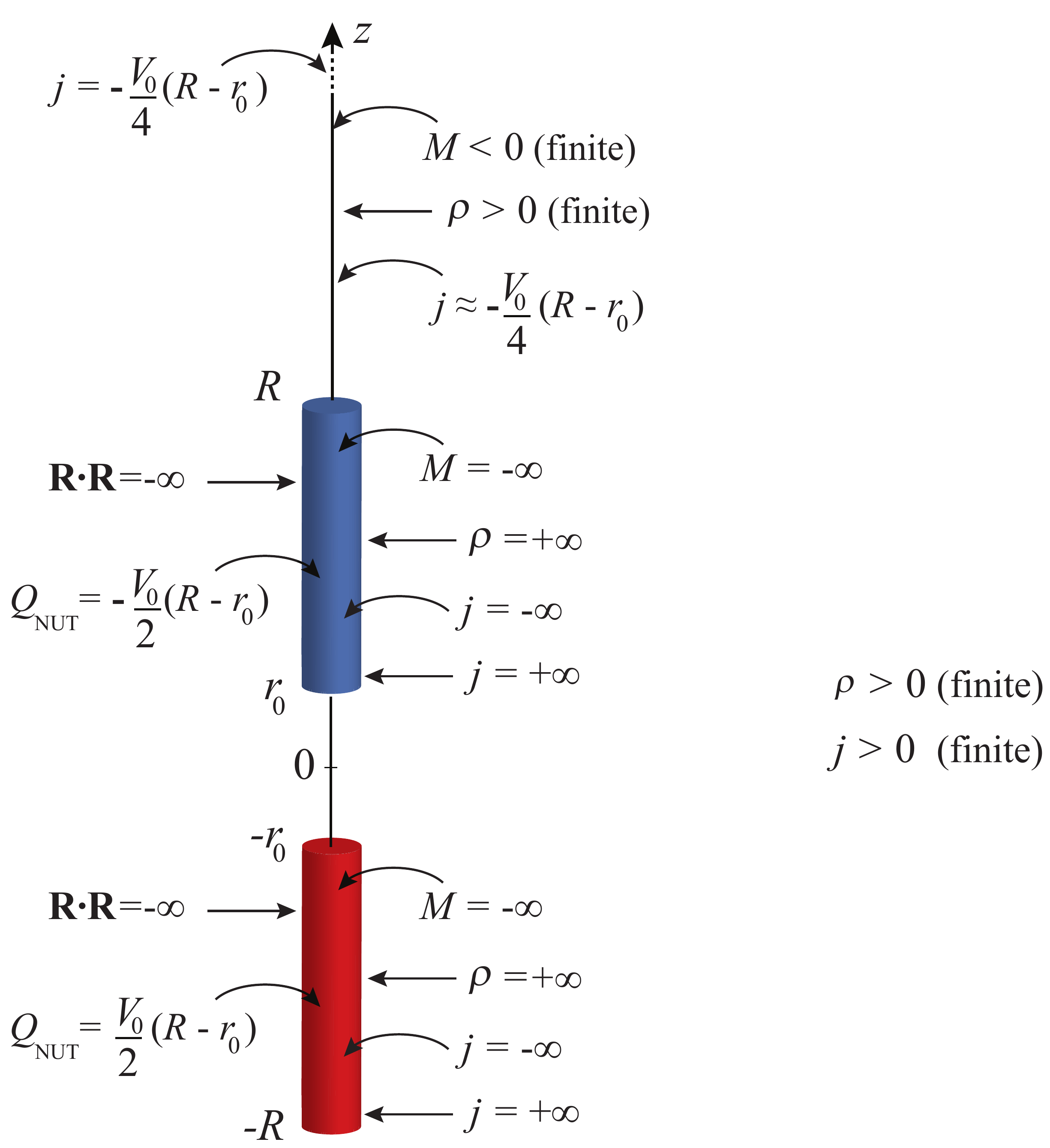}

\caption{\label{fig:BGRod}Rod structure and singularities of the Balasin-Grumiller
solution. Singularities along the axis possess mass, angular momentum,
and, in the case of the rods located at $-R<z<-r_{0}$ and $r_{0}<z<R$,
also NUT charges $Q_{{\rm NUT}}=\pm V_{0}(R-r_{0})/2$, respectively.
The Komar mass $M$ of the rods is infinite negative, and finite elsewhere
along the axis. The dust's energy density $\rho$ is positive everywhere,
approaching $+\infty$ along the rods. The axis possesses an infinite
negative angular momentum; its value per unit length is nearly constant
outside the rods for $|z|>R$, very close to its asymptotic value
$j_{z\rightarrow\pm\infty}=-V_{0}(R-r_{0})/4$. The dust possesses
a (much smaller) positive angular momentum, which likewise is infinite
along the rods, and has finite density $\rho N(r,z)$ elsewhere. The
Kretschmann scalar ${\bf R}\cdot{\bf R}$ of the metric \eqref{eq:BGmetric}
also diverges approaching the rods.}
\end{figure}

In Fig. \ref{fig:BGKomarvsRho} the Komar mass of infinitely long
solid and hollow cylinders are plotted as functions of their outer
and inner radii, respectively. The mass of the solid cylinders is
computed through the surface integral in Eq. \eqref{eq:KomarMass},
taking $\partial\mathcal{V}=\mathcal{\mathcal{L}}\cup\mathcal{B}_{{\rm t}}\cup\mathcal{B}_{{\rm b}}$
to be their boundary, where $\mathcal{L}$ is the cylinder's lateral
surface, parametrized by $\{\phi,z\}$, and $\mathcal{B}_{{\rm t}}$
and $\mathcal{B}_{{\rm b}}$ its top and bottom bases, lying in planes
orthogonal to the $z$-axis and parametrized by $\{r,\phi\}$. Equation
\eqref{eq:KomarMass} then becomes 
\begin{align}
M & =\frac{1}{8\pi}\left[\int_{\mathcal{B}_{{\rm t}}\sqcup\mathcal{B}_{{\rm b}}}(\star d\xi)_{r\phi}{\bf d}r\wedge{\bf d}\phi+\int_{\mathcal{L}}(\star d\xi)_{\phi z}{\bf d}\phi\wedge{\bf d}z\right]\label{eq:KomarCylinder}\\
 & =\frac{1}{4}\int_{0}^{r_{{\rm c}}}\left[(\star d\xi)_{r\phi}|_{z=z_{{\rm t}}}-(\star d\xi)_{r\phi}|_{z=z_{{\rm b}}}\right]dr\nonumber \\
 & +\frac{1}{4}\int_{z_{{\rm b}}}^{z_{{\rm t}}}(\star d\xi)_{\phi z}|_{r=r_{{\rm c}}}dz\ ,\label{eq:Komar2}
\end{align}
where $r_{{\rm c}}$ is the cylinder's radius, and the components
$(\star d\xi)_{r\phi}$ and $(\star d\xi)_{\phi z}$ read

\begin{align*}
(\star d\xi)_{r\phi}= & \frac{V_{0}^{2}}{4r}(2R-2r_{0}-d_{R}-d_{-R}+d_{r_{0}}+d_{-r_{0}})\\
 & \times\left[\frac{R-z}{d_{R}}-\frac{R+z}{d_{-R}}-\frac{r_{0}-z}{d_{r_{0}}}+\frac{r_{0}+z}{d_{-r_{0}}}\right]\ ;\\
(\star d\xi)_{\phi z}= & \frac{V_{0}^{2}}{4}(2R-2r_{0}-d_{R}-d_{-R}+d_{r_{0}}+d_{-r_{0}})\\
 & \times\left[\frac{1}{d_{r_{0}}}+\frac{1}{d_{-r_{0}}}-\frac{1}{d_{R}}-\frac{1}{d_{-R}}\right]\ .
\end{align*}
For finite radius $r_{{\rm c}}$, the mass of the solid cylinders
is negative; it becomes larger for increasing $r_{{\rm c}}$, approaching
$M=0$ as $r_{{\rm c}}\rightarrow\infty$. This is because the larger
the cylinder, the more dust (whose mass $M_{{\rm dust}}$ is positive)
it encloses. The mass of the hollow cylinders, as expected, increases
as its inner radius $r_{{\rm in}}$ decreases; however it approaches
$+\infty$ as $r_{{\rm in}}\rightarrow0$. This is because the mass
density peaks to infinity as $\rho\sim r^{-2}$ approaching the rods
$-R<z<-r_{0}$, $r_{0}<z<R$ along the axis, see Fig. \eqref{fig:Rho3D}.
Since any cylinder enclosing such rods has finite mass, this implies
as well that the rods have an infinite negative Komar mass. We thus
have the structure depicted in Fig. \ref{fig:BGRod}. 

Analogously to Eq. \eqref{eq:KomarCylinder}, the Komar angular momentum
\eqref{eq:JKomarSurface} contained within a cylinder of boundary
$\partial\mathcal{V}=\mathcal{\mathcal{L}}\cup\mathcal{B}_{{\rm t}}\cup\mathcal{B}_{{\rm b}}$
is 
\begin{align}
J= & -\frac{1}{16\pi}\left[\int_{\mathcal{B}_{{\rm t}}\sqcup\mathcal{B}_{{\rm b}}}(\star d\zeta)_{r\phi}{\bf d}r\wedge{\bf d}\phi+\int_{\mathcal{L}}(\star d\zeta)_{\phi z}{\bf d}\phi\wedge{\bf d}z\right]\label{eq:Jcylinder}\\
= & -\frac{1}{8}\int_{0}^{r_{{\rm c}}}\left[(\star d\zeta)_{r\phi}|_{z=z_{{\rm t}}}-(\star d\zeta)_{r\phi}|_{z=z_{{\rm b}}}\right]dr\nonumber \\
 & -\frac{1}{8}\int_{z_{{\rm b}}}^{z_{{\rm t}}}(\star d\zeta)_{\phi z}|_{r=r_{{\rm c}}}dz\ .\label{eq:JCylinder2}
\end{align}
The lengthy full expressions for the components $(\star d\zeta)_{r\phi}$
and $(\star d\zeta)_{\phi z}$ are given in the Supplemental Material
\cite{EPAPSPRD}; they have the asymptotic limits 
\begin{align*}
 & \lim_{z\rightarrow\infty}(\star d\zeta)_{r\phi}=\lim_{r\rightarrow\infty}(\star d\zeta)_{r\phi}=0\ ;\\
 & \lim_{z\rightarrow\infty}(\star d\zeta)_{\phi z}=\lim_{r\rightarrow\infty}(\star d\zeta)_{\phi z}=2V_{0}(R-r_{0})\ .
\end{align*}
Hence, for cylinders located at large $|z|$, the first term of the
integral \eqref{eq:JCylinder2} is negligible, and so $J\approx-\int_{z_{{\rm b}}}^{z_{{\rm t}}}(\star d\zeta)_{\phi z}|_{r=r_{{\rm c}}}dz/8\approx-V_{0}(R-r_{0})\Delta z/4$,
independent of the cylinder's radius. This tells us that, therein,
the angular momentum is essentially contained in the axis $r=0,$
the contribution of the dust being negligible, and the spacetime therein
corresponding to that of a spinning cosmic string with uniform (negative)
angular momentum per unit length $j=-V_{0}(R-r_{0})/4$, cf. Eq. \eqref{eq:Jstring}.
Since the integral \eqref{eq:Jcylinder} is moreover finite for \emph{any}
finite cylinder, then the total angular momentum $J$ of the spacetime
is infinite \emph{negative}. The negative angular momentum is entirely
contained in singularities along the axis. The static dust, in turn,
possesses a positive angular momentum, manifest in the Komar angular
momentum of hollow cylinders; the latter can be computed either from
the surface integral \eqref{eq:Jcylinder}-\eqref{eq:JCylinder2}
(subtracting solid cylinders), or from the volume integral \eqref{eq:JKomarVolume},
which reads here
\begin{equation}
J_{{\rm dust}}=-\int_{\mathcal{V}}T_{\ \beta}^{\alpha}\zeta^{\beta}n_{\alpha}d\mathcal{V}=\int_{\mathcal{V}}\rho u_{\phi}e^{\nu}rdrd\phi dz\ ,\label{eq:Jdust}
\end{equation}
where the second equality follows from steps analogous to those in
\eqref{eq:MassBGVol}. The component $u_{\phi}=N(r,z)$ is, as is
well known (e.g. \cite{Misner:1974qy}), the angular momentum per
unit mass of a particle of 4-velocity $u^{\alpha}=\partial_{t}^{\alpha}\equiv\delta_{0}^{\alpha}$,
at rest in the coordinates of \eqref{eq:BGmetric}; it is nonzero
due to the off-diagonal term $g_{0i}=N$ in \eqref{eq:BGmetric}.
Observe that $N>0$ under the defining assumption $R>r_{0}$. Since,
as we have seen in Sec. \ref{subsec:The-dust-is-static}, such particle
is static with respect the asymptotic inertial frame, such angular
momentum arises from the frame-dragging effect (``dragging of the
ZAMOS'' \cite{PaperDragging}) associated to the gravitomagnetic
potential $\mathcal{A}_{\phi}=N$. The latter, in turn, is generated
by the singularities along the axis --- namely the combined effect
of the above mentioned spinning string (of negative angular momentum)
plus a pair of NUT rods, as we shall see in Sec. \ref{subsec:OriginGM}.
Therefore, the angular momentum that the \emph{static} dust possesses
is a purely general relativistic effect, originated by the dragging
of the ZAMOs produced by the singularities along the axis. Since $u_{\phi}$
is finite everywhere, the angular momentum density $\rho u_{\phi}$,
similarly to $\rho$ in Fig. \ref{fig:Rho3D}, is infinite along the
axial rods $-R<z<-r_{0}$, $r_{0}<z<R$, where it diverges as $r^{-2}$,
and finite elsewhere. Hence, the dust's angular momentum contained
in any cylinder enclosing the rods is infinite; and since the cylinder's
total angular momentum $J$ is finite (negative), this implies the
angular momentum of the rods to be infinite negative. Outside the
rods, both the dust's angular momentum density $\rho u_{\phi}$ and
the angular momentum per unit length of the axis, $j$, are finite.
Numerical inspection shows that, for $|r|>R$, $j$ has a very nearly
constant value $j\approx-V_{0}(R-r_{0})/4$ {[}for $R=100{\rm kpc}$,
$r_{0}=1{\rm kpc}$, $j=-V_{0}(R-r_{0})/4+O(10^{-6}j)${]}. As expected
from being a purely general relativistic effect created by the axis,
the dust's angular momentum is very small comparing to that contained
in the axis for any finite cylinder (amounting to an almost negligible
correction to the total angular momentum therein). The angular momentum
contained in the axis, $J_{{\rm axis}}=J-J_{{\rm dust}}$, is computed
from Eqs. \eqref{eq:JCylinder2} and \eqref{eq:Jdust}. For instance,
for a cylinder of radius $r_{{\rm c}}=200{\rm kpc}$ and height $101{\rm kpc}<z<200{\rm kpc}$,
not enclosing the rods, and taking $V_{0}=220\,{\rm km/s}$, we have
$J_{{\rm axis}}=-6.91\times10^{7}J_{{\rm MW}}\approx J$ and $J_{{\rm dust}}=4.9J_{{\rm MW}}$,
where $J_{{\rm MW}}=10^{67}{\rm kg}{\rm m}^{2}{\rm s}^{-1}$ is the
Milky Way's actual angular momentum. The angular momentum distribution
in this spacetime has thus the structure depicted in Fig. \ref{fig:BGRod}.

\subsubsection{Singularities\label{subsec:Singularities}}

The expression for the Kretschmann scalar of \eqref{eq:BGmetric}
is given in Appendix \ref{sec:Kretschmann-scalar-BG}. Along the rods,
it yields ${\bf R}\cdot{\bf R}\rightarrow-\infty$; hence the rods
are curvature singularities of the exterior metric \eqref{eq:BGmetric}.
The model thus unphysically predicts the existence of a region close
to the axis, and along an extension similar to the galaxy's diameter,
of arbitrarily large curvature (whose tidal effects would destroy
any astrophysical object) around a naked singularity. It should be
noted, however, that these are \emph{not} the only curvature singularities
that lie along the axis, but just those already manifest in the limit
$r\rightarrow0$ of \eqref{eq:BGmetric}. The metric \eqref{eq:BGmetric}
is defined only outside the axis $r=0$ {[}given the singularity of
the component $g_{0\phi}$ at $r=0$, and similarly to the situation
in the NUT \eqref{eq:NUTmetric} and cosmic string \eqref{eq:String}
metrics{]}. As seen in the preceding section, the axis $r=0$ possesses
nonvanishing Komar mass (actually infinite along the rods $r_{0}<|z|<R$)
and angular momentum \emph{everywhere}. Unless one assumes a nontrivial
topology $\mathbb{R}^{1}\times\mathbb{R}^{3}\backslash\{r=0\}$, it
follows from the Stokes theorem that (like for an infinitely thin
cosmic string, discussed in Sec. \ref{subsec:Cosmic-string}) it must
be assumed to be matched to the singular limit of an ``interior''
solution along the axis having a singular Ricci tensor, so that the
Komar volume integrals \eqref{eq:JKomarVolume} and \eqref{eq:KomarMassVolume}
match the Komar surface integrals \eqref{eq:JKomarSurface} and \eqref{eq:KomarMass},
respectively. Thus, even from the curvature point of view, the whole
axis is singular.

\subsubsection{CTC's and the \textquotedblleft forbidden\textquotedblright{} ZAMO
region\label{subsec:CTC's-and-the-forbidden}}

\begin{figure}
\includegraphics[width=1\columnwidth]{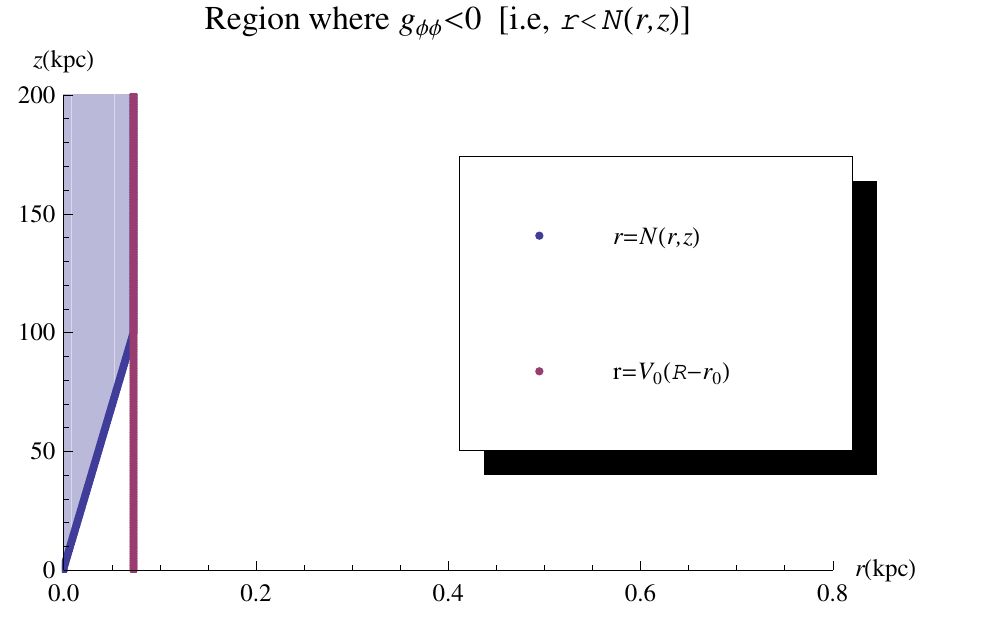} \caption{\label{fig:Forbidden}Plot of the \textquotedblleft time machine\textquotedblright{}
region $r<N(r,z)$ (shadowed), in which closed timelike curves arise,
and ZAMOs are forbidden, for $r_{0}=1\,{\rm kpc}$, $R=100\,{\rm kpc}$,
$V_{0}=220\,{\rm km}/{\rm s}$. Only the half-space $z\ge0$ is represented.
It is pencil-shaped, and contained within the cylinder $r<V_{0}(R-r_{0})$.}
\end{figure}

When $r^{2}<N^{2}$, the $\phi$ coordinate in \eqref{eq:BGmetric}
becomes timelike, $g_{\phi\phi}<0$, and so the circles tangent to
$\partial_{\phi}$ (circles of constant $r$ and $z$) are closed
timelike curves. Observers following such worldlines would travel
to their own past; in other words, the spacetime contains a ``time
machine'' in this region, plotted in Fig \ref{fig:Forbidden}, which
is very narrow (of radius smaller than $r_{0}$) and consists of two
``pencil-like'' shapes, one above and the other below the $z=0$
plane. The situation is somewhat analogous to that of the van Stockum
spacetime, except that therein the region where $g_{\phi\phi}<0$
is the infinite hollow cylinder $r>1/w$, see Sec. \ref{subsec:The-van-Stockum}.

The region $r^{2}<N^{2}$ is also one where ZAMOs cannot exist. The
ZAMOs' angular velocity with respect to the system of coordinates
in \eqref{eq:BGmetric} (fixed to the asymptotic inertial rest frame),
is, from Eq. \eqref{eq:OmegaZamo}, 
\begin{equation}
\Omega_{{\rm ZAMO}}(r,z)=-\frac{N}{r^{2}-N^{2}}\ .\label{eq:OmegaZamoBG}
\end{equation}
Hence $\lim_{r\rightarrow N}\Omega_{{\rm ZAMO}}=-\infty$. The magnitude
of the ZAMOs' velocity relative to the static observers $u^{\alpha}=\delta_{0}^{\alpha}$
is given by Eq. \eqref{eq:vMaggamma} with $\gamma_{{\rm Z}}=-u^{\alpha}(u_{{\rm Z}})_{\alpha}=r(r^{2}-N^{2})^{-1/2}$,
\begin{equation}
|v_{{\rm Z}}|\equiv\sqrt{v_{{\rm Z}}^{\alpha}(v_{{\rm Z}})_{\alpha}}=\frac{N}{r}\ ,\label{eq:vZAMO}
\end{equation}
approaching the speed of light when $r\rightarrow N$. Indeed, the
ZAMOs cease to be timelike for $r<N$: 
\begin{align}
u_{{\rm Z}}^{\alpha}g_{\alpha\beta}u_{{\rm Z}}^{\beta} & =(2g_{0\phi}\Omega_{{\rm ZAMO}}+\Omega_{{\rm ZAMO}}^{2}g_{\phi\phi}-1)(u_{{\rm Z}}^{0})^{2}\nonumber \\
 & =(u_{{\rm Z}}^{0})^{2}\frac{r^{2}}{N^{2}-r^{2}}\ \ (>0\ {\rm if\ }r^{2}<N^{2})\ .\label{eq:ZamoLimit}
\end{align}
The static observers $u^{\alpha}=\delta_{0}^{\alpha}$, in turn, remain
timelike and exist everywhere. This contrasts with the situation in
the ergosphere of, e.g., the Kerr spacetime, which is a forbidden
region for the static observers, but not for the ZAMOs (possible everywhere
outside the outer horizon $r_{+}$).

\subsubsection{Origin of the gravitomagnetic effects\label{subsec:OriginGM}}

The gravitomagnetic field \eqref{eq:HBG} is well defined everywhere
outside the axis $r=0$, being both curl and divergence free, $\tilde{\nabla}\cdot\vec{H}=0$,
$\tilde{\nabla}\times\vec{H}=0$. Along the axis the metric \eqref{eq:BGmetric}
is not defined; therefore, in rigor, it contains no information about
$\vec{H}|_{r=0}$. From Eqs. \eqref{eq:HFieldEq}, $\tilde{\nabla}\cdot\vec{H}=0$
is a consequence of the fact that $\vec{G}=0$ in the reference frame
associated to \eqref{eq:BGmetric}, whereas $\tilde{\nabla}\times\vec{H}=0$
implies $\vec{J}=0$ (i.e., the mass-energy current to be zero), consistent
with the fact that the dust is at rest. Hence, it is clear that $\vec{H}$
is not originated by mass-energy currents (or any sort of source)
outside the axis. Its source must thus be singularities contained
in the axis. As shown in detail in Appendix \ref{sec:Electromagnetic-analogue-of},
the covariant components $H_{i}$ of the gravitomagnetic field \eqref{eq:HBG}
are formally identical to the electric field produced by a pair of
oppositely charged rods located at $r_{0}<|z|<R$, with charges $Q=\mp V_{0}(R-r_{0})/2$,
in flat spacetime; or, equivalently, to the magnetic field $\vec{B}$
of a pair of rods of oppositely charged magnetic monopoles {[}the
contravariant components $H^{i}$ in Eq. \eqref{eq:HBG} exhibiting
the extra factor $e^{-\nu}$, due to the nonflat spatial metric{]}.
Indeed, $\vec{H}$ is originated by gravitomagnetic monopoles, as
we will now see.

For the metric \eqref{eq:BGmetric}, since $\Phi=0$ {[}cf. Eq. \eqref{eq:StatMetric}{]},
it follows from Eqs. \eqref{eq:GEM1forms} and \eqref{eq:QNUT} that
$\vec{H}=\tilde{\nabla}\times\vec{\mathcal{A}}$ and 
\begin{equation}
Q_{{\rm NUT}}=\frac{1}{4\pi}\int_{\mathcal{S}}\vec{H}\cdot\vec{d}\mathcal{S}\ ;\label{eq:QNUTrepeat}
\end{equation}
that is, similarly to the electromagnetic analogue, the NUT charge
reduces to the flux of $\vec{H}$. The gravitomagnetic field has the
following limits along the axis: 
\begin{equation}
H_{r}\stackrel{r\rightarrow0}{=}\left\{ \begin{array}{cc}
0 & \quad{\rm for}\ |z|<r_{0}\ {\rm or\ }|z|>R\\
-V_{0}/r & \quad{\rm for}\ r_{0}<z<R\\
V_{0}/r & \quad{\rm for}\ -R<z<-r_{0}
\end{array}\right.\ ;\label{eq:HrAxial}
\end{equation}
\begin{equation}
\lim_{r\rightarrow0}H_{z}=\frac{V_{0}}{2}\sum_{\pm}\left[\frac{1}{|r_{0}\pm z|}-\frac{1}{|R\pm z|}\right]\ .\label{eq:HzAxial}
\end{equation}
It thus diverges along the rods, the component $H_{r}$ diverging
along the whole rods ($r_{0}<|z|<R$), and $H_{z}$ diverging at the
points $z=\pm R,\pm r_{0}$. Elsewhere along the axis $\lim_{r\rightarrow0}\vec{H}$
is finite. Assuming $\vec{H}$ to be continuous therein, $\vec{H}|_{r=0}=\lim_{r\rightarrow0}\vec{H}$
(and since $\vec{H}$ is well defined everywhere off the axis), then
this, together with the equation $\tilde{\nabla}\cdot\vec{H}=0$,
implies, by application of Gauss theorem (see Sec. \ref{subsec:NUT-spacetime}),
that: (i) the NUT charge within any closed 2-surface $\mathcal{S}=\partial\mathcal{V}$
on $\Sigma$ not enclosing the rods is zero, $Q_{{\rm NUT}}=\int_{\mathcal{S}}\vec{H}\cdot\vec{d}\mathcal{S}/(4\pi)=\int_{\mathcal{V}}\tilde{\nabla}\cdot\vec{H}/(4\pi)=0$;
(ii) it has the same (nonzero) value for all 2-surfaces $\mathcal{S}$
enclosing one of the rods. Its value is easily computed by taking
$\mathcal{S}$ as the boundary of a cylinder enclosing the rod, and
taking the limit where its radius goes to zero: 
\begin{align}
Q_{{\rm NUT}} & =\frac{1}{4\pi}\left[\int_{\mathcal{B}_{{\rm t}}\sqcup\mathcal{B}_{{\rm b}}}H_{z}d\mathcal{S}^{z}+\int_{\mathcal{L}}H_{r}d\mathcal{S}^{r}\right]\nonumber \\
 & =\frac{1}{2}\int_{z_{{\rm b}}}^{z_{{\rm t}}}\lim_{r\rightarrow0}(H_{r}r)dz=\mp V_{0}(R-r_{0})/2\,,\label{eq:QNUTBG}
\end{align}
the minus sign applying to the upper rod in Fig. \ref{fig:HBG}, located
at $r_{0}<z<R$, and plus sign to the lower rod, located at $-R<z<-r_{0}$.
Here, again, $\mathcal{B}_{{\rm t}}$ and $\mathcal{B}_{{\rm b}}$
are the cylinder's top and bottom bases, parametrized by $\{r,\phi\}$,
and $\mathcal{L}$ its lateral surface, parametrized by $\{\phi,z\}$,
and in the second equality we noticed that $d\mathcal{S}^{z}=rdrd\phi$,
$d\mathcal{S}^{r}=rd\phi dz$, and that, since $\lim_{r\rightarrow0}H_{z}$
is finite, the term $\int_{\mathcal{B}_{{\rm t}}\sqcup\mathcal{B}_{{\rm b}}}H_{z}d\mathcal{S}^{z}$
vanishes when the cylinder's radius goes to zero. We see thus that,
under the continuity assumption $\vec{H}|_{r=0}=\lim_{r\rightarrow0}\vec{H}$,
the source of $\vec{H}$ is the pair of oppositely charged NUT rods
depicted in the right panel of Fig. \ref{fig:HBG}. We must note,
however, like in the NUT spacetime (Sec. \ref{subsec:NUT-spacetime}),
or in the electromagnetic analogue in Appendix \ref{sec:Electromagnetic-analogue-of},
that, since the metric is not defined along the axis $r=0$, one must
also admit the possibility of a Dirac-delta type $\vec{H}$ along
the axis canceling out the flux of $\vec{H}$ in \eqref{eq:QNUTrepeat}
(making $Q_{{\rm NUT}}=0$). In this case there would be no gravitomagnetic
charges; from the electromagnetic analogue based on solenoids in Appendix
\ref{subsec:EM2} and Fig. \ref{fig:Solenoids}(b), and the discussion
of the NUT solution in Sec. \ref{subsec:NUT-spacetime}, one expects
the monopoles to consist then of the tips of semi-infinite spinning
cosmic strings along the axis. For simplicity, however, we will still
henceforth refer to rods the in Fig. \ref{fig:HBG} as NUT rods of
``charge'' \eqref{eq:QNUTBG}, regardless of the actual origin of
the gravitomagnetic monopoles. What is important to emphasize here
is that it is these unphysical singularities along the axis, \emph{not}
the static dust, that source the gravitomagnetic field $\vec{H}$
in the BG solution \eqref{eq:BGmetric}.

Transforming to spherical-like coordinates $(\varrho,\theta,\phi)$
such that $r=\varrho\sin\theta$, $z=\varrho\cos\theta$, we have
the asymptotic limit 
\begin{align}
 & \vec{H}\stackrel{\rho\rightarrow\infty}{=}-\frac{p}{\re^{3}}(2\cos\theta\vec{e}_{\hat{\re}}+\sin\theta\vec{e}_{\hat{\theta}})\ ;\label{eq:HBGasymptotic}\\
 & p\equiv e^{-\nu(\infty,\theta)/2}V_{0}\frac{R^{2}-r_{0}^{2}}{2}\ ,\nonumber 
\end{align}
where $\vec{e}_{\hat{\re}}=e^{-\nu/2}\partial_{\re}$ and $\vec{e}_{\hat{\theta}}=e^{-\nu/2}\re^{-1}\partial_{\theta}$
are unit vectors. If the angular dependence of $e^{-\nu(\infty,\theta)}$
is negligible (as assumed in \cite{BG} for practical purposes), then
\eqref{eq:HBGasymptotic} corresponds, as expected, to the gravitomagnetic
field of gravitomagnetic dipole, of dipole moment $p$. In such a
regime, the field is thus indistinguishable from that of a spinning
source of angular momentum $J=p$ (cf. e.g. Eq. (31) of \cite{PaperDragging}).

As for the gravitomagnetic potential 1-form $\boldsymbol{\mathcal{A}}=N(r,z){\bf d}\phi$,
the second (nonconstant) term in \eqref{eq:N} is likewise sourced
by the NUT rods (so that $\vec{H}=\tilde{\nabla}\times\vec{\mathcal{A}}$);
but it contains also the constant term $V_{0}(R-r_{0})$, which corresponds
to the potential of an infinite spinning cosmic string along the $z$-axis,
of angular momentum per unit mass $j=-V_{0}(R-r_{0})/4$, cf. Sec.
\ref{subsec:Cosmic-string} and Eq. \eqref{eq:Jstring} therein. Hence,
the constant part of the gravitomagnetic potential $\boldsymbol{\mathcal{A}}$
is sourced by the angular momentum that, as we have shown in Sec.
\ref{subsec:Mass-and-angular}, is contained along the axis $r=0$.
Such constant contribution is formally analogous to the external magnetic
potential of an infinitely long solenoid; indeed, as shown in Appendix
\ref{subsec:EM1}, the field of the BG solution is fully mirrored
by the electromagnetic field produced by the combination of an infinite
solenoid with a pair of rods of opposite magnetic charges.

\subsubsection{Staticity mechanism\label{subsec:Staticity-mechanism}}

Since, as we have seen in Sec. \ref{subsec:The-dust-is-static}, the
dust is static with respect to the asymptotic inertial frame (or to
the distant quasars), the question then arises as to what holds it
in place, preventing it from gravitationally collapsing. In other
words, why does the gravitoelectric field $\vec{G}$ vanish in such
frame? The answer is in the first of Eqs. \eqref{eq:GFieldEq}, which,
for the metric \eqref{eq:BGmetric}, reduces to 
\begin{equation}
\tilde{\nabla}\cdot\vec{G}=-4\pi\rho+\frac{1}{2}{\vec{H}}^{2}=0\ ,\label{eq:DivGBG}
\end{equation}
telling us that the term $\vec{H}^{2}/2$ acts as an effective negative
``energy'' source for $\vec{G}$ \cite{ZonozBell1998}, exactly
canceling out the attractive contribution from the dust's mass density
$\rho$, and allowing for $\vec{G}=\vec{0}$.

We must however remark that the negative energy interpretation has,
in rigor, the status of an analogy, since, by virtue of the equivalence
principle, there is no such thing as energy of the gravitomagnetic
field: unlike its magnetic counterpart $\vec{B}$ (which is a physical
field, contributing to the energy-momentum tensor $T^{\alpha\beta}$),
$\vec{H}$ is an \emph{inertial} field, that can always be made to
vanish by a suitable choice of reference frame (e.g. a locally inertial
frame). At a more fundamental level, the vanishing of $\vec{G}$ stems
from an aspect of frame-dragging which, albeit well known, is commonly
overlooked. First, recall that the dragging of the compass of inertia
can be cast as the fact that, close to a moving of rotating body,
a system of axes that is fixed with respect to inertial frames at
infinity, actually rotates with respect to a locally nonrotating frame
(i.e., to a Fermi-Walker transported frame, physically realized by
the spin axes of local guiding gyroscopes, defining the local ``compass
of inertia''). Such rotation generates not only a gravitomagnetic
(or Coriolis) field $\vec{H}$, but also a contribution to the gravitoelectric
field $\vec{G}$, as originally noticed by Thirring-Lense \cite{LenseThirringTranslated}
and Einstein \cite{EinsteinMeaning} considering the interior field
of a spinning shell. Actually, in vacuum, by the first of Eqs. \eqref{eq:GFieldEq},
$\vec{H}$ cannot exist without $\vec{G}$. A familiar example is
a rigidly rotating frame with angular velocity $\vec{\Omega}$ in
flat spacetime, where both a Coriolis $\vec{H}=2\vec{\omega}$ and
a centrifugal field $\vec{G}=\gamma^{-2}\vec{\omega}\times(\vec{r}\times\vec{\omega})$
arise. {[}Here $\vec{\omega}=\gamma^{2}\vec{\Omega}$ is the rotating
observers' vorticity, cf. Eqs. \eqref{eq:GEM Fields Cov}, and $\gamma=(1-\Omega^{2}r^{2})^{-1/2}$.{]}
In the presence of matter, the situation changes, in that it is possible
for the gravitational attraction to exactly balance the centrifugal
action caused by the frame's rotation, $16\pi\rho=\vec{\omega}^{2}$,
so that \eqref{eq:DivGBG} be obeyed. This is the case of the van
Stockum rotating dust cylinder, where, as we have seen in Sec. \ref{subsec:The-van-Stockum},\emph{
from the point of view of the comoving coordinate system}, there is
an exact cancellation between the cylinder's gravitational attraction
and the centrifugal forces created by the frame's rotation, resulting
in a vanishing gravitoelectric field (so that particles can remain
at rest in such coordinates, while being geodesic). The situation
for the BG metric \eqref{eq:BGmetric} is formally similar, but with
the additional subtlety that, in this case, the associated reference
frame is not rotating with respect to inertial frames at infinity:
as shown in Sec. \ref{subsec:OriginGM}, it is the NUT rods along
the axis $r=0$ that drag the compass of inertia, endowing the rest
observers with a vorticity $\vec{\omega}=\vec{H}/2$ given by Eq.
\eqref{eq:HBG}. In other words, causing the dust's rest frame to
rotate with respect to the \emph{local} compass of inertia. It is
the repulsive action of this (purely general relativistic) rotation
with respect to the local compass of inertia that allows Eq. \eqref{eq:DivGBG}
to be obeyed, and the dust to remain static.

\subsubsection{Origin of the claimed \textquotedblleft flat rotation curves\textquotedblright}

The fact that in the BG solution the dust is static with respect to
the asymptotic inertial frame at infinity has seemingly gone unnoticed
in the literature \cite{BG,Crosta2018,RuggieroBG} (in spite of some
authors having realized that it is rigid \cite{RuggieroBG,Astesiano_Rigid_Rotation},
or tangent to a Killing vector field \cite{BG}, while, surprisingly,
not immediately ruling it out as a viable galactic model). It is actually
claimed\footnote{To back the claim that the velocity relative to the ZAMOs represents
a velocity ``an asymptotic observer at rest with respect to the rotation
axis'', van Stockum's paper \cite{Stockum1938} and the textbook
by Stephani \emph{et al} \cite{StephaniExact} are cited; however,
nowhere in those references such an incorrect claim is made.} \cite{BG,Crosta2018} that the dust rotates with respect to ``an
asymptotic observer at rest with respect to the rotation axis'' with
a velocity consistent with the observed Milky Way rotation curve,
using, as reference observers, the ZAMOs. These, however, as we have
seen in Sec. \ref{subsec:ZAMOs}, are unsuitable for such purposes
when (non-negligible) frame-dragging is present. In particular, it
is \emph{false} that they are at rest with respect to the axis's asymptotic
rest frame (or to any observer at rest with respect to the axis, by
any cogent definition of ``rest''). We have exemplified with the
Kerr, van Stockum, and spinning cosmic string solutions (Secs. \ref{subsec:Kerr-spacetime}
and \ref{subsec:The-van-Stockum}-\ref{subsec:Cosmic-string}) the
absurdities one runs into by making this confusion. For the BG metric,
as shown in Sec. \ref{subsec:The-dust-is-static}, the frame rigidly
fixed to the axis' asymptotic rest frame is precisely the coordinate
system in \eqref{eq:BGmetric}, where the dust is at rest (forming
actually the only Killing congruence of worldlines globally defined
in this spacetime). It is the ZAMOs that rotate with angular velocity
\eqref{eq:OmegaZamoBG} relative to this frame; since $\Omega_{{\rm ZAMO}}(r,z)$
is not constant, they are moreover a shearing congruence, cf. Eq.
\eqref{eq:ShearZAMO}, so they cannot actually be at rest in \emph{any}
rigid frame, and, as discussed in Secs. \ref{subsec:Generalization-of-IAU}
and \ref{subsec:ZAMOs}, their connecting vectors do not define directions
fixed to any distant reference objects (or to inertial frames at infinity).
The velocity of the dust relative to the ZAMOs, as computed in \cite{BG,Crosta2018,RuggieroBG}
{[}and can be obtained from Eqs. \eqref{eq:OmegaZamoBG} and \eqref{eq:velwrtZamo},
with $\Omega_{{\rm circ}}=0$, $g^{00}=-1${]}
\begin{equation}
v_{{\rm rZ}}=v_{{\rm rZ}}^{\hat{\phi}}=\frac{N(r,z)}{r}\ ,\label{eq:vZAMOBG}
\end{equation}
is thus actually just \emph{minus} the velocity of the ZAMOS with
respect to the static frame anchored to inertial frames at infinity.
Suggesting that \eqref{eq:vZAMOBG} represents a galactic rotation
curve, is thus nonsense.

The use of the ZAMOs has also been advocated based on the fact that
their worldlines are orthogonal to the $t=const.$ hypersurfaces \cite{BG,Crosta2018,RuggieroBG},
and that they are said to be locally nonrotating \cite{Crosta2018,RuggieroBG}
--- which is correct, but is however misinterpreted: the fact that
they are hypersurface orthogonal means just that they have no vorticity,
i.e., do not \emph{locally} rotate relative to the \emph{local} compass
of inertia (see Sec. \ref{subsec:ZAMOs}); it tells us nothing about
the motion with respect to distant reference objects (or to the axis'
asymptotic rest frame). They are sometimes also said that to be nonrotating
with respect to the ``local geometry'' \cite{Bardeen1970ApJ,Misner:1974qy,Semerak_Stationary,PaperDragging}
--- but in the sense of measuring no Sagnac effect (see Sec. \ref{subsec:ZAMOs}).
In \cite{RuggieroBG} it is also asserted (referring to \cite{BardeenPressTeukolsky})
that ``their motion compensates, as much as possible, the dragging
effect''; this is, however, a misleading statement: the ZAMOs are
crucially affected by frame-dragging, precisely in that a nonvanishing
$g_{0\phi}$ implies that they describe circular motions about the
$z$- axis, instead of being at rest in the static frame (as would
be the case if $g_{0\phi}=0$). In \cite{RuggieroBG}, moreover, the
static observers were discarded as reference observers for rotation
curves based on the misguided argument that they not always exist,
citing the example of the ergosphere in the Kerr spacetime. We note
that, as shown in Sec. \ref{subsec:CTC's-and-the-forbidden}, actually
in the BG spacetime the exact opposite occurs: static observers are
well defined everywhere, whereas the ZAMOs are possible only for $r>N$,
i.e., outside the pencil-shaped region in Fig. \ref{fig:Forbidden}.

We can summarize the BG galactic model, and the basic misconceptions
at its origin, as follows: 
\begin{itemize}
\item the dust is static with respect to the asymptotic inertial frame (so
its actual rotation curve is $v_{{\rm c}}=0$). 
\item The metric contains unphysical singularities along the axis --- NUT
rods of charges \eqref{eq:QNUTBG}, plus a spinning cosmic string
--- whose frame-dragging effects hold the dust static, are \emph{solely}
responsible for the artificially large gravitomagnetic potential $\boldsymbol{\mathcal{A}}=N(r,z){\bf d}\phi$
and field $\vec{H}=\tilde{\nabla}\times\vec{\mathcal{A}}$, and drag
(via $\boldsymbol{\mathcal{A}}$) the ZAMOs, causing them to rotate
relative to inertial frames at infinity with angular velocity \eqref{eq:OmegaZamoBG}
about the $z$-axis, thereby making them unsuitable as reference observers
for rotation curves. 
\item The flat rotation curves obtained in \cite{BG,Crosta2018} are but
an artifact of such an invalid choice of reference frame --- being
just minus the velocity of the ZAMOs with respect \emph{to the rigid
asymptotic inertial frame} that corresponds to the generalization
of the IAU reference system for this spacetime. 
\end{itemize}
We conclude this section with the following remarks: (i) for more
realistic galactic models (free from the BG axial singularities and
the artificially large gravitomagnetic potential $\boldsymbol{\mathcal{A}}$
they produce, but still axisymmetric), the ZAMOs \emph{would be} suitable
reference observers for rotation curves since, as is well known, and
recently shown explicitly in \cite{Ciotti:2022inn}, the gravitomagnetic
potential produced by galaxies is negligible, given their relatively
low rotation speeds ($v\simeq220{\rm km}{\rm s}^{-1}=7\times10^{-4}c$,
for the Milky Way). Hence, therein, the ZAMOs nearly coincide with
the static observers, that are rigidly fixed to inertial frames at
infinity. (ii) If one entertains singularities along the $z$-axis,
with extension comparable or larger than the galactic diameter, then
infinitely many rotation curves can be produced, including flattened
ones with respect to the correct (static) frame. For instance, an
infinite line mass along the $z$-axis, described by the Levi-Civita
solution, leads to circular geodesics with constant velocity relative
to the static observers (see e.g. \cite{Cilindros}, Sec. 5.1.1);
a line mass segment, or a pair of black holes along the $z$-axis,
held apart, at a distance much larger than the galactic diameter,
by a Misner string or by the repulsion of NUT charges, also produce
a flattened velocity profile, as recently shown in \cite{GovaertsGMDipole}.
Likewise, the ``homogeneous'' linearized theory solutions for $\boldsymbol{\mathcal{A}}$
recently claimed in \cite{AstesianoRuggieroPRDII} to yield flat velocity
profiles, have been shown in \cite{LasenbyHobsonBarker} to actually
be sourced by axial singularities. (iii) It has been claimed, in the
literature \cite{CooperstockTieu,CarrickCooperstock,BG,RuggieroBG}
on the BG and akin models, that nonlinear gravitomagnetic effects
can, partially or totally, replace the role of dark matter in flattening
the rotation curves; for the metric \eqref{eq:BGmetric}, however,
the only nonlinear contribution to the field equations \eqref{eq:GFieldEq}-\eqref{eq:HFieldEq},
is the term $\vec{H}^{2}$ in \eqref{eq:GFieldEq}, which, as seen
in Sec. \ref{subsec:Staticity-mechanism}, has the opposite effect,
countering (actually exactly canceling out) the attractive effect
of the dust mass density $\rho$.

\section{Conclusions}

In this work we explored the extension of the IAU reference system
to the exact theory preserving some of its most important features,
namely defining fixed directions with respect to distant reference
inertial objects (``stellar'' or ``quasar'' compass); we found
it to be possible in spacetimes admitting shear-free observer congruences,
which are also asymptotically vorticity and acceleration-free. We
obtained the general form of the metric in the coordinates adapted
to such observers. It yields natural results in the examples considered:
the (star-fixed) Boyer-Lindquist coordinate system in the Kerr spacetime,
the (not so well known) star-fixed coordinate system for the van Stockum
cylinder, the rotating cosmic string's inertial rest frame, and the
coordinate system comoving with the cosmological fluid in the FLRW
models. We then debunked the BG galactic model, which exemplifies
the grave consequences of failing to set up an appropriate reference
frame: a static dust, held in place by unphysical singularities along
the symmetry axis, which has been confused with a dust that rotates
with a velocity profile consistent with that of the Milky Way, due
to an unsuitable choice of reference observers --- the ZAMOs, which,
due to the frame-dragging effects created by the singularities at
the axis, undergo circular motions with respect to inertial frames
at infinity, making it seem that it is the dust that is rotating.
In the above mentioned well-known solutions, we further exemplified
the confusions one may run into by using the ZAMOs as reference observers
in spacetimes where frame-dragging is important: one would conclude
that Kerr black holes do not rotate after all, that circular geodesics
would exist around a cosmic string (where there is no gravitational
attraction), or massively underestimate the actual rotation velocity
of the van Stockum cylinder with respect to the distant stars. The
case of the BG model serves also as a reminder that singularities
should not be overlooked, as they may dramatically impact the behavior
of the whole solution.

In a growing number of works, based both on linearized theory \cite{RuggieroGalatic,AstesianoRuggieroPRDII}
and exact models \cite{CooperstockTieu,CarrickCooperstock,BG,Crosta2018,RuggieroBG},
it has been asserted that gravitomagnetic effects can have a significant
impact \cite{RuggieroBG,RuggieroGalatic,AstesianoRuggieroPRDII,BG},
or even totally account for the galactic flat rotation curves \cite{Crosta2018,CooperstockTieu,CarrickCooperstock}.
In the framework of a weak field slow motion approximation, this has
been shown to be impossible, and such claims addressed, in \cite{Ciotti:2022inn,LasenbyHobsonBarker,GlampedakisJones_Pitfalls}.
It has however been argued \cite{CooperstockTieu,CarrickCooperstock,BG,RuggieroBG}
that in the exact theory this is possible, due to nonlinear effects
not captured in linearized theory (and not manifest ``locally''
\cite{CooperstockTieu,CarrickCooperstock,BG}), basing such claims
on the BG model, or variants of it. Our exact analysis shows such
claims to be also unfounded, the conclusion extending to the akin
models\footnote{The akin Cooperstock-Tieu model \cite{CooperstockTieu} has been shown
in \cite{Korzynski2007} to be also plagued with singularities, in
this case not along the symmetry axis, but in the equatorial plane.
(The erroneous use of the ZAMOs as reference observers, however, went
unnoticed therein.) } in \cite{CooperstockTieu,CarrickCooperstock,RuggieroBG} (where the
same misguided choice of reference observers is made). The exact approach
proves also useful in determining the nature of the sources: we were
able to easily identify a pair of oppositely charged rods of gravitomagnetic
monopoles as the source of the gravitomagnetic field $\vec{H}$; in
the far field region, however, this field is, to leading (i.e., dipole)
order, indistinguishable from that of a spinning source.\footnote{It could not, in principle, be achieved either by consider higher
order multipole moments, in the likes of e.g. Geroch-Hansen \cite{GerochMultipole,HansenMultipole},
Thorne \cite{ThorneMultipole} or Gürlebeck \cite{GurlebeckMultipole}
expansions, since not only such methods require asymptotic flatness
and an isolated source (see however \cite{MayersonMultipoles}, where
this limitation was recently asserted to be lifted), as infinite multipole
moments would be needed to reconstruct the gravitomagnetic field of
the rods.}

\subsection*{Acknowledgments}

We thank C. Will, A. Barnes, N. Van den Bergh, S. Klioner, D. Grumiller,
and M. Crosta for correspondence and very useful discussions, and
the anonymous Referee for valuable suggestions. L.F.C. and J.N. were
supported by FCT/Portugal through projects UIDB/MAT/04459/2020 and
UIDP/MAT/04459/2020. F.F.A. was supported by the Research Vice Rectory
of the University of Costa Rica.

\appendix

\section{Electromagnetic analogue of the BG solution\label{sec:Electromagnetic-analogue-of}}

The Balasin-Grumiller solution \eqref{eq:BGmetric} has two direct
electromagnetic analogues, corresponding to the two types of sources
for its gravitomagnetic monopoles --- NUT charges, or semi-infinite
spinning strings.

\subsection{An infinite solenoid plus a pair of rods with opposite magnetic charges\label{subsec:EM1}}

\begin{figure}
\includegraphics[width=0.5\textwidth]{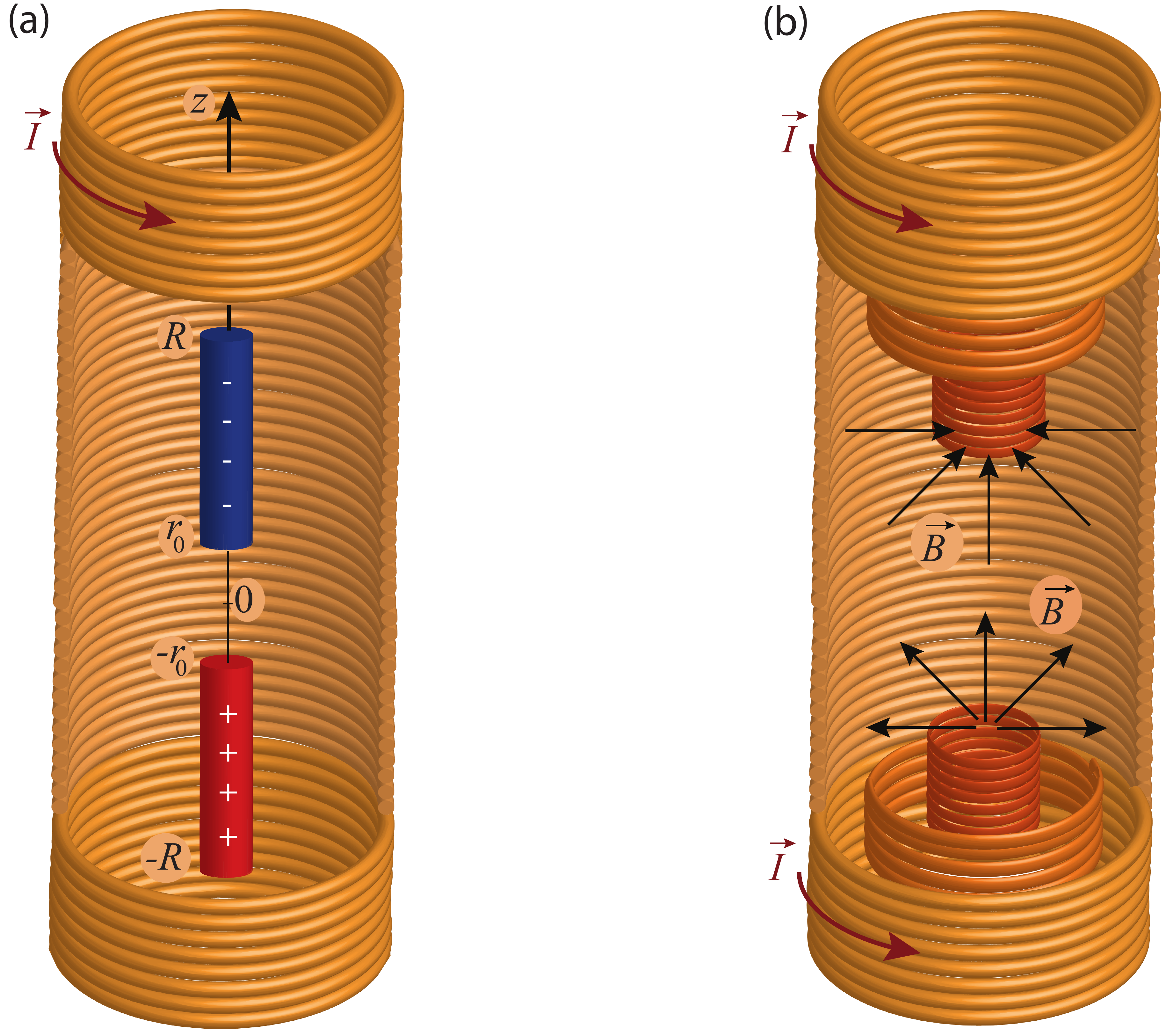}

\caption{\label{fig:Solenoids}Electromagnetic analogues of the BG solution:
(a) a thin infinite solenoid enclosing two rods of opposite magnetic
charges $\mp\lambda_{{\rm M}}$ per unit length; (b) a thin infinite
solenoid enclosing two continuous (upper and lower) sets of semi-infinite
thin solenoids, whose endpoints span the same location of the rods
in (a), and where their magnetic moment per unit length $\mathfrak{m}_{{\rm set}}$
is such that $d\mathfrak{m}_{{\rm set}}/dz=\mp\lambda_{{\rm M}}$.
The magnetic field $\vec{B}$ in the two settings is the same, except
along the axis $r=0$, where the setting in (b) contains the Dirac-delta
type term in Eq. \eqref{eq:Bsolenoids}. Identifying the appropriate
parameters, the exterior magnetic potential and field also exactly
match (up to the function $e^{\nu(r,z)}$, in the case of $\vec{H})$
the gravitational counterparts $\boldsymbol{\mathcal{A}}$ and $\vec{H}$
of the BG solution everywhere outside the axis (where the solution
is not defined).}
\end{figure}

Consider, as depicted in Fig. \ref{fig:Solenoids}(a), a pair of magnetically
(uniformly) charged rods, of opposite charges, located along the $z$-axis,
the positive rod at $-r_{0}>r>-R$, and the negative rod at $r_{0}<r<R$.
Let $\lambda_{{\rm M}}$ and $-\lambda_{{\rm M}}$, respectively,
be their uniform magnetic charges per unit $z$-length. The magnetic
scalar potential $\Psi(r,z)$ produced by the rods is 

\begin{align*}
\Psi(r,z) & =\lambda_{{\rm M}}\left[\int_{-R}^{-r_{0}}\frac{1}{d_{z'}}dz'-\int_{r_{0}}^{R}\frac{1}{d_{z'}}dz'\right]\\
 & =\lambda_{{\rm M}}\left[\ln\left[\frac{r_{0}-z+d_{r_{0}}}{R-z+d_{R}}\right]+\ln\left[\frac{r_{0}+z-d_{-r_{0}}}{R+z-d_{-R}}\right]\right]\ ,
\end{align*}
where 
\[
d_{z'}\equiv\sqrt{(z'-z)^{2}+r^{2}}\ ,
\]
and the distances $d_{R}$, $d_{-R}$, $d_{r_{0}}$, $d_{-r_{0}}$
are defined by Eqs. \eqref{eq:dR}-\eqref{eq:dr0}. The magnetic field
$\vec{B}_{{\rm rods}}=-\nabla\Psi$ reads 
\begin{align}
\vec{B}_{{\rm rods}} & =\lambda_{{\rm M}}\left[\frac{1}{d_{r_{0}}}+\frac{1}{d_{-r_{0}}}-\frac{1}{d_{R}}-\text{\ensuremath{\frac{1}{d_{-R}}}}\right]\vec{e}_{z}\nonumber \\
 & +\frac{\lambda_{{\rm M}}}{r}\left\{ \frac{z-R}{d_{R}}+\frac{z+R}{d_{-R}}-\frac{z-r_{0}}{d_{r_{0}}}-\frac{z+r_{0}}{d_{-r_{0}}}\right\} \vec{e}_{r}\ ,\label{eq:Brods}
\end{align}
which, apart from the factor $e^{-\nu}$, has the same form as the
gravitomagnetic field \eqref{eq:HBG} of the BG solution, identifying
$\lambda_{{\rm M}}\leftrightarrow V_{0}/2$; their covariant counterparts
actually exactly match: 
\[
(B_{{\rm rods}})_{i}\stackrel{\lambda_{{\rm M}}\rightarrow V_{0}/2}{=}H_{i}\ .
\]
With such an identification, it follows also that the total charge
in each rod is $Q_{{\rm M}}=\mp V_{0}(R-r_{0})/2$, exactly matching
the NUT charge obtained in Eq. \eqref{eq:QNUTBG}. This field has,
of course, also the same form as the electric field produced by a
pair of rods of opposite electric charges, identifying $\lambda_{{\rm M}}$
with the electric charge density $\lambda$, cf. Eqs. (2)-(3) of \cite{ZuoEMRod}.\footnote{In \cite{ZuoEMRod} the field of a single rod along the $x$ axis
is computed in the $x_{O}y$ plane; to relate it with the rods in
Fig. \ref{fig:Solenoids}(a), one needs to substitute therein $\{x,y\}\rightarrow\{r,z\}$,
and identify $\{a,b\}\leftrightarrow\{r_{0},R\}$ (upper rod) or $\{a,b\}\leftrightarrow\{-r_{0},-R\}$
(lower rod).} At every point outside the rods, where $\nabla\cdot\vec{B}=0$ holds,\footnote{Admitting the existence of magnetic charges requires the modification
of Maxwell's equations $\nabla\cdot\vec{B}=4\pi\rho_{{\rm M}}$, where
$\rho_{{\rm M}}$ is the magnetic charge density, which is nonzero
along the rods.} this field can also be cast as the curl of a magnetic vector potential
$\vec{A}_{{\rm rods}}$: $\vec{B}=\nabla\times\vec{A}_{{\rm rods}}$,
whose corresponding 1-form reads 
\[
{\bf A}_{{\rm rods}}=\lambda_{{\rm M}}(d_{r_{0}}+d_{-r_{0}}-d_{R}-d_{-R}){\bf d}\phi\ .
\]
Hence, if one considers the rods to be inside a thin infinite solenoid
of magnetic moment per unit length $\mathfrak{m}$, whose exterior
magnetic potential 1-form reads ${\bf A}_{{\rm sol}}=2\mathfrak{m}{\bf d}\phi$,
we have ${\bf A}={\bf A}_{{\rm sol}}+{\bf A}_{{\rm rods}}$, 
\begin{equation}
{\bf A}=\left[2\mathfrak{m}+\lambda_{{\rm M}}(d_{r_{0}}+d_{-r_{0}}-d_{R}-d_{-R})\right]{\bf d}\phi\ ,\label{eq:AEMAnalogue}
\end{equation}
which has exactly the same form as the gravitomagnetic potential $\boldsymbol{\mathcal{A}}=N{\bf d}\phi$
of the BG solution, cf. Eq. \eqref{eq:N}, identifying $\{\mathfrak{m},\lambda_{{\rm M}}\}\leftrightarrow\{V_{0}(R-r_{0})/2,V_{0}/2\}$.

\subsection{An infinite solenoid plus an array of semi-infinite solenoids\label{subsec:EM2}}

The electromagnetic field \eqref{eq:AEMAnalogue} can be reproduced
everywhere outside the axis without invoking magnetic charges. Consider
a semi-infinite solenoid along the $z$-axis with ``tip'' (i.e.,
endpoint) at $z=z_{{\rm tip}}$, and extending to $+\infty$. The
magnetic field it produces is (cf. \cite{KitanoSolenoid}) 
\begin{equation}
\vec{B}_{{\rm s}}(\vec{\re},\vec{\re}_{{\rm tip}})=-\frac{\mathfrak{m}_{{\rm s}}(\vec{\re}-\vec{\re}_{{\rm tip}})}{|\vec{\re}-\vec{\re}_{{\rm tip}}|^{3}}+4\pi\mathfrak{m}_{{\rm s}}\delta^{2}(\rc)\Theta(z-z_{{\rm tip}})\vec{e}_{z}\ ,\label{eq:FieldSemi}
\end{equation}
where $\mathfrak{m}_{{\rm s}}$ is the solenoid's magnetic moment
per unit length, $\delta^{2}(\rc)\equiv\delta(x)\delta(y)$ is the
2-dimensional delta function, $\Theta(x)$ the Heaviside function,
and $\re\equiv\sqrt{\rc^{2}+z^{2}}=\sqrt{x^{2}+y^{2}+z^{2}}$ is the
spherical radial coordinate. Everywhere outside the solenoid, this
is a monopolar field identical to that of a magnetic point charge
$Q_{{\rm M}}=-\mathfrak{m}_{{\rm s}}$ placed at $\vec{\re}_{{\rm tip}}$
(see Fig. 2 in \cite{KitanoSolenoid}); the fields are distinguishable
only in the second term of \eqref{eq:FieldSemi} that arises inside
the solenoid. The field for the case where the solenoid extends from
$-\infty<z<z_{{\rm tip}}$ is obtained from \eqref{eq:FieldSemi}
by changing the sign of the first term, and replacing $\Theta(z-z_{{\rm tip}})$
by $\Theta(z_{{\rm tip}}-z)$.

Consider now two sets of semi-infinite solenoids, an upper one extending
from $+\infty$ with endpoints $z_{{\rm tip}}\in[r_{0},R]$, and a
lower one extending from $-\infty$ with endpoints $z_{{\rm tip}}\in[-R,-r_{0}]$,
see Fig. \ref{fig:Solenoids}(b). Let $n$ be the number of endpoints
per unit $z$-length; if it is such that $n\mathfrak{m}_{{\rm s}}=\lambda_{{\rm M}}$,
so that, in the continuous limit $n\rightarrow\infty$, $\mathfrak{m}_{{\rm s}}\rightarrow0$,
the magnetic field produced by such setting is, everywhere \emph{outside
the axis} $r=0$, the same as that of the rods in Fig. \ref{fig:Solenoids}(a).
This can be checked by computing explicitly $\vec{B}=\int_{r_{0}}^{R}d\vec{B}_{{\rm up}}+\int_{-R}^{-r_{0}}d\vec{B}_{{\rm low}}$,
where $d\vec{B}_{{\rm up}}(\vec{\re},\vec{\re}\,')=n\vec{B}_{{\rm s}}(\vec{\re},\vec{\re}\,')$,
and so
\begin{align*}
d\vec{B}_{{\rm up}}(\vec{\re},\vec{\re}\,') & =\lambda_{{\rm M}}\left[-\frac{(\vec{\re}-\vec{\re}\,')}{|\vec{\re}-\vec{\re}\,'|^{3}}+4\pi\delta^{2}(\rc)\Theta(z-z')\vec{e}_{z}\right]dz';\\
d\vec{B}_{{\rm low}}(\vec{\re},\vec{\re}\,') & =\lambda_{{\rm M}}\left[\frac{(\vec{\re}-\vec{\re}\,')}{|\vec{\re}-\vec{\re}\,'|^{3}}+4\pi\delta^{2}(\rc)\Theta(z'-z)\vec{e}_{z}\right]dz'\ ;
\end{align*}
\begin{align}
 & \vec{B}=\lambda_{{\rm M}}\left[\int_{-R}^{-r_{0}}-\int_{r_{0}}^{R}\right]\left[\frac{z-z'}{d_{z'}^{3}}\vec{e}_{z}+\frac{r}{d_{z'}^{3}}\vec{e}_{r}\right]dz'\nonumber \\
 & +4\pi\lambda_{{\rm M}}\delta^{2}(\rc)\left[\int_{-R}^{-r_{0}}\Theta(z'-z)+\int_{r_{0}}^{R}\Theta(z-z')\right]dz'\vec{e}_{z}\nonumber \\
 & =\vec{B}_{{\rm rods}}+4\pi\lambda_{{\rm M}}\delta^{2}(\rc)\sum_{\pm}\left[\mathcal{R}(-r_{0}\pm z)-\mathcal{R}(-R\pm z)\right]\vec{e}_{z},\label{eq:Bsolenoids}
\end{align}
where $\mathcal{R}(x):=\{0,\ x<0;\,x,\ x>0\}$ is the ramp function
and, in the first line, we noted that $\vec{\re}-\vec{\re}\,'=(z-z')\vec{e}_{z}+r\vec{e}_{r}$
and $|\vec{\re}-\vec{\re}\,'|=d_{z'}^{3}$. The field thus differs
from that of the magnetically charged rods in Eq. \eqref{eq:Brods}
only in the Dirac delta term along the axis. Notice that this term
ensures the vanishing of the flux $\int_{\mathcal{S}}\vec{B}\cdot\vec{d}\mathcal{S}$
along any closed surface: consider, for simplicity, a cylindrical
surface $\mathcal{S}$ enclosing one of the rods but not the other,
i.e., its bottom and top bases lie at $-r_{0}<z_{{\rm b}}<r_{0}$
and $z_{{\rm t}}>R$, respectively. An integration analogous to that
in Eq. \eqref{eq:QNUTBG} yields $\int_{\mathcal{S}}\vec{B}_{{\rm rods}}\cdot\vec{d}\mathcal{S}=-4\pi\lambda_{{\rm M}}(R-r_{0})$,
and thus 
\[
\int_{\mathcal{S}}\vec{B}\cdot\vec{d}\mathcal{S}=4\pi\lambda_{{\rm M}}\left[\mathcal{R}(z_{{\rm t}}-r_{0})-\mathcal{R}(z_{{\rm t}}-R)-R+r_{0}\right]=0\ .
\]
This ensures consistency with the Maxwell equation $\nabla\cdot\vec{B}=0$
(which must hold in the absence of magnetic charges) since, via the
Stokes theorem, $\int_{\mathcal{S}}\vec{B}\cdot\vec{d}\mathcal{S}=\int_{\mathcal{V}}\nabla\cdot\vec{B}d\mathcal{V}$,
for $\mathcal{S}=\partial\mathcal{V}$.

If the set of semi-infinite solenoids is enclosed in an infinite solenoid
as depicted in Fig. \ref{fig:Solenoids}(b), then, outside the axis,
the magnetic vector potential $\vec{A}$ is given by \eqref{eq:AEMAnalogue},
the same as in the setting in Fig. \ref{fig:Solenoids}(a). Since,
as discussed in Secs. \ref{subsec:Singularities} and \ref{subsec:OriginGM},
the BG solution is not defined along the axis, both the settings (a)
and (b) can be regarded as its electromagnetic analogues.

\section{Kretschmann scalar of the BG solution\label{sec:Kretschmann-scalar-BG}}

The Kretschmann scalar, $R^{\alpha\beta\gamma\delta}R_{\alpha\beta\gamma\delta}\equiv{\bf R}\cdot{\bf R}$,
of the dust solution \eqref{eq:BGmetric}, reads, 
\begin{widetext}
\begin{align*}
{\bf R}\cdot{\bf R} & =\frac{e^{-2\nu}V_{0}^{2}}{4r^{4}}\left\{ \frac{1}{4}{\Delta_{5}}^{2}\left[\frac{1}{2}{V_{0}}^{2}{\Delta_{1}}\left(-\frac{3}{8}{V_{0}}^{2}{\Delta_{1}}^{3}+6{\Delta_{1}}+8r{\Delta_{2}}-4r{\Delta_{4}}\right)-16\right]+{\Delta_{1}}^{2}\left[\frac{1}{4}{V_{0}}^{2}r^{2}\left({\Delta_{2}}+{\Delta_{4}}\right)^{2}-2\right]\right.\\
 & \left.-\frac{1}{2}r{V_{0}}^{2}{\Delta_{1}}^{3}{\Delta_{2}}+4r{\Delta_{1}}{\Delta_{2}}+2r{\Delta_{3}}{\Delta_{5}}\left[2-\frac{3}{4}{V_{0}}^{2}{\Delta_{1}}^{2}\right]+\frac{1}{16}{V_{0}}^{2}{\Delta_{5}}^{4}\left[4-\frac{3}{4}{V_{0}}^{2}{\Delta_{1}}^{2}\right]\right.\\
 & \left.-\frac{1}{64}{V_{0}}^{4}{\Delta_{1}}^{6}+\frac{3}{4}{V_{0}}^{2}{\Delta_{1}}^{4}-4r^{2}\left[\frac{1}{2}{\Delta_{2}}^{2}+{\Delta_{3}}^{2}\right]+\frac{1}{2}r{V_{0}}^{2}{\Delta_{3}}{\Delta_{5}}^{3}-2r^{2}{\Delta_{4}}^{2}-\frac{1}{64}{V_{0}}^{4}{\Delta_{5}}^{6}\right\} 
\end{align*}
where
\begin{align*}
 & \Delta_{1}=-\frac{r}{{d_{-R}}}+\frac{r}{{d_{-r_{0}}}}-\frac{r}{{d_{R}}}+\frac{r}{{d_{r_{0}}}}\ ;\qquad\ \Delta_{5}=-\frac{R+z}{{d_{-R}}}+\frac{{r_{0}}+z}{{d_{-r_{0}}}}-\frac{z-R}{{d_{R}}}+\frac{z-{r_{0}}}{{d_{r_{0}}}}\\
 & \Delta_{2}=\frac{r^{2}}{{d_{-R}}^{3}}-\frac{1}{{d_{-R}}}-\frac{r^{2}}{{d_{-r_{0}}}^{3}}+\frac{1}{{d_{-r_{0}}}}+\frac{r^{2}}{{d_{R}}^{3}}-\frac{1}{{d_{R}}}-\frac{r^{2}}{{d_{r_{0}}}^{3}}+\frac{1}{{d_{r_{0}}}}\\
 & \Delta_{3}=\frac{r(R+z)}{{d_{-R}}^{3}}-\frac{r({r_{0}}+z)}{{d_{-r_{0}}}^{3}}+\frac{r(z-R)}{{d_{R}}^{3}}-\frac{r(z-{r_{0}})}{{d_{r_{0}}}^{3}}\\
 & \Delta_{4}=\frac{(R+z)^{2}}{{d_{-R}}^{3}}-\frac{1}{{d_{-R}}}-\frac{({r_{0}}+z)^{2}}{{d_{-r_{0}}}^{3}}+\frac{1}{{d_{-r_{0}}}}+\frac{(z-R)^{2}}{{d_{R}}^{3}}-\frac{1}{{d_{R}}}-\frac{(z-{r_{0}})^{2}}{{d_{r_{0}}}^{3}}+\frac{1}{{d_{r_{0}}}}
\end{align*}
and the function $\nu(r,z)$ in \eqref{eq:BGmetric} was eliminated
using the components $R_{rr}=R_{rz}=R_{zz}=0$ of the Einstein field
equations $R_{\alpha\beta}-Rg_{\alpha\beta}/2=8\pi\rho u_{\alpha}u_{\beta}$,
for dust of 4-velocity $u^{\alpha}=\delta_{0}^{\alpha}$. A computer-ready
\emph{Mathematica} version of this expression is provided in the Supplemental
Material \cite{EPAPSPRD}. 
\end{widetext}

\appendix

\bibliographystyle{utphys}
\bibliography{RefarXivv3}

\end{document}